\documentclass[range,a4paper]{ar2eGN}
\usepackage{ARAstroBib}
\usepackage{amsmath}
\usepackage[pdftex]{graphicx}

\def\Msun{M$_\odot$}

\begin{document}

\input epsf.tex    

\input psfig.sty

\jname{..}
\jyear{2000}
\jvol{}
\ARinfo{1056-8700/97/0610-00}
\def\lesssim{\mathrel{\hbox{\rlap{\hbox{\lower4pt\hbox{$\sim$}}}\hbox{$<$}}}}
\def\gtrsim{\mathrel{\hbox{\rlap{\hbox{\lower4pt\hbox{$\sim$}}}\hbox{$>$}}}}

\title{Observational Clues to the Progenitors of Type Ia Supernovae}

\markboth{Maoz, Mannucci, \& Nelemans}{Type Ia Supernova Progenitors}

\author{Dan Maoz 
\affiliation{School of Physics and Astronomy, Tel-Aviv University} 
Filippo Mannucci\affiliation{INAF, Osservatorio Astrofisico di Arcetri}
Gijs Nelemans\affiliation{Department of Astrophysics/IMAPP, Radboud
  University Nijmegen\\Institute for Astronomy, KU Leuven}
}


\begin{abstract}
Type-Ia supernovae (SNe~Ia) are important
distance indicators, element factories, cosmic-ray accelerators,
kinetic-energy sources in galaxy evolution, and
endpoints of stellar binary evolution. It has long been clear that a SN Ia
must be the runaway thermonuclear 
explosion of a degenerate carbon-oxygen stellar
core, most likely a white dwarf (WD). However,
the specific progenitor systems of SNe~Ia, and the processes that lead to their
ignition, have not been identified. Two broad classes of progenitor
binary systems have long been considered: single-degenerate (SD), in which a
WD gains mass from a non-degenerate star; and double-degenerate (DD),
involving the merger of two WDs. 
New theoretical work has enriched 
these possibilities with some interesting updates and variants. 
We review the significant recent observational 
progress in addressing the progenitor problem. We consider clues 
 that have emerged from the
observed properties of the various proposed progenitor populations,
from studies of their sites -- pre- and post-explosion, from analysis 
of the explosions themselves, and from the measurement of event rates.
The recent nearby and well-studied event, SN~2011fe, 
has been particularly revealing.
The observational results are not yet conclusive, and sometimes prone to 
competing theoretical interpretations. Nevertheless, it appears that
DD progenitors, long considered the underdog option, could be behind
some, if not all, SNe~Ia. We point to some directions that may lead to
future progress.

\end{abstract}

\maketitle

\begin{quote}
  {\it 
  \noindent
  Because I have contemplated explaining
  what I think, not only about the location and movement of
  this light, but also about its substance and origin, and believing
  that I have found an explanation that, for lack  
  of evident contradictions, may well be true,  
    I have finally arrived
  at the belief of being capable of knowing something about this wonder,
  beyond the point where pure conjecture ends.\\
  }
  Letter by Galileo to O. Castelli regarding the {\it Stella Nova} of 1604 
\end{quote}

\section{Introduction}

\subsection{Type-Ia supernovae -- a brief overview} 
\label{sec:overview}

The observations of the supernovae (SNe) of 1572 and 1604, and the
attempts, by Tycho, Kepler, Galileo, and others, to understand their
natures and locations, 
were transformational events in the history of science. 
These attempts can be viewed as  
signaling the beginnings of modern astrophysics.
Among the handful of historical SN events in the Galaxy
recorded  over the past two millenia,
 we now know that at least some were core-collapse (CC) SNe, involving
the explosion of massive ($M\gtrsim 8 M_\odot$) stars, while
others were the thermonuclear explosions of lower-mass stars, now known as  
Type-Ia supernovae (SNe~Ia). 
In modern-day surveys that search for stellar explosions,  
SNe~Ia are often the ones most frequently
discovered, owing mainly to
their large optical luminosities, $\sim 10^{43}$~erg~s$^{-1}$ near
the time of maximum light. 

SNe~Ia are classified
(see review by \citealt{Filippenko97}) by a lack of hydrogen and helium
signatures in their spectrum, and by
kinematically broadened and blueshifted ($\sim 10^3 - 10^4$~km~s$^{-1}$) 
features, mainly of silicon,
iron, and calcium, around the time of maximum light.  
Contrary to CC~SNe (e.g., \citealt{Smartt09}), SNe~Ia are
observed to explode both in young and in old stellar populations. The
spectra and light curves of most SNe~Ia are remarkably uniform.
The light curves rise to
maximum light within  $\sim 15-20$~d, decline by about 3~mag in one
month, and then by a further mag each month, with the more specific
behavior depending on the observed photometric band. Based on the light-curve
shape, optical spectroscopy, and X-ray analysis of SN~Ia remnants,
the observed
luminosity of SNe~Ia is driven by inverse-$\beta$ and electron-capture
radioactive decays, 
first of $^{56}$Ni to $^{56}$Co, and then of $^{56}$Co to stable
$^{56}$Fe, with exponential timescales of 9~d and 114~d, respectively. 
Gamma-rays and positrons from these decays are reprocessed by the
optically thick ejecta into the optical photons that dominate the
luminosity \citep{Colgate69}. 
SNe~Ia are a decisively optical-range phenomenon, with 
$\sim 85\%$ of the luminosity emitted between the $U$
and $I$ bands (e.g. \citealt{Howell09b}).

The tight correlation between the luminosity of SNe~Ia at maximum light
and various measures of light-curve development speed 
(\citealt{Phillips93}, see \citealt{Howell11} for a review) 
have made them excellent cosmological distance
indicators. As such, they provided the first evidence for
acceleration of the cosmic expansion 
\citep{Riess98,Perlmutter99},
and hence for the current,
dark-energy-dominated cosmic mass-energy inventory. 
SNe~Ia continue being a major
cosmographic tool, with future efforts focusing on determining the
evolution, on cosmological timescales, of the dark-energy equation of
state \citep{Howell09b}.
The yet-unknown identity of the progenitor systems of SNe~Ia, the subject of
this article, is a concern for SN~Ia cosmography, given that changes
with cosmic time or environment, in the progenitor populations or in 
some details
of the explosions, could, in principle, lead to systematic errors in
distances deduced based on calibrations of nearby SNe~Ia. 
For example, systematic differences have been reported 
in the residuals in the Hubble diagram,
 depending on host galaxy mass and metallicity 
(e.g., \citealt{DAndrea11,Hayden12,Pan13}).
These differences could impact the measurements of cosmological parameters,
as low-mass and low-metallicity galaxies dominate the production of SNe~Ia
at progressively higher redshifts.

Every normal SN~Ia enriches the interstellar medium with roughly
 0.7~M$_\odot$ of iron (to within a factor of a few), and a similar 
total amount in other elements,
including mainly carbon, oxygen, neon, magnesium, silicon, sulfur, 
argon, and calcium. 
Combined with their large numbers, this means that SNe~Ia 
play an important role in chemical evolution,
and understanding such evolution 
requires knowing the dependences, on cosmic time
and environment, of SN~Ia rates and metal yields 
(e.g. \citealt{Wiersma11}).

The kinetic energy of SN~Ia ejecta also affects gas dynamics and
star formation in galaxies, contributing to the ejection of matter
from galaxies to the intergalactic medium, and thus SNe~Ia are a
factor in galaxy evolution (e.g. \citealt{Dave11b}).  
As is the case
for other SN types, the remnants of SNe~Ia are the likely sites for
the acceleration of cosmic rays up to energies of $\sim 10^{15}$~eV
(e.g. \citealt{Drury12}).  And, as detailed further in
this review, SNe~Ia may or may not signal two important end points in
binary stellar evolution -- accretion onto, and/or the mergers of,
white dwarfs (WDs). The final stages of the latter events are expected
to be the main source of foreground, but also interesting signals, for
space-based gravitational-wave observatories 
(e.g. \citealt{Amaro-Seoane13}).
For all of these reasons,
it is important to understand the physics of all stages of the
development of SNe~Ia, and particularly, to identify their progenitor
systems.

\subsection{Basic physics of SNe~Ia} 
\label{sec:physics}

Physical models of SN~Ia explosions have been reviewed by
\cite{Hillebrandt00}, with updates in \cite{Hillebrandt13}. Here we
only skim the surface.  The energetics and chemical
composition of SN~Ia explosions, as measured in the course of the
explosions and in their remnants, have long led to the inference
\citep{Hoyle60} that they involve the thermonuclear combustion of a
degenerate stellar core. The absence of hydrogen and helium in the
spectrum, and the occurrence of some SNe~Ia in old stellar
populations, indicate that this core is most likely a WD of
carbon-oxygen (C/O) composition.  The burning is partly into iron-group
elements and partly into intermediate-mass elements, with some
residual unburnt carbon and oxygen, as indicated by the stratification
and composition deduced from observations.  The spectral evolution of
SNe~Ia indicates the presence of $\sim 0.6$~M$_\odot$ of radioactive
~$^{56}$Ni~ in a typical explosion, along with a similar amount of
total mass distributed among stable iron-group elements, and among the
lighter elements mentioned above, for a total ejecta mass similar to
the Chandrasekhar WD mass limit, $M_{\rm Ch}=1.44$~M$_\odot$
(e.g. \citealt{Mazzali07}). This chemical makeup and stratification is
also deduced from X-ray analysis of the ejecta of historical SN~Ia
remnants (e.g. \citealt{Badenes06}) such as Tycho's SN of 1572.

Looking at energetics, the nuclear binding energy released by burning
$0.6$~M$_\odot$ of C and O into $^{56}$Ni is $1.1\times
10^{51}$~erg. A comparable mass of C and O burning to
intermediate-mass elements releases additional energy.  The
gravitational binding energy of a WD of mass close to $M_{\rm Ch}$ is
$\sim 0.5\times 10^{51}$~erg.  The thermonuclear energy release is
thus sufficient to unbind the WD, and to give the ejecta the kinetic
energy indicated by the deduced ejecta mass and its observed expansion
velocities, of order $10^4$~km~s$^{-1}$, i.e. $\sim10^{51}$~erg.
Furthermore, the equation of state of the degenerate electron gas in
a WD is just what is needed for an unstable thermonuclear runaway,
once carbon is somehow ignited. Nuclear reaction rates have steep
rising dependences on temperature. A highly degenerate gas, when
heated, does not expand and cool as a classical gas would. As a
result, an explosive burning front can, in principle, quickly consume
a WD.  The carbon ignition threshold can be crossed via an increase
in some combination of pressure and temperature. That increase has
generally been attributed to the accumulation of accreted mass on the
WD, up to the neighborhood of $M_{\rm Ch}$, with a corresponding
decrease in radius. Specifically, the carbon in the core of a
non-rotating C/O WD will ignite at a WD mass of $\sim 1.38 M_\odot$
\citep{Arnett69,Nomoto82}. 
The fact that the ``bomb'' is
always a C/O WD with $M\sim M_{\rm Ch}$ would naturally explain the
limited range of SN~Ia luminosities, which has been so instrumental
for cosmology.

Beyond this simplified picture, however, much about SNe~Ia is still
poorly understood. At the most fundamental level, it is unknown what
are the initial conditions and evolutionary paths that, in practice,
take a WD away from stable equilibrium and
lead it to mass growth, ignition, and explosion. In other words, we
do not know the identity and nature of the systems that explode as
SNe~Ia. This is the ``SN~Ia progenitor problem''.

\section{Progenitor models and open questions}
\label{sec:models}

Progenitor scenarios have traditionally focused on getting a
C/O WD to ignite, by having it approach or exceed 
$M_{\rm Ch}$. This can be done either through accretion from a
non-degenerate star in the single-degenerate (SD) model
\citep{Whelan73}); or by merging two WDs in a tight binary, through loss
of energy and angular momentum to gravitational waves,
in the double-degenerate (DD) model 
(\citealt{Tutukov81,Webbink84,Iben84a}). 
See \citealt{Wang12b} for a recent review. 

\subsection{Single-degenerate models}
\label{sec:SDmodels}

Accretion from a non-degenerate secondary can assume many
configurations. Mass transfer can proceed through Roche-lobe overflow
or through a strong wind from the companion \citep{Li97}. 
 The secondary can be a 
main sequence star (e.g. \citealt{van-den-Heuvel92}), a subgiant
(e.g. \citealt{Han04}), a helium star (\citealt{Tutukov96},
 often called a ``hot subdwarf'', 
i.e. a star that has been
stripped of its hydrogen envelope by binary interaction,
e.g. \citealt{Geier13}), or a red giant (sometimes called a
``symbiotic'' companion, e.g. \citealt{Patat11}).
For each of these cases, different configurations are possible, 
depending on previous common enevelope and mass transfer phases (e.g., \citealt{Wang12b}).

In the SD scenario, the
challenge is to get an accreting WD to actually
grow in mass. In a narrow range of about a factor of three in
 mass accretion rate, centered around $\dot M = 10^{-7}
~M_\odot~{\rm yr}^{-1}$ for a 0.8 \Msun~ WD, and around $\dot M = 5
\times 10^{-7} ~M_\odot~{\rm yr}^{-1}$ for a 1.4 \Msun~ WD, stable
nuclear burning of hydrogen to helium on the WD surface can take
place \citep{Nomoto82}.
WDs in such an accretion mode should resemble an observed class of 
objects known as supersoft X-ray sources (\citealt{van-den-Heuvel92} 
-- see Section~\ref{sec:supersoft}, below). Accretion above this range likely
leads to expansion of the accretor to a red-giant-like configuration,
engulfment of the donor in a common envelope, and thus cessation of
the growth process toward $M_{\rm Ch}$ and toward explosion as a
SN~Ia \citep{Iben84a}.  
In this high-accretion-rate regime, however, 
\cite{Hachisu96, Hachisu99} have proposed that self-regulation 
of the accretion flow
occurs, keeping the mass flow that actually accumulates on the WD at
the steady-burning rate. This is achieved by means of an
optically-thick wind emerging from the accretor, driven by the
luminosity of the nuclear burning on the WD surface.  Questions,
however, have been raised as to whether this does not require too much
fine tuning (e.g., \citealt{Piersanti00,Shen07,Woosley11}), 
or contradictory assumptions
about the spherical symmetry of the accretion flow -- on the one hand
an asymmetric Roche-lobe overflow, on the other hand a spherical
inflow that can be regulated by the spherical outflowing wind
\citep{Hillebrandt00}. In any case, the efficiency of the accretion mode
is limited to the ratio of the steady burning rate to the mass
transfer rate.

Accretion at rates lower than the stable-burning range will lead to
accumulation of the accreted hydrogen in a cold degenerate layer on
the accretor, up to its ignition and
burning in a thermonuclear-runaway ``nova'' eruption \citep{Starrfield72}.
Because hydrogen burning produces more energy per unit mass
than required to eject that mass from the surface of a WD, 
most of the accreted material is
expected to leave the accretor. Over the accumulation timescale,
hydrogen can diffuse inward and mix with the WD's carbon and oxygen
layer. As a result, during the burning, some of the original WD
material is blown away, along with the ashes of the accreted material
(e.g. \citealt{Yaron05}). However, with increasing accretion rates and
WD masses, ignition conditions are reached at lower accumulated mass
and less degenerate conditions, so there may be a parameter range
where mass gain is possible after all.

Recent studies of the long term evolution of WDs accreting hydrogen at
different rates have had some seemingly conflicting
conclusions. \cite{Idan13}, studying accretion at a high rate ($\dot M
= 10^{-6} ~M_\odot~{\rm yr}^{-1}$, i.e. above the steady burning
regime) find, rather than steady hydrogen burning, recurrent
thermonuclear runaway flashes on ten-year timescales, during which
most of the accreted mass is retained. However, the accumulated helium
ash is ignited after several thousand hydrogen flashes, and ejects
essentially all of the accreted mass 
At a lower accretion rate of ($\dot
M = 10^{-7} ~M_\odot~{\rm yr}^{-1}$), some mass is retained, and the
WD mass grows, albeit inefficiently.  \cite{Newsham13} and 
\cite{Wolf13}, 
on the other hand, do find steady burning,
and WD mass growth, but again that a hot helium
layer ignites at some point, possibly ejecting much
of the accumulated mass.

Alternatively, the donor may be a helium-rich star, avoiding the
challenges involved in mass gain through hydrogen-rich accretion (see
\citealt{Wang09a}).  The accumulated helium layer ignites at a larger
accumulated mass, but may suffer from the same problem as described
above, of a helium nova in which much of the accreted material is
ejected. The efficiency of hydrogen and helium accretion
is thus theoretically uncertain,
leading to a large range of possibilities for WD growth (see
\citealt{Yungelson05,Bours13}) 

\subsection{Double-degenerate models}
\label{sec:DDmodels}

In the DD scenario, the more-massive WD is generally thought to
tidally disrupt and accrete the lower-mass WD, initially through an
accretion disk configuration, that then likely takes on a more
spherical geometry (e.g.,
\citealt{Loren-Aguilar09,Pakmor12,Raskin12,Shen12,Schwab12,
Moll13}).
Accretion of carbon and oxygen at a high rate should prevent the
problems of inefficient mass growth, encountered in novae, eventually
leading to carbon ignition in the core. Against this picture, however,
it has long been argued that the large accretion rate will lead to off-center
ignition and burning of carbon in the accretor, to oxygen and neon.
Further accretion and approach to $M_{\rm Ch}$
will drive electron capture on Mg and Ne, leading to gravitational
instability and to an ``accretion-induced collapse'' to a neutron star
\citep{Nomoto85,Saio98,Shen12}. 
 The collapse could manifest itself as some kind of non-SN~Ia,
 low-luminosity, hydrogen-free, transient (\citealt{Darbha10} and 
\citealt{Piro13e}
 have considered the signatures of such an event).  On the other hand,
 rotation of the merger remnant has been considered as a way to slow
 down the accretion and to avoid off-center ignition
 (\citealt{Piersanti03,Tornambe13}; see also \ref{sec:ALTmodels},
 below).

Recent 3D hydrodynamical merger models 
have cast doubt on the picture of a merger remnant as an unperturbed
primary-mass WD that accretes the remains of the secondary
WD. Instead, complex structures with regions of high temperature and
density are found, in which explosive carbon ignition may occur. 
  Such ``violent mergers'' may lead to a
SN Ia, even if the total mass of the merger is below $M_{\rm Ch}$
\citep{Pakmor13,
van-Kerkwijk10,Guillochon10,Zhu13,Dan13,Raskin13b}. 
Details depend on the
WD masses (a primary WD mass $\gtrsim 0.9$~\Msun~ is likely required) 
and on the presence of He, which can facilitate ignition.

\subsection{Collisional double-degenerate models}
\label{sec:collisions}

A special case of a WD merger configuration is the ``collisional''
one, whereby two WDs collide head-on, rather than spiraling in due to
gravitational-wave losses. \cite{Benz89,Raskin09,Raskin10,
  Rosswog09,Loren-Aguilar10} and \cite{Hawley12} have invoked this
scenario, generally in the context of some SNe~Ia possibly arising in
dense stellar environments such as globular clusters and galactic
nuclei.  A different implementation of the collisional DD scenario was
raised by \cite{Thompson11}, who proposed that many binary WDs may
actually be in ``hierarchical triple'' systems, with a third, low-mass
star in orbit about the inner WD binary. The tertiary could induce
\cite{Kozai62} - \cite{Lidov62} oscillations in the eccentricity of
the inner binary. A high eccentricity increases gravitational wave
losses, and thus
decreases the time until merger, and so enlarges the
population of tight WD binaries that merge 
(leading to an enhanced rate of SNe~Ia). Rare head-on collisions
would sometimes also occur among the most eccentric systems. \cite{Prodan13} considered also
tidal dissipation of orbital
energy of the inner binary \citep{Mazeh79}, further enhancing the
merger rate. However, it is not clear that triple systems, with the
post-common-envelope orbit ratios needed for this mechanism to
operate, can be realized.

\cite{Katz12}
discovered that one of the approximations used in previous Kozai-Lidov
calculations is invalid for ``mild'' hierarchical triples, i.e. those
in which the tertiary orbit is $\sim 3-10$ times the inner
orbit. Among such cases, they found, through direct numerical
integration for a range of initial tertiary orbit inclinations, a
rather high probability, $\sim 5\%$, for a Kozai-Lidov-induced head-on
collision between the inner WD pair. \cite{Katz12} argued that, if a
large fraction of all intermediate-mass stars are in the appropriate
triple systems, most or all SNe Ia could come from such collisions.
 \cite{Kushnir13a} and 
\cite{Garcia-Senz13} have
performed the latest hydrodynamic thermonuclear simulations of such collisions,
and obtained promising agreement with some of the phenomenology of
observed SN~Ia explosions. \cite{Kushnir13a}, 
using high spatial resolution ($\sim 1$~km)
in their
simulations, and avoiding some numerical artifacts that
affected previous work,  
find that a collision between, e.g., two 0.7~\Msun~ WDs produces an
explosion with a $^{56}$Ni yield of 0.56~\Msun, similar to a typical
SN~Ia. 

However, it seems unlikely that a large fraction of all WDs are in
such triples --  only $\sim 10-20\%$
of stars are triples \citep{Leigh13, Duchene13},
and probably only a fraction of those have the required
special orbit ratios. A further problem with this
model is how to avoid a collision of the inner binary early on, when
the components are still on the main sequence, and their geometrical
cross-sections for collision are 100 times greater than when they have
become WDs. \cite{Hamers13} have used ``binary population synthesis''
calculations (see Section~\ref{sec:linking}, 
below) to confirm that, for this reason alone, the enhancement of
SN~Ia rates is already minimal for initially wide inner binaries that
would not interact without a tertiary component. \cite{Katz12} have
suggested that the relative orientations of the inner and outer orbits
could be ``reset'' by small kicks, during the mass loss phases of the stars
of the inner binary, or by a passing perturbing star. Colliding WD cases
could then arise from among those systems that avoided collisions
while on the main-sequence. 

\subsection{Double detonations and rotating super-Chandrasekhar-mass models}
\label{sec:doubledet}

WDs accreting helium at low rates (from either non-degenerate or
helium-WD donors) can accumulate a helium layer that is sufficiently
massive and degenerate such that the resulting helium ignition becomes
explosive \citep{Taam80}. This detonation drives a shock into the
underlying C/O WD, which then ignites carbon at or
near the center, even in sub-$M_{\rm Ch}$ WDs
\citep{Shen09,Livne90,Fink10}.  Originally, these
``double-detonation'' or ``edge-lit detonation'' models were
considered for non-degenerate helium donors, in which the ignition
masses of the helium layers were found to be $\sim0.2 $~\Msun. Models
for such explosions typically predicted strong signatures of carbon
and oxygen (from the burned helium) in the SN spectra, which were
inconsistent with observed spectra. However, \cite{Bildsten07} found
that, for helium WD donors, the higher mass transfer rates 
lead to smaller helium layers at explosion time, and better
agreement
 with
observations (e.g. \citealt{Fink07}). Recently, 
 \citet{Guillochon10,Pakmor13}, and \cite{Shen13a} have
  found that also in ``violent mergers'' (see
  Section~\ref{sec:DDmodels}, above) there may first be a surface
  He detonation, triggering the C detonation, if the merger is between
  a He and a C/O WD, or in mergers between two C/O WDs where there is a
 thin surface layer of He present.  

At the other extreme, the nickel mass deduced for some very bright
SN~Ia explosions is suggestive of a super-Chandrasekhar-mass
progenitor, with mass $\gtrsim 2 M_\odot$
(e.g. \citealt{Howell06,Silverman11,
  Scalzo10,Scalzo12,Kamiya12,Taubenberger11,Taubenberger13a}).  WD
rotation has been proposed as a means of supporting such massive
progenitors. On the other hand, \cite{Hillebrandt07} and
\cite{Hachinger13} have argued that the explosion of a rotating
super-$M_{\rm Ch}$ WD will not necessarily produce the inferred nickel
mass, or other characteristics, of such bright events, and they have
suggested asymmetric explosions instead (see also \citealt{Moll13}). 
\cite{Chamel13}
have shown that another proposed means of supporting super-$M_{\rm
  Ch}$ WDs, via ultra-strong quantizing magnetic fields, is impractical,
due to electron-capture reactions that would make the WD unstable.

More generally, WD rotation has been invoked in several so-called
``spin-up/spin-down'' scenarios. \cite{Yoon04,Yoon05,Di-Stefano11,
Justham11}, and \cite{Hachisu12b} have argued, in
the context of the SD model, that a WD that has grown in mass, even
beyond $M_{\rm Ch}$, could be rotation-supported against collapse and
ignition, perhaps for a long time, during which the accretion could
run its course and end, and the traces of the process (or even of the
donor itself) could disappear.  DD mergers whose explosions are
delayed by rotational support have also been proposed
\citep{Piersanti03,Tornambe13}.  A WD undergoing maximal solid-body
rotation is stable against carbon ignition up to a mass of
$1.47~M_\odot$, i.e., only $\approx 0.1 M_\odot$ more than in the case
of a non-rotating WD. In the context of SD models, where a typical WD
needs to accrete $\sim 1~M_\odot$ before exploding, it would appear
that this extra $0.1 M_\odot$ provides not much of an opportunity for
the accretion process to conclude. \cite{Yoon05} have
therefore constructed models of differentially rotating WDs in which,
for some radially increasing profiles of angular rotation speed, the
maximal stable WD mass reaches $\sim 2 M_\odot$. 
\cite{Saio04} and \cite{Piro08b},
however, have argued that ``baroclinic''
instabilities, and/or the shear growth of small magnetic fields,
provide torques that will quickly bring a differentially rotating WD
to solid-body rotation, with its limited rotational support. 
\cite{Hachisu12a} have countered that these instabilities
might not occur, as only the necessary conditions for them, but not
the sufficient ones, are satisfied.  The theoretical viability of
massive, rotation-supported, WDs is thus still
unclear. Observationally, a spin-up/spin-down scenario could
potentially ``erase'' many of the clues that we discuss in this
article.

\subsection{Alternative models} \label{sec:ALTmodels}

\cite{Kashi11} have introduced a ``core-degenerate'' model, in
which a WD and the core of an asymptotic-giant-branch (AGB) star merge
already in the common-envelope phase. After ejection of the envelope,
the merged core is supported by rotation, potentially for long times,
until it spins down, e.g. via magnetic dipole radiation, and only then
explodes \citep{Ilkov11}.
\cite{Wheeler12} has sketched a SD progenitor model consisting of a WD
and an M-dwarf donor, where accretion by the WD and its growth toward
$M_{\rm Ch}$ are enhanced by magnetic channeling, self-excited mass
loss from the M dwarf, and magnetic inhibition of mixing of the WD
surface layer, thus avoiding excessive mass loss in nova events.
Single-star SN~Ia progenitor models have also been occasionally
attempted \citep{Iben83,Tout05}, in which the degenerate carbon-oxygen
core of an AGB star is somehow ignited after it has lost its hydrogen
envelope (as it must, if the SN is to appear as a Type-Ia). 

\subsection{Linking theory and observations}
\label{sec:linking}

As already noted briefly above, apart from the issue of the
identity of the progenitors, and the intrinsic problems of each of the
progenitor scenarios, many gaps remain in our understanding of the
phases that precede each ``progenitor setup'' that will eventually
lead to an observed SN~Ia. These gaps of knowledge include the details
of binary evolution, and particularly the enigmatic common-envelope phases that a
pre-SN~Ia binary system undergoes in almost all models 
(see \citealt{Ivanova13}, for a review). 

An important theoretical tool for obtaining
observational predictions from the various scenarios, despite these
obstacles, has been the calculation of binary population synthesis
(BPS) models. In BPS, one begins with a large population of binaries
with a chosen distribution of initial parameters (component masses,
separations), and one models the various stages of their stellar and
binary evolution, including transfer and loss of both mass and angular
momentum, and possibly multiple common-envelope phases, with the
physics of each stage parametrized in some way 
(e.g., \citealt{Jorgensen97, Yungelson00,Wang10a,Meng11b,Bogomazov11,
Mennekens12a,Ruiter13, Toonen12} Claeys et al. 2013b).
At the
end of such a simulation, one can see what fraction of the initial
stellar population, and from which specific progenitors, has ended up
at the conditions thought to lead to a SN~Ia explosion through a
particular channel (e.g. DD, Chandrasekhar-mass SD,
etc.). Furthermore, as discussed in Section~\ref{sec:rates}, below,
for each progenitor channel one can obtain the distribution of ``delay
times'' between star formation and SN~Ia explosion, a distribution
that can be compared to observations. 

Even after the conditions for a SN~Ia explosion have been met, many
additional questions remain as to the ensuing phases: the trigger and
the locations of WD ignition; the mode in which the burning front
consumes the WD (see \citealt{Hillebrandt00}, and
Section~\ref{sec:correlations}, below); uncertainties in nuclear
cross sections \citep{Parikh13}; and the transfer of
radiation through the expanding ejecta (e.g.
\citealt{Pinto00,Piro13a,Piro13b,Mazzali13,Dessart13a}). 
There is thus a multitude of
theoretical paths to a SN~Ia explosion, some of this multiplicity
arising from real physical possibilities, and some of it due to
uncertainties in the often-complex physics.  Among these many
theoretical paths, the ones that are actually realized and seen in
Nature likely encrypt observational clues to the solution of the
progenitor question.  With this in mind, we attempt an overview of the
state of the art of observations that may provide such clues. We
arrange our review according to the various possible observational
approaches to the problem.

\section{Evidence from the observations}

A variety of observational approaches have been brought to bear on the
SN~Ia progenitor problem. The existing populations of potential
progenitors can be studied (Sec.~\ref{sec:populations}); pre-explosion
data at the sites of nearby events may reveal the progenitors
(Sec.~\ref{sec:preexplosion}); the observed properties of the events
themselves may carry clues to the progenitors (Sec.~\ref{sec:events});
the remnants of presumed SNe~Ia can be searched for remaining traces
of the progenitor systems (Sec.~\ref{sec:remnants}); and the rates at
which SNe~Ia explode as a function of 
time and environment provide another avenue to address the problem
(Sec.~\ref{sec:rates}).  We review each of these approaches in turn.

\subsection{Clues from potential progenitor populations}
\label{sec:populations}

Since a typical galaxy hosts of order $10^7$ SNe~Ia over a Hubble time
(see Section~\ref{sec:ratesize}), a viable type of progenitor system
should be present in significant numbers and therefore observable,
whether as individual objects, or through their collective emission.
Thus, the first observational approach to the progenitor problem that
we consider is to look for specific potential progenitor systems, to
measure their properties and numbers,
and to see if those conform to expectations, given known SN~Ia
properties.

\subsubsection{Recurrent novae}
\label{sec:recurrent}

The Milky Way and similar large galaxies have populations of $\sim
10^6$ ``cataclysmic variables'' -- WDs accreting from non-degenerate
donor stars through Roche-lobe overflow or through a wind, with
orbital periods down to about 80~min 
(e.g. \citealt{Warner03}).
Among the
ways in which these systems reveal themselves are ``nova eruptions''
(see Sect.~\ref{sec:SDmodels}, above), which occur in the Galaxy and
in M31 at rates of $\sim 35$~yr$^{-1}$ and $\sim 65$~yr$^{-1}$,
respectively \citep{Della-Valle94,Darnley06}. 
As already noted, it is thought that few, if any,
novae gain more mass during accretion than the mass they lose in the
eruptions, 
in which case such systems 
are not SN~Ia progenitors
\footnote{See, however, \cite{Zorotovic11}, who
  find that accreting WDs, on average, have higher masses than WDs
  with low-mass companions in post-common-envelope binaries that are
  destined to become semi-detached systems with stable mass transfer.
  This suggests either that WD masses do grow, or that the majority of
  cataclysmic variables form from systems with higher mass companions,
  that by some selection effect, harbor more massive WDs.}.

However, members of a subclass of novae called ``recurrent novae'', have long
been suspected as possible SN~Ia progenitors 
(e.g. \citealt{Starrfield85,Della-Valle96,Schaefer10a,Kato12}).
Recurrent novae are defined as novae with
more than a single outburst over the past century or
so. There are only 10 recurrent novae known in the Galaxy.  Some of
the donor stars are main-sequence, some sub-giants, and some red
giants. 
Recurrent novae have
eruptions every few decades, on an irregular basis. Because the mass
needed for ignition scales inversely with both WD mass and accretion
rate \citep{Fujimoto82,Truran86,Shen09},
these short recurrence times have been interpreted to
indicate that the WD mass is close to $M_{\rm Ch}$, specifically
$M\gtrsim 1.2 M_\odot$, and that the accretion rate is relatively
high, the same parameters for which mass gain may be possible. A
number of observational mass estimates are indeed suggestive of large
WD masses, 
but uncertainties are large, and there
are only two double-lined eclipsing systems that have 
reliable mass estimates: U Sco, with a WD mass of $1.55\pm0.24 M_\odot$
\citep{Thoroughgood01}, and CI Aql, with $1.00\pm0.14 M_\odot$
\citep{Sahman13}. 
Observationally, little is known about the
formation of massive WDs in binaries, so it is unclear if
these high masses indicate that the WDs have accumulated
mass. Recurrent nova eruptions may in fact be caused by
instabilities in the accretion discs, leading to periodic accretion at
the steady-burning rate, during which the WDs do grow in mass
\citep{Alexander11}. Then again, \cite{Hachisu12c}
point out that it is unclear if steady burning
can ignite for such short accretion episodes.

Yet another view is that recurrent novae are systems in which
the WD mass is decreasing with time, and which will thus never become
SNe~Ia (\citealt{Schaefer13a,Patterson13}, see below).  
As noted, there are few known
recurrent novae, and even fewer that have had well-studied multiple
outbursts. The debate regarding their being
SN~Ia progenitors has therefore focused on individual objects and on
individual outbursts.
  
 For example, after the
  latest outburst of U Sco in 2010 (e.g. \citealt{Schaefer10b}), 
\cite{Schaefer13a} deduces, based on the period
  change, that much more material was ejected
  than had been previously accreted, challenging the idea of WD mass growth
  (e.g.  \citealt{Diaz10}). Whether or not the massive
  WD has a  C/O composition (as required of a SN~Ia), 
  or an oxygen-neon one,
   is also debated 
  (e.g. \citealt{Mason13,Kato12}).
RS Oph is a symbiotic binary consisting of a massive WD and a
  red giant in a 454~d orbit (see \citealt{Kato12}, and
  references therein). Its latest outburst in 2006, from which 
  \cite{Hachisu06} infer a high WD mass, also revealed a
  shock from the interaction of the ejecta with the red-giant wind
  \citep{Sokoloski06},
  and radio emission suggesting the launching of
  a jet \citep{Rupen08}. 
  \cite{Patat11} have studied the evolution of the absorption
  features, and find similarities with those found in the spectra of some
  SNe~Ia (see Section~\ref{sec:absorption}).
T Pyx 
  is a short period system ($P=1.8$~hr, \citealt{Uthas10}),
  implying a low donor mass (if it is to fit in the
    Roche lobe at this period), and more typical of systems harboring
  classical novae, with its most recent outburst in
  2011. \cite{Schaefer10c} suggested
a high mass-transfer rate that would imply a high WD
mass. However, \cite{Uthas10} derive a mass ratio of 0.2, that
  for a realistic donor mass implies a WD mass of only 0.7 \Msun. In
any case, \cite{Selvelli08} and
  \cite{Patterson13}
  conclude that this system, again, is ejecting more mass than it accretes.
T CrB is another wide symbiotic recurrent nova, with two known
  outbursts (see \citealt{Anupama13}), for which  \cite{Luna08}
  argue a high WD mass based on the detection of hard X-ray emission.
V407 Cyg  
is formally not a recurrent nova, as
  only one nova outburst (in 2010) has been observed, 
but  the similarity of the eruption to that of RS Oph 
  (e.g. \citealt{Shore11b})
  implies a massive WD  
  \citep{Nelson12b,Hachisu12c}.
  On the other hand, \cite{Chomiuk12b} use
  extensive radio observations to argue that the environment 
  of V407 Cyg is not one typical of SNe Ia.

Could recurrent novae be the phase during which accreting WDs 
 achieve a significant
fraction of their growth toward $M_{\rm Ch}$ and explosion as SNe~Ia? 
The answer is no, unless the number of recurrent novae is
  orders of magnitude larger than estimated. Beyond the 10
  known systems, \cite{Schaefer10a} estimates that as many as
  60-100 Galactic sources that have been classified as classical novae
  are, in reality, recurrent novae whose repeated outbursts have been
  missed. Furthermore, his analysis of the Galactic spatial
  distribution of both types of novae suggests that their true numbers
  are several times larger than the numbers currently known. Thus, the
  Galactic population of recurrent novae could conceivably number
  $\sim 300$. However, to get a Galactic SNIa rate of at least once
  per 200~yr (see Section~\ref{sec:ratesize}) and accrete at the
  very least $\sim 0.2 M_\odot$ at the limiting accretion rate onto a
  WD for a recurrent nova, of $\dot M < 3 \times 10^{-7}~ M_\odot~
  {\rm yr}^{-1}$, over 3300 systems are needed.
 The recurrent nova phase can constitute $\lesssim
 9\%$ of the time of the final $0.2~M_\odot$ of WD mass growth.
 If we consider a more-realistic $0.4-0.9~M_\odot$ mass gain
 required of C/O WDs (which have masses of $\sim 0.5-1 M_\odot$) 
 in order to reach $M_{\rm Ch}$, then only 2-5\% of that mass
 gain can take place during recurrent nova phases. 

\subsubsection{Supersoft X-ray sources}
\label{sec:supersoft}

In the evolutionary scenario proposed by \cite{Hachisu99}, the
recurrent nova phase, with accretion rates estimated at $\dot M \sim
1-3 \times 10^{-7}~ M_\odot~ {\rm yr}^{-1}$, follows a phase when a SD
progenitor system is a supersoft X-ray source, growing via a larger
accretion rate, $\dot M \sim 3-6 \times 10^{-7}~ M_\odot~ {\rm
  yr}^{-1}$. The high accretion rate (and hence temperature), compared
to those of novae, lifts the degeneracy of the accreted hydrogen
layer, and it burns stably to helium on the WD surface.

Observationally, ``persistent'' or ``permanent'' supersoft X-ray
sources are identified, as their name implies, by their soft X-ray
spectra, peaking at 30-100~eV, with typical luminosities of
$10^{37-38}$~erg~s$^{-1}$.  Optical and X-ray followup of the best
studied of these objects (e.g. \citealt{Lanz05,Rajoelimanana13}) has
shown that they indeed consist of a hot WD in a close orbit with a
non-degenerate donor, where hydrogen accreted from the donor, at
roughly the above rates, burns more-or-less stably into helium on the
WD surface \citep{van-den-Heuvel92}. However, not all supersoft source
spectra are easily interpreted in this way, as some show P-Cygni
profiles indicative of a wind rather than a WD atmosphere
\citep{Bearda02}.  Because of interstellar absorption of their soft
X-ray spectrum, supersoft sources have been discovered mainly in
external nearby galaxies. Only two are known in the Galaxy (MR Vel and
Q And; e.g. \citealt{Simon03}) but there are 15 in the Magellanic
Clouds, and of order 10 in M31 \citep{Orio10}. Classical and
recurrent novae, discussed in Section~\ref{sec:recurrent} above, also
have supersoft phases, but these are transient, occurring after their
outbursts, and lasting of order a month. As a result, only a minority
 of the supersoft sources that turn up in X-ray surveys of
nearby galaxies are of the persistent kind, rather than being the
transient supersoft phases of post-outburst novae \citep{Orio10}.
 
Since the persistent supersoft X-ray sources in an $\sim L_*$ galaxy
such as M31 again number only in the few tens of objects, the same
argument, used above in the case of recurrent novae, holds as well: if
supersoft sources are accreting WDs growing toward $M_{\rm Ch}$, and
such SD systems are to explain the bulk of the SN~Ia rate, then $<1\%$
of the WD's growth time is spent in this phase. This is an order of
magnitude less than the mean $\sim 5\%$ fraction of the time in the supersoft
phase found by model calculations \citep{Meng11b}.
This paucity of supersoft sources has been pointed out by
\cite{Di-Stefano10}, based on the numbers of such sources observed in
six nearby galaxies, and by \cite{Gilfanov10}, based on the
integrated X-ray flux observed from nearby elliptical galaxies.
\cite{Hachisu10} and \cite{Meng11b} (see also \citealt{Lipunov11}) 
have countered that the
nuclear-burning accreting WDs spend the
majority of the time in a third possible phase, hidden within
optically thick outflows (e.g \citealt{Nielsen13a,Wheeler13,Woods13a}).
We discuss this possibility in Section \ref{sec:rapidaccretion}.
Although proto-DD systems, at the time that only the first WD has
formed, may also undergo 
mass transfer and a SD-like supersoft phase, its duration is much
shorter than in SD progenitor systems
\citep{Nielsen13b}. 
As a result, fewer supersoft
sources are expected in the DD scenario, by at least an order of
magnitude, and consistent with their observed rarity.
Considering also the delay, between the supersoft phase of proto-DD systems 
and the eventual WD merger and SN Ia explosion, further lowers the number
of expected supersoft sources from DD progenitors. 

\subsubsection{Rapidly accreting WDs}
\label{sec:rapidaccretion}

As noted above, \cite{Hachisu96,Hachisu99} have proposed that 
the rapid accretion phase is self-regulated by an
optically thick wind.  The wind drives away the excess mass,
effectively keeping the WD growth and the stable nuclear burning on
the WD surface at the same rates as during the supersoft phase
(but see \citealt{Idan13}, Section~\ref{sec:SDmodels}, above).  
The excess mass
that is blown off the WD could create an optically thick photosphere
that reprocesses the X-rays to UV emission.
\cite{Nomoto07,Wheeler13,Nielsen13a}, and \cite{Lepo13b} have
considered the possible appearances of such ``rapidly accreting WDs'',
and argued that they might appear as undersized OB stars, i.e. with
effective temperatures of $ 10^{4.5-5} K$, but with sizes of only a
few $R_\odot$, dictated by the WD's Roche lobe radius.  Another
possibility is an appearance similar to Wolf-Rayet (WR) stars or WR
planetary nebulae.
  
Among the objects considered possibly to be such rapidly accreting WDs
are V-Sagittae-type nova-like cataclysmic variables, a class
consisting of
a handful of known objects \citep{Steiner98}.  The prototype, V~Sge, 
is a double-lined eclipsing binary with a rich high-ionization
emission-line spectrum, a period of 0.5~day, quasi-periodic high and
low states lasting $\sim 180$~days, and supersoft X-ray emission in
its low state. \cite{Hachisu03b} have modeled V Sge as a $1.25
M_\odot$ WD accreting at a high rate from a Roche-lobe-filling $\sim 3
M_\odot$ companion. Other moderately well-studied objects in the V Sge
class are RX 0513.9-6950 \citep{Hachisu03a} and QU Car
\citep{Kafka12}. 
A different type of object which could be a
rapidly accreting WD is the peculiar planetary nebula LMC N66 in the
Large Magellanic Cloud, whose WR-like core has undergone two optical
outbursts over the past 60 years, each lasting several years, during
which its luminosity was comparable to WN-type WR stars.  There is
no evidence yet for binarity in LMC N66, and no X-ray emission has
been detected. \cite{Hamann03} have modeled it, again, as a WD
with mass inflow from a companion at rates of $10^{-6}$ to
$10^{-5}~M_\odot~{\rm yr}^{-1}$ during its low and high states.

\cite{Lepo13b} have performed a search in the Small
Magellanic Cloud for the $\sim 100$ rapidly accreting WD systems that
are expected there, if such progenitors are to produce the bulk of the
SN~Ia rate.  They obtained optical spectra for about 750 sources,
selected to be UV-bright at 1600~\AA\ and to have unusual optical
colors.  They failed to find any object with unusual spectral
signatures such as strong He~II lines, or with optical variability,
based on the long term OGLE monitoring database \citep{Udalski97}.
 From this, they deduce a $\lesssim 10\%$ contribution to the SN~Ia
progenitor population, of systems that are similar to LMC N66 or to WR
stars. The color selection of targets employed by \cite{Lepo13b}
included only a limited number of objects with colors similar
to V~Sge, and therefore such systems cannot yet be ruled out as SN~Ia
progenitors by this experiment.

Taking a more integral approach, \cite{Woods13a} have argued that, in
any scenario where the X-ray emission from nuclear-burning WDs is
shielded by $T= 10^{5-6} K$ photospheres, the emerging UV radiation
will still be quite hard, and capable of ionizing interstellar He
atoms.  In elliptical galaxies, a substantial mass of 
extended neutral hydrogen is often detected, as well as diffuse regions of 
low-ionization line emission, thought to arise through 
ionization of some of the gas by evolved stars.
 The integrated radiation from a population of rapidly accreting WDs
could lead to ionization of some of the He in this gas to He$^{++}$, 
and to a
detectable signal of diffuse He~II$~\lambda 4686$ recombination line
emission.  \cite{Woods13a} combine
stellar spectral synthesis models with photoionization models to
calculate the line emission as a function of galaxy age, gas covering
fraction, and WD photospheric temperature.  They predict, for an
elliptical galaxy with a SD-dominated progenitor population with $T=
2\times 10^{5} K$ photospheres, a typical
line equivalent width of a few tenths \AA.  In the absence of
nuclear-burning SD systems, with ionization only by evolved stellar
populations, a line flux one order of magnitude lower is
expected. \cite{Woods13b} 
extend these predictions to excitation  of forbidden emission lines 
of C, N, and O by 
nuclear-burning WDs with photospheres in the upper part of the 
$T=10^{5-6} K$ range. 
Johansson et al. (2013, submitted) have searched for the He~II$~\lambda 4686$
signal in high-signal-to-noise stacks of 11,600  
Sloan Digital Sky Survey (SDSS; \citealt{Abazajian09}) spectra of
early-type galaxies. The galaxies are selected to have weak but 
measurable line emission, in order to include galaxies with gas, but
to exclude, based on the line diagnostics, galaxies with ongoing star
formation or with active nuclei.
In all four galaxy age-group stacks that they produce, they detect 
HeII$~\lambda4686$ emission 
only at the level expected from ionization by the known
evolved stellar populations, but not from ionization by rapidly
accreting WDs. This sets a 10\% upper limit on the contribution to
the SN Ia rate by rapidly accreting WDs. 
 
\subsubsection{Helium-rich donors}
\label{sec:heliumdonors}

No clear-cut cases are known of systems that could be SN~Ia
progenitors through the helium-rich SD donor channel. However, the
only known helium nova, V445 Pup, shows a pre-outburst magnitude
\citep{Ashok03} that suggests it is a WD
accreting from an evolved helium star 
\citep{Kato08,Woudt09}. Modeling
the nova outburst, \cite{Kato08} find a high WD
mass, but the predicted distance disagrees somewhat with the one
determined from the expansion parallax of the bi-polar outflow from
\cite{Woudt09}.
At least one case is known of a close detached binary consisting of a 
C/O pre-WD and a lower-mass helium
star, CD-30$^\circ$~11223 \citep{Geier13}.
Its future helium accretion rate is
expected to be in the right range to set off an edge-lit detonation,
possibly leading to a double-detonation SN~Ia 
(\citealt{Yungelson08}, 
see
Sect.~\ref{sec:doubledet}). The same could be true for some fraction
of the so-called AM CVn systems, a
class of objects in which a WD accretes helium from a very low-mass
companion (see \citealt{Solheim10} for a review), 
although the majority of the observed systems have evolved
to too-low mass transfer rates for a double detonation.

\subsubsection{Binary WDs}
\label{sec:binarywd}

As a test of the DD scenario, one can search
our Galactic neighborhood, within a few kpc of the Sun, for
short-period WD binaries whose orbits will gravitationally decay and
merge within, say, a Hubble time, surpassing (or not) the
Chandrasekhar mass and perhaps thus producing SNe~Ia.
 For the classic DD scenario to work, the
Galactic SN~Ia rate should match the WD merger rate (or some fraction
of it involving mergers of the suitable masses).  For the range of
masses of WDs with a C/O composition, $0.5-1.0 ~ M_\odot$, pairs
merging within a Hubble time will have separations of $a \lesssim
0.015$~AU, orbital periods of $P \lesssim 12$~hr, and circular orbital
velocities $v\gtrsim 100$~km~s$^{-1}$.

\cite{Marsh11} and \cite{Kaplan12}
present recent compilations from
searches for binary WDs. In the current century, the first large
survey to search systematically for such WD pairs was SPY
\citep{Napiwotzki04, Nelemans05b,Geier07}. SPY obtained high-resolution
($\sim 2$~km~s$^{-1}$ radial velocity [RV] resolution) VLT spectra for
a sample of $\sim 1000$ WDs, with two epochs per target, separated by
at least one day. Given the $\lesssim 12$~hr periods of close WD
binaries, this essentially gives two samplings at random phases in the
orbit. They searched for RV variations between epochs, that could
result from the motion of an observed WD about a binary center of
mass\footnote{Note that a WD's luminosity can have a large range,
  depending on the WD effective temperature, which decreases with WD
  age, and on its surface area, which decreases with increasing WD
  mass. Therefore, in a WD binary, one of the WDs, not necessarily the
  least or more massive one, will often be much fainter than the other
  WD, and thus will remain undetected.}.  Candidate systems
discovered by the survey were then followed up to obtain radial
velocity curves and constraints on the binary parameters.  SPY
discovered $\sim 100$ candidate binary WDs, and among the $\sim 10$
systems
for which parameters are measured, a handful will merge
within a Hubble time.  None of these future mergers have total masses
$M_{\rm tot}$ that are unambiguously above $M_{\rm Ch}$, although for
one system (WD 2020-425) the masses are difficult to measure and the
best estimate is $M_{\rm tot}=1.35$~\Msun~
\citep{Napiwotzki07}.
Then again,  $M_{\rm tot} \ge M_{\rm Ch}$ may not
necessarily be a pre-condition for a DD-scenario SN Ia  (see
Sections~\ref{sec:DDmodels} and \ref{sec:doubledet}, above). 
Furthermore, a statistical interpretation of these results that
accounts for the selection effects and efficiencies of the survey is
still lacking. It is therefore yet unclear what these results imply
for the merger rate of WDs in general, and for $M_{\rm tot} \ge M_{\rm
  Ch}$ mergers in particular.

A number of tight binaries, found by other searches, some that are not
quite yet bona-fide double WD systems, may
possibly become $M_{\rm tot} \ge M_{\rm Ch}$, Hubble-time-merging, WD
systems \citep{Maxted00,Geier07,Geier10,
Rodriguez-Gil10,Tovmassian10}. \cite{Brown11}
have even discovered a binary WD system with a 13-min orbit
that will merge within $< 1$~Myr, 
and \cite{Hermes12} have
detected, over a $\sim 1$-year baseline, the gravitational decay of
its orbit. However, with component WD masses of only 0.26 and
0.50~$M_\odot$, this merger is unlikely to result in a SN~Ia.
 
In terms of analysis of the observed binary WD population,
\cite{Nelemans01} and \cite{Toonen12} have used BPS to simulate, under the
various parametrizations for the physics of the common-envelope phase,
the properties of WD binaries that would be observed, such as masses
and periods. They have qualitatively compared them to the masses and
periods of a compilation of known binary WD systems, as well as
quantitatively with the WD space density and WD production rate,
finding reasonable correspondence. 
The Galactic WD merger rates indicated by
these BPS models, $1-2 \times 10^{-2}~\rm{ yr}^{-1}$ (or $\sim2-3 \times
10^{-13}$~yr$^{-1} M_\odot^{-1}$ for a Milky Way stellar mass of $6\times
10^{10} M_\odot$), agree with other BPS studies 
(e.g. \citealt{Iben97,Han98}). 
This merger rate is similar to recent estimates of the SN~Ia rate
per unit mass in Sbc-type galaxies of Milky-Way mass, $\sim 1\times
10^{-13}$~yr$^{-1} M_\odot^{-1}$ 
(\citealt{Mannucci05,Li11b}, see Section~\ref{sec:ratesize}, below).
However, only about one-third of the BPS mergers involve
two C/O WDs, and an even smaller fraction 
have $M_{\rm tot}\ge M_{\rm Ch}$. 
Among known binary WDs, indeed only a handful are C/O+C/O,
but this is partly due to the fact that some surveys are
designed specifically to discover low-mass WD binaries 
(e.g. \citealt{Kilic12}).
The merger rates predicted by BPS thus may or may not be at the level 
required to explain SN~Ia
rates, depending on the ranges of $M_{\rm tot}$ assumed to lead to
a SN~Ia. More on this issue in Section~\ref{sec:ratenorm},
below.

\cite{Badenes09d} have been searching for close binary WDs among the
spectra in the SDSS. The large number of WDs in SDSS, and the fact
that SDSS spectra are always split into at least two sub-exposures, 
separated by at least 30~min, make this possible. With a $\sim
70$~km~s$^{-1}$ RV resolution possible with SDSS spectra of WDs (the
resolution varies with WD type and brightness), this permits
discovering candidate short-period WD binaries. The observed
distribution of maximum RV differences between any two epochs,
$\Delta_{\rm RVmax}$, can be compared to model distributions predicted
for a range of binary population parameters 
\citep{Maoz12b}.  \cite{Badenes12} have applied the method to a sample
of 4000 DA-type WDs in SDSS with the best signal-to-noise
ratio. The model combinations of binary fraction and
separation-distribution parameters that are consistent with the
observed $\Delta_{\rm RVmax}$ distribution have a WD merger rate, per
unit stellar mass\footnote{The analysis naturally gives the specific
  merger rate per WD, i.e. the reciprocal of the mean time until
  merger. This can be converted to a rate per unit stellar mass,
  using estimates for the stellar mass density and the WD
    number density in the Solar neighborhood.
}, 
of $\sim 1\times
10^{-13}$~yr$^{-1} M_\odot^{-1}$, a bit below the BPS estimates, 
and again quite similar to the expected
 SN~Ia rate
per unit mass in the Milky-Way. The fraction of 
$M_{\rm tot} \ge M_{\rm Ch}$ mergers is unconstrained by these
observations -- $\Delta_{\rm RVmax}$ is sensitive foremost to the WD
separation, rather than to $M_{\rm tot}$. As noted above, based on
known double WD systems and from BPS calculations, the fraction
may be small, but  $M_{\rm tot} \ge M_{\rm Ch}$ may not
be required for a  SN Ia.

\subsubsection{Rapidly rotating massive WDs}
\label{sec:spinningWDs}

In the various spin-up/spin-down scenarios (see
Section~\ref{sec:models}, above), one could expect the existence of a
progenitor population of rapidly rotating, super-$M_{\rm Ch}$ WDs,
with rotation periods of order seconds,
either single or in post-interaction binaries.  To achieve stability
at significantly super-$M_{\rm Ch}$ masses, the WDs would need to have
differential rotation profiles.

WD rotation is difficult to measure but,
with few or no exceptions, WDs with measured rotations 
are observed to spin extremely slowly
(see, e.g. \citealt{Kawaler04}). 
The typical period is $\sim 1$~d, and most
periods range from a few hours to a few days. For example,
\cite{Berger05}, using the core of the Ca~II K line to measure
rotational broadening in 38 DAZ-type WDs, find
a projected rotation velocity $v_{\rm rot} \sin i < 30$~km~s$^{-1}$,
and typically just a fraction of this velocity. This shows that there
is efficient radial angular momentum transport in WDs, at least
between the core and the pre-WD envelope, or that the WDs were formed
slowly spinning \citep{Spruit98,Suijs08}.
\cite{Charpinet09} and \cite{Fontaine13}
have used asteroseismology
to measure the inner rotation profiles of four newly formed pulsating
WDs of the GW Vir class. They find slow, solid-body, rotation
throughout $\sim 99\%$ of the WD mass. While still few, these
cases suggest strong coupling and efficient angular-momentum transfer between
the layers of the WD, which again argues against the differential WD
rotation that is essential for the spin-up/spin-down scenario.  
\cite{Corsico11}, applying the same methods to
another GW-Vir-type WD, find that the WD core may be spinning up to 4 times
faster than the surface, but within the uncertainties, solid-body
rotation cannot be excluded in this case.

Magnetic fields in WDs are also difficult to measure, unless the
fields are strong. Some 10-30\% of WDs have fields of $10^4-10^9$~G
\citep{Kawka07}, and $\sim 10$\% have  $10^3-10^4$~G 
\citep{Jordan07,Landstreet12}. \cite{Tout08,Nordhaus11}
and \cite{Garcia-Berro12} have argued that strong
magnetic fields in WDs are always the result 
of binary interactions, be it through accretion, 
common-envelope evolution, or mergers. In magnetic WDs, 
photometric variability due to 
spots and variable dichroic polarization 
open two additional avenues to measure rotation. 
While magnetic WDs tend to be significantly more massive than normal WDs 
(e.g. \citealt{Kepler13}), 
they rotate just as slowly \citep{Kawaler04}, suggesting that
WDs are unable to spin persistently, even after binary interactions.
 Several cases of non-variable magnetic WDs, suspected
as possibly unrecognized
fast rotators, have been shown to actually be 
very slow rotators \citep{Friedrich01,
Beuermann02}. However, \cite{Boshkayev13} 
have proposed that
soft-gamma-ray repeaters and anomalous X-ray pulsars could actually be
massive, rapidly rotating, magnetic WDs.

If spin-up/spin-down is a dominant SN~Ia progenitor channel, we can
estimate the number of WDs in Solar-neighborhood samples that could
reveal evidence of being such progenitors, 
again using a SN~Ia rate per unit mass in Sbc spirals 
of Milky Way mass, $1 \times 10^{-13}$~yr$^{-1}~M_\odot^{-1}$.
The  local stellar density is $0.085~M_\odot$~pc$^{-3}$
\citep{McMillan11}. 
If spin-down requires a Hubble time to
occur, there should be, within a distance of 100~pc, about 400 WDs 
with super-$M_{\rm Ch}$ masses, and rotations of order
1000~km~s$^{-1}$. If spin-down and explosion occur on
timescales of $\lesssim 10^8$~yr, only about four such
nearby fast rotators are expected. Nonetheless, in the latter case, 
being young and hence hot and
luminous, they should still be quite conspicuous \citep{Di-Stefano11}.
Upcoming large and complete WD samples, such as
those that will be discovered by {\it Gaia}, will further test this scenario.
  
\subsection{Pre-explosion evidence}
\label{sec:preexplosion}

The most direct way to resolve the progenitor problem would be to see
the progenitor system of an actual SN~Ia, before it exploded.
Unfortunately, and contrary to the situation for CC SNe,
where a good number of progenitor stars have been identified in
pre-explosion images (see review by \citealt{Smartt09}), no such
progenitor has ever been convincingly detected for a
SN~Ia. Pre-explosion optical images for several Virgo-distance SNe~Ia,
at the pre-explosion sites, have set upper limits on progenitor
luminosities, corresponding to supergiants evolved from stars of
initial mass $\gtrsim 8$~M$_\odot$ 
(\citealt{Maoz08b,Voss08,Nelemans08,Graur12a,Graur12b};
see also summary in \citealt{Li11c}).
\cite{Nielsen12} and \cite{Nielsen13c}  placed upper limits on the pre-explosion progenitor
X-ray luminosities at the sites of 13 nearby SNe~Ia, in several cases
limits that are comparable to the luminosities of supersoft X-ray
sources.  \cite{Voss08} did find a possible pre-explosion X-ray source at
the site of the SN~Ia 2007on.  However, the best-fit position of the
X-ray source (based on 14 photons) is offset by 1.1 arcsec from the
optical SN position, making this detection ambiguous. On the other
hand, the source was not detected after the event (albeit in shallower
images) and chance alignment of such a soft source, that almost
certainly originates in the same galaxy as the SN, is unlikely as
well \citep{Roelofs08}.

\subsubsection{The case of SN~2011fe}

The strongest pre-explosion limit on the presence of a SD donor star
was set in the case of SN~2011fe in the nearby (6.4~Mpc;
\citealt{Shappee13a}) galaxy M101.  Being the nearest SN~Ia event in
25 years,
and having been discovered very early, SN~2011fe was
extremely well studied, and has permitted many tests relevant to the
progenitor question 
(see \citealt{Chomiuk13} 
and \citealt{Kasen13}
for reviews).  
SN~2011fe was about as typical as a SN~Ia can
be, in all of its observed characteristics \citep{Mazzali13}, making
it particularly relevant for addressing the question of the
progenitors of SNe~Ia, as a population. As a final bonus, it had
negligible Milky Way and intrinsic line-of-sight extinction
\citep{Nugent11}.

\cite{Li11c} analyzed deep (2$\sigma$ limiting Vega magnitudes $\sim
27-28$) pre-explosion {\it Hubble Space Telescope} (HST) images of the
site of the event in four optical bands, corresponding roughly to $B$,
$V$, $I$ (all from 2002), and $R$ (from 1998). They used
adaptive-optics imaging of the SN, obtained with the Keck II
telescope, to determine the SN location on the HST images to 21~mas
precision. {\it Spitzer} images at 3.6~$\mu$m to 8~$\mu$m, and 142~ks
of {\it Chandra} images in the 0.3-8 keV band, both from 2004, were
also examined. No source was detected at the SN position.  In
addition, they searched about 3000 epochs of shallower, ground-based,
monitoring data of M101, from 12 years preceding the explosion. These
data set optical flux limits, reaching $\sim 20-23.5$~mag, on any pre-SN
variable or transient events at the SN location.

\begin{figure}[th!]
\centerline{
  \includegraphics[angle=0,width=0.6\textwidth]{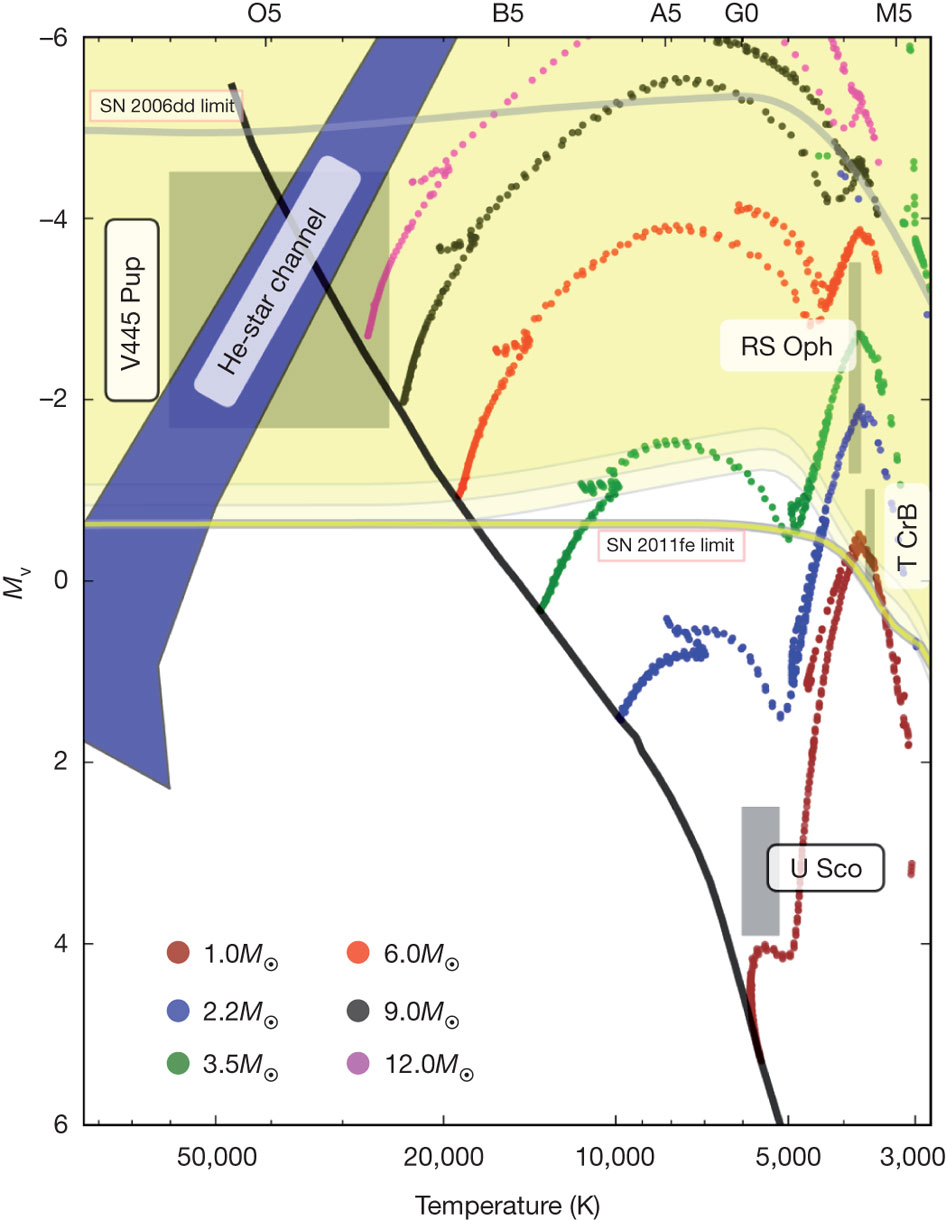}
}
\caption{
 HR diagram (absolute $V$ magnitude versus effective
 temperature) showing the $2\sigma$ upper limits (thick-yellow line) on the presence
 of progenitors in pre-explosion HST images of SN~2011fe in M101, from
 \cite{Li11c}. Also shown are theoretical evolution tracks of
 isolated stars with a range of masses, theoretical location of a SD
 He-star donor, and location on the diagram of
 several known recurrent novae. The data rule out red giants, and any
 evolved star more massive than $3.5~M_\odot$, as well as the
 recurrent nova systems above the limit. Gray curve is the 
corresponding limit  by \cite{Maoz08b} for 
the more distant SN~2006dd.   
}
\label{fig:Li11c}
\end{figure}

 From the HST data, \cite{Li11c} strongly rule out the presence of a
red giant at the location of the SN in the decade prior to the
explosion. Also excluded are any stars more massive than $3.5~M_\odot$
that have evolved off the main sequence. 
Two Galactic symbiotic recurrent novae, RS Oph and T
CrB, would have been detected in the data if placed, at their
quiescent luminosities, at the distance of M101. The ``helium nova''
V445 Pup in quiescence, where the donor is a helium star, would have
likewise been detected in the HST data. Conversely, main-sequence and
subgiant donors with masses below $3.5~M_\odot$ are allowed, and  
 a system like the recurrent nova U~Sco, in
which the donor is a main sequence star, would be undetected in
quiescence. Figure~\ref{fig:Li11c} shows these results. 

\cite{Li11c} found that a pre-explosion eruption at the site of
SN~2011fe, with a luminosity typical of nova outbursts, could have
been detected by the decade of optical monitoring data, but there is a
37\% probability that such an eruption would have been missed, given
the actual cadence of those observations. Analysis of the X-ray data
sets an upper limit on the pre-explosion bolometric X-ray luminosity
of $(4-25)\times 10^{36}$~erg~s$^{-1}$ (depending on the assumed
spectrum).  \cite{Nielsen12} used a deeper stack of of $800$~ks {\it
  Chandra} data to set a somewhat stronger X-ray upper limit in the
case of SN2012fe. A weaker limit was set by \cite{Liu12a}, based on
these {\it Chandra} data, but \cite{Nielsen12} point out possible
errors in the analysis. In any event, the X-ray data rule out the
pre-explosion presence, at the location of SN~2011fe, of typical
persistent supersoft X-ray sources, but allows the presence of
somewhat fainter ones.

\cite{Shara13} imaged M101 with HST, including the location of SN
2011fe, in 2010, about 1 year prior to the event, in a narrow band
centered on the He~II $\lambda 4686$ line. The hard photon flux
 from a nuclear-burning
accreting WD (see Section~\ref{sec:rapidaccretion}, should, in principle,
produce a 1-30-parsec-sized
(i.e. unresolved or resolved, depending on the outflow rate and the
surrounding gas density) He~III
Stromgren sphere in the interstellar medium (ISM) of the galaxy
\citep{Rappaport94}.
 Graur, Maoz, \& Shara (2013, in preparation) use
the HST data to set a $2\sigma$ upper limit on the pre-explosion He~II
line luminosity at the SN position, of $L_{\rm HeII}< 3\times 10^{34}~{\rm
  erg~s}^{-1}$, for an unresolved He~III region. From photoionization
models for a supersoft X-ray source ionizing a surrounding ISM of
density $n=10~{\rm cm}^{-3}$ \citep{Rappaport94},
this implies a limit on the X-ray luminosity of
$L_x < 2\times 10^{37}$~erg~s$^{-1}$. As opposed to the
limits based on direct X-ray observations, 
detailed above, on the presence of supersoft 
progenitors during the decade before the explosion, 
the He~II line 
emission limit is an indicator of the mean ionizing luminosity over
 $\sim 10^5$~yr 
(the He~III region's recombination time) prior to the explosion.

On the other hand, there is only one known case, CAL~83, of a 
supersoft X-ray source that has a detected ionization nebula, while
nine others that have been searched for such extended 
line emission have yielded only upper limits, at luminosity levels an order
magnitude lower than that of CAL~83 \citep{Remillard95}. 
Furthermore, the X-ray luminosity of CAL~83 
is $L_x=3\times 10^{37}$~erg~s$^{-1}$, but
its He~II
line luminosity is only $L_{\rm HeII}\approx 2\times 10^{33}~{\rm
erg~s}^{-1}$,  an order of magnitude below model
expectations \citep{Gruyters12}. Contrary to the H$\alpha$ and [O~III]
emission, which is roughly symmetrical around the source, the He~II
emission is concentrated on one side, within $\sim 1$~pc. 
The reasons are unclear for the
discrepancy between the observed supersoft ionization nebula
luminosities and model
expectations, but, in any case, 
an He-III region similarly
faint to that seen in CAL~83 would be undetected in the HST M101 data
at the site of SN~2011fe. 
Furthermore, if the
WD progenitor of SN~2011fe were in a rapidly accreting mode, the fast
wind would carve-out in the ISM a rarified cavity of order 30~pc 
(e.g. \citealt{Badenes07}). Such an extended, low-surface-brightness, 
emission-line region would be undetected in the HST image. 

In summary, the pre-explosion data for SN~2011fe rule out SD
systems with red giant and helium-star donors, bright supersoft
X-ray sources, and accreting WDs that produce significant
He~II ISM ionization, 
but allow main-sequence and sub-giant SD donors, and faint
supersoft sources. 
DD progenitors are of course not limited by these data, nor
spin-up/spin-down models with long delays between the end 
of accretion and explosion.

\subsection{Clues from during the SN events}
\label{sec:events}

The observed characteristics of the emission of a 
SN~Ia explosion itself, at various 
wavelengths and times, contains information on the exploding system,
including its progenitor aspect.

\subsubsection{Early light curve and spectral evolution}
\label{sec:earlylc}

Among the clues to the progenitor that, in principle, can be obtained
from observables obtained during a SN event itself, the early development
of the light curve and the spectra can be particularly revealing.  The
early light-curve evolution of a SN~Ia, starting from the time of
explosion, has been recently studied in a number of theory papers
\citep{Hoeflich09b,Piro10,Nakar10,Nakar12,
Kasen10,Rabinak11,Rabinak12,Piro13a,Piro13b}.
Briefly, the expected behavior is as follows.

Assuming that the explosion, which begins near the center of the WD,
at some point expands outward as a supersonic detonation (see
Section~\ref{sec:correlations}, below), the shock wave propagates
toward the surface, heating and igniting the WD material and giving it
an outward bulk velocity. The shock propagates most of the way at
$\sim 10^4$~km~s$^{-1}$, approaching the surface after $\sim 1$~s.
The shock accelerates to relativistic velocities in the dropping
density profile near the WD surface. When it emerges, it produces a
ms-timescale gamma-ray flash of energy $E_\gamma\sim 10^{40-41}$~erg,
constituting ``shock breakout'' \citep{Nakar12}.  Following
breakout, the radiation emerging from the explosion comes from the
matter that is outside of a ``diffusion front'', advancing from the
surface into the expanding ejecta below it. The radius of the
diffusion front is defined by the diffusion time of photons through
the outer envelope, at the time of observation.  The expanding
shock-heated ejecta quickly cools adiabatically. In the outer regions,
where pressure is radiation dominated, the temperature decreases
with growing radius $R$ as $T\propto R^{-1}$, while in the inner 
matter-dominated regions, the cooling is faster, $T\propto
R^{-2}$. The luminosity drops steeply for about 1~s after breakout,
until the diffusion front reaches material with sub-relativistic bulk
velocity. From this point on, the escaping bolometric luminosity
initially falls with time as $L_{\rm bol}\sim t^{-0.4}$, while the effective
temperature observed in the ejecta falls as $T_{\rm
  eff}\sim t^{-0.6}$.

The peak of the spectral energy distribution thus transits within
minutes from gamma-rays and X-rays to the UV band.  Optical-UV-band
observations will, at first, be on the Rayleigh-Jeans side of this
thermal spectrum, and will therefore see an optical luminosity that
rises as $L_{\rm opt}\sim t^{1.5}$, peaks after $\sim 1000$~s
at $L_{\rm opt}\sim 10^{39}$~erg~s$^{-1}$, and then falls as the Wien
peak enters the bandpass. After a time that is proportional to the
initial stellar radius, ($\sim 1$~hr in the case of a WD), the
diffusion front has traversed the outer $\sim 10^{-4}~M_\odot $ of the
WD mass, and enters the cooler, matter-dominated region of the
adiabatically expanding shock-heated ejecta, and therefore there is a
sharp drop-off in luminosity \citep{Rabinak12}.  The luminosity may
drop below detection limits, initiating a ``dark phase'' in the
light curve, lasting until photons powered by radioactive decay can
diffuse to the surface. This dark phase can last up to a few days,
depending on the depth of the radioactive material.
 
As the diffusion front enters the ejecta, the material contributing to
the observed luminosity includes progressively larger amounts of
$^{56}$Ni from the explosion. The energy from the gamma-rays formed by
the radioactive decay of $^{56}$Ni diffuses through the line-blanketed
ejecta and emerges in the optical.  This causes the classic observed
rise in the optical light curve, all the way until the peak of the
bolometric light curve at $\sim 18$~d, which occurs roughly when the
diffusion front has traversed all of the ejecta (and therefore the
luminosity from all of the $^{56}$Ni is observable). Beyond maximum
light, the fall of the light curve is determined by the exponentially
decreasing number of $^{56}$Ni nuclei and their radioactive $^{56}$Co
daughter nuclei, the changing UV opacity of the ejecta, and the
increasing fraction of gamma rays and positrons that can escape the
remnant due to the falling density.

Observations of the early light curve of any type of SN can provide
constraints on the progenitor in several ways. First, the time
between explosion and shock breakout, presuming the epoch of the
explosion can be estimated, provides a measurement of the
pre-explosion radius of the exploding star. (Shock breakout has been
seen, to date, in several CC SNe, see e.g.
\citealt{Tominaga11}). Non-detection of the
post-breakout thermal emission from shock-heated material, can also set
upper limits on the pre-explosion radius, as further detailed
below. Second, the observed time dependence of the $^{56}$Ni-powered
rise indicates the amount of $^{56}$Ni as a function of depth into the
ejecta, and can thus constrain the ignition and combustion scenarios,
and the amount of mixing in the ejecta. Third, in any
configuration of accretion onto a WD via Roche-lobe overflow from a
donor star, the L1 Lagrange point, and hence the donor, will be within
a few donor-star radii of the explosion.  Impact of the ejecta on the
donor is therefore expected within minutes to hours (for main-sequence
and giant donors, respectively). The consequences of the impact
should be visible for up to a few days. This third possibility is
discussed in Section~\ref{sec:shock}, below.

\begin{figure}[ht!]
\centerline{
  \includegraphics[angle=0,width=0.6\textwidth]{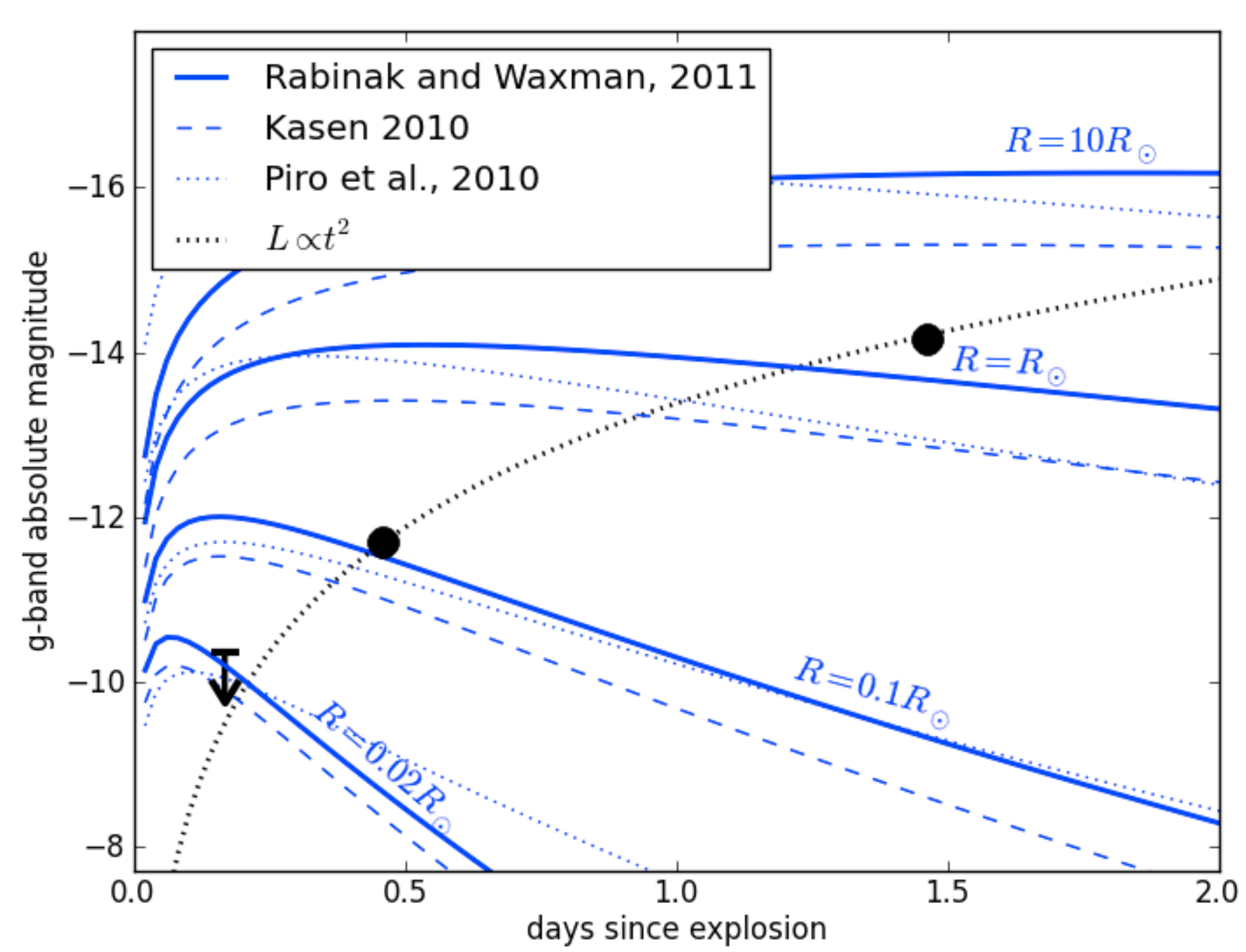}
}
\caption{
Early optical light curve for SN~2011fe, compared
to theoretical models. The two detections are from \cite{Nugent11}, and the
$5\sigma$ upper
limit is from \cite{Bloom12}. Black dotted curve shows the 
$t^2$ dependence due to radioactive heating
 expected in the simplified ``fireball'' approximation. 
The \cite{Rabinak11} and \cite{Piro10} models are for the 
shock-heated luminosity from an object of a given radius, as marked.
\cite{Kasen10} models are for a shock impacting a companion (see Section~\ref{sec:shock}, below), with
$R$ being the separation distance, for the case of an observer aligned
with the collision axis. The data indicate a small exploding star {\it
  and} the absence of a shocked companion, but details depend on
the assumed time of explosion. From  \cite{Bloom12}. 
}
\label{fig:Bloom12}
\end{figure}

The actual gamma-ray flash of the shock breakout in a SN~Ia is
expected to be too dim, too brief, and at too-high photon energies, to
be detected anywhere but in our Galaxy and perhaps in the Magellanic
Clouds. However, the UV/optical emission from the adiabatically
cooling, shock-heated, ejecta, on timescales $\lesssim 1$~hr, as well
as the earliest parts of the $^{56}$Ni-powered light curve, are
detectable in more distant events.  Among the best observations to date, in
this context, have again been those of SN~2011fe, which was
detected 18 days before maximum light, and quickly followed up in many
wavebands. \cite{Nugent11} found that the $g$-band luminosity of the
event is well fit by a $L_g\propto t^2$ dependence. Such a dependence
has often been used to
characterize the $^{56}$Ni-powered phase of the light curve, both 
theoretically, under some simplifying assumptions, 
\citep{Arnett82}, 
and observationally (e.g. \citealt{Hayden10a}, who found $t^{1.8\pm
  0.2}$ by fitting the average light curves of SDSS-II SNe).
Assuming also that this dependence can be extrapolated
back in time, the actual time of explosion was determined by
\cite{Nugent11} to $\pm 20$~min accuracy. They argued that their
earliest detection of the SN, corresponding to 0.5~day after the
derived explosion time, was already dominated by the $^{56}$Ni-powered
rise, and was therefore past the shock-heated phase of the light
curve.  By comparing their photometry to models by \cite{Piro10},
\cite{Kasen10},
and \cite{Rabinak11}, they translated the
limit on the duration of the shock-heated phase to a limit on the
radius of the exploding star of $R_*<0.1~R_\odot$. \cite{Bloom12} used
upper limits on the observed luminosity from a non-detection at the SN
location obtained 8~hr earlier, i.e. just 4~hr after the presumed
explosion time (see Figure~\ref{fig:Bloom12}), to argue that
the light curve was past the point of the drop-off in the shock heated
phase, occurring when the diffusion front reaches the shells with
gas-dominated pressure. This strengthened the limit on the initial
stellar radius to $R_*<0.02~R_\odot$. Setting aside, for a moment, the
progenitor scenarios being discussed in this review (SD, DD, etc.),
these limits are particularly interesting, as they constitute the most
direct evidence that the exploding object in a SN~Ia is, in fact, a
WD.


However, \cite{Piro13a,Piro13b} and \cite{Mazzali13} have questioned
the accuracy of these conclusions. \cite{Piro13a} show that, in
general for a SN~Ia, the $L_{\rm opt}\propto t^2$ ``fireball model''
dependence is not expected under realistic conditions for the
photospheric velocity (assumed to be constant in time in the fireball
model) and for the radial distribution of $^{56}$Ni in the ejecta
(assumed to be uniform by \citealt{Nugent11}).  \cite{Piro13a} show
that, for a range of possible conditions, the light curve is poorly
approximated by a single power law, and the back-extrapolated $L_{\rm
  opt}\propto t^2$ dependence of the optical luminosity does not
necessarily give the correct explosion time. \cite{Piro13b} fit the
observed photospheric velocity evolution in SN~2011fe (as measured by
\citealt{Parrent12}) with a $v\propto t^{-0.22}$ dependence (shown to
be expected theoretically in \citealt{Piro13a}) in order to constrain
the explosion time, and they fit models with varying $^{56}$Ni radial
distributions to the observed optical light curve. They conclude that
the true time of explosion is uncertain by $\sim \pm 0.5$~d, and could
be $\sim 1$~d earlier than concluded by \cite{Nugent11}. This, in
turn, weakens the upper limit on the stellar radius, obtained by
\cite{Bloom12} to $R_*<0.1~R_\odot$.  \cite{Mazzali13} reach a similar
conclusion by estimating the explosion time from their detailed
fitting of photospheric temperatures, abundances, and velocities to
early-time spectra (see Section~\ref{sec:spectra}, below). They also
favor an explosion time 0.9~d earlier than deduced by
\cite{Nugent11}, and find a limit of $R_*<0.06~R_\odot$.  These
somewhat weaker limits on the stellar radius are still quite useful
in ruling out most options for the exploding object in
SN~2012fe. \cite{Bloom12} show that, apart from a WD, only pure
carbon-oxygen stars, burning carbon stably, could have radii
consistent even with these weaker limits. 

The modeling of SN~2011fe 
by both \cite{Mazzali13} and \cite{Piro13b} lead them to 
conclude that significant amounts of
$^{56}$Ni, $\sim 0.1$ of the total amount, existed in the outer $0.1
M_\odot$ of the ejecta. 
\cite{Piro13b} find this not only for SN 2011fe, but also
for SN2012cg, another SN~Ia with early time data. The presence or
absence of $^{56}$Ni in the outer layers is relevant to understanding
the explosion details, especially the possible transition from a
deflagration to a detonation (see Section~\ref{sec:correlations},
below).

\cite{Zheng13}
have discovered a SN Ia, 2013dy, caught even earlier in its light 
evolution. With good light curve coverage in the first few days, they 
find that, indeed, a $t^2$ behavior fits the data poorly. Instead,
they obtain a best fitting broken power law that begins with an index 0.9
and transits to an index 1.8 after about 3 days. Extrapolating this
fit back in time to zero flux implies that the SN's ``first light''
occurred
 just  $2.4\pm1.2$~hr before the first detection. The non-detections at
earlier times again argue for
upper limits of $\lesssim 0.25~R_\odot$ on the progenitor radius.
 
\subsubsection{Shocks from ejecta impacting companion stars and debris}
\label{sec:shock}

As already noted, in the classic SD picture, shocks from the SN ejecta hitting
the companion star are unavoidable, and may be observable.
\cite{Kasen10} has calculated the observational signatures of such
events, as a function of the companion size (giants naturally produce
the largest signals), and as a function of the angle between the
observer, the WD, and the donor. When a giant star is between the
observer and the WD, the signatures of the impact are expected in
X-rays, within minutes to hours of the explosion. 
Emission from the shocked region in
ultraviolet/blue bands develops on a timescale of several days.
\cite{Hayden10b, 
Bianco11,Tucker11}, and \cite{Ganeshalingam11}
have set observational upper limits in blue optical bands in
the light curves of SNe~Ia with good early-time optical coverage, and
have used the \cite{Kasen10} models to limit the presence of red-giant
donors to $<20\%$ of SN Ia progenitor systems.  
\cite{Foley12e} and \cite{Brown12a} 
have extended this approach using space-UV
measurements to obtain such limits in several cases.
 
This type of limit is strongest, here again, in the case of the nearby
and promptly observed SN~2011fe in M101.  \cite{Nugent11} and
\cite{Bloom12} used the same early optical observations (discussed in
Section~\ref{sec:earlylc} above, in the context of
emission from shock-heated ejecta) to exclude also shocks
from ejecta hitting a companion. \cite{Brown12b} use prompt UV
observations of SN~2011fe, obtained with {\it Swift} starting 18~hr
after the first \cite{Nugent11} epoch, to do the same.  All three
studies rule out red giants donors, \cite{Brown12b} exclude also
main-sequence companions of radius $R_c\gtrsim R_\odot$, and
\cite{Bloom12} sharpen this limit to $R_c\lesssim 0.1 R_\odot$.
Figure~\ref{fig:Bloom12} shows also 
these limits. The conclusions, like those in
Section~\ref{sec:earlylc} above, are sensitive to the exact time of
the explosion.

 This limitation has been circumvented by Olling, Mushotzky \& Shaya
 (2013, submitted) and Olling et al. (2014, in preparation), 
who used the Kepler Mission to monitor continuously
 over several years, with 30-min exposures, 400 galaxies in the Kepler
 field. Among these galaxies, they discovered four SNe, two or three of
 them likely  SNe Ia, based on their light curves. The continuous
 coverage reveals no signatures of companion impacts within the few
 days before first light, when the actual explosion must have occurred.
 Olling et al. again use the Kasen (2010) models to limit the presence
 of red giant and main-sequence companions for a range of viewing
 angles. An extension of the Kepler mission, particularly in sky
 directions that would allow ground-based followup and classification,
 would permit obtaining such data for many additional events, which
 would set strong constraints on the fraction of SNe Ia in which a SD
 donor star is present.

Interestingly, \cite{Nugent11} have used the absence of impact
signatures in SN~2011fe to argue also against a DD progenitor
system. This argument is based on calculations by \cite{Fryer10},
who simulated DD mergers and found that a potentially large mass of
debris, $0.1-0.7 ~M_\odot$, forms an inhomogeneous envelope around the
merged object, with a steeply falling density profile extending out to
$\sim 1~R_\odot$. This is based, in turn,
 on BPS calculations \citep{Ruiter09}
that predict the majority of DD mergers to have masses around
2 \Msun, whereas most other BPS studies predict lower masses (see
also Section~\ref{sec:correlations}).  \cite{Fryer10} calculated
the shock breakout and early-time spectral evolution in this model,
and found behavior very distinct from that expected of a ``bare''
exploding WD, with UV emission lasting for days, and similar to the
early-phases of Type II SNe. However, \cite{Fryer10} also found
that, in general, the spectra and light curves produced in the context
of this model bear little resemblance to those of normal SNe~Ia. The
results of the calculations, if taken at face value, can therefore be
considered evidence against the DD scenario, irrespective of the lack
of shock signatures in the \cite{Nugent11} data for a very typical
SN~Ia.  However, DD merger simulations differ regarding the amount of
debris at small distances at the time of explosion
\citep{Dan12,Shen13b,Raskin13a}.
 As explained by \cite{Pakmor12}, one critical
ingredient is the time elapsed from merger until explosion. In the
violent-merger scenarios, where ignition occurs within minutes of the
merger, any merger debris, moving at the escape velocity of a few
thousand km~s$^{-1}$, will be overtaken by
the ejecta, and will cease to interact with it,
minutes after the explosion (see, in this context, also
\citealt{Soker13c}).
Also important, for the amount and location of debris, are the WD mass
 ratio and the assumed WD spin initial conditions \citep{Dan13}.
       
\subsubsection{Spectral and light curve modeling}
\label{sec:spectra}

As the ejecta of a SN expands, the photosphere recedes into
progressively deeper layers, revealing the chemical stratification and
velocity structure of the ejecta. These in turn, may help distinguish
among progenitor models.  A way to reconstruct this physical structure
is through calculation of synthetic spectra and light curves that
match the observations.

In one such modeling approach 
(e.g. \citealt{Stehle05,Mazzali07,Tanaka11,Hachinger13}), an
ejecta density profile is assumed, based on a hydrodynamical explosion
model. A Monte Carlo radiative transfer calculation is used to
propagate photon packets from a blackbody spectrum
 through the model photosphere. The element abundances
and the velocity at each radius are varied in order to obtain the best
fit to the set of observed spectra and photospheric velocities.
\cite{Mazzali13} have recently applied the method to the combined
UV spectra from {\it HST} and ground-based optical spectra for
SN~2011fe. A good match to the data is found using a hybrid 
between the density profiles from the SD one-dimensional (1D)
fast-deflagration W7 model of \cite{Nomoto84}, and from a 1D
delayed-detonation model from \cite{Iwamoto99} (more on deflagrations
and detonations in Section~\ref{sec:correlations},
below).

\cite{Blondin11b} have carried out a more
``first-principles'' approach by taking a set of 44 hydrodynamical 2D
$M_{\rm Ch}$ SD delayed-detonation models from \cite{Kasen09},
and comparing them to observed light curves and
multi-epoch spectra of 251 SNe Ia.  The details of the nucleosynthesis
and the resulting element distributions were calculated by means of
tracer particles distributed throughout the volume, with nuclear
reaction networks calculated for the conditions at each
location. Finally, photon packets were again Monte-Carlo-propagated through
the ejecta, to obtain synthetic light curves and spectra at the epochs
corresponding to the observations. Overall good agreement is found
between the data and some of the models, particularly those that
satisfy the \cite{Phillips93} relation, but some details do not match, such as
too-high model velocities in the maximum-light spectra.

\cite{Ropke12} have extended this approach, to test example
models from the two main progenitor scenarios, SD and DD,
 against optical spectra of SN 2011fe.  One model is a $M_{\rm Ch}$ SD 
delayed-detonation
model from \cite{Seitenzahl13a}, and the other is a violent-merger
DD model from \cite{Pakmor12}, merging $1.1~M_\odot$ and
$0.9~M_\odot$ WDs.  For each model, \cite{Ropke12} used a 3D
hydrodynamics code to calculate the evolution of the density and the
temperature. Nucleosynthetic abundances are again calculated via
tracer particles and Monte-Carlo radiative transfer is performed.
Both progenitor models produce reasonable matches to the observations
but, again, both also have shortcomings in matching the
details. 
Overall, the DD model fares somewhat better in this
comparison.  The differences in the predictions of the two models, and
the discriminating power of the data, suggest that future applications
of this approach, with a larger range of 3D models, should be
very useful. 

The amount and velocity distribution of unburnt carbon observed in SN
Ia ejecta can be a diagnostic that distinguishes among scenarios.
\cite{Parrent11,Thomas11b,Folatelli12}, and \cite{Silverman12} analyze
several samples with pre- or near-maximum spectra, and estimate that
the C~II~$\lambda 6580$ absorption feature is present in about 30\% of
SNe Ia.  In all studies, the feature tends to appear in events with
``low velocity gradients'' (see Section~\ref{sec:correlations},
below). The implied carbon mass is $10^{-3}-10^{-2} M_\odot$.  In
SN~2011fe, \cite{Nugent11} clearly detect this line in the earliest
high-resolution spectrum, along with high-velocity O~I
absorption. Both features almost disappeared in a spectrum obtained
only 8 hours later. From their modeling of these data,
\cite{Mazzali13} deduce the existence of an almost-pure carbon outer
layer of $\sim 0.01 M_\odot$. They speculate that this layer reflects
the properties of the accreted material, whether hydrogen or helium
that has burned to carbon. \cite{Soker13c} propose, instead, that the
outer-layer carbon enrichment is a byproduct of crystallization in a
WD that has been rotation-supported against collapse and has cooled,
 for $\sim 1.4$~Gyr, in a spin-up/spin-down scenario, such as core-degenerate.
  \cite{Zheng13} similarly see strong C~II
absorption in SN~2013dy, within the first few days after explosion,
weakening and becoming undetectable within a week. 

\subsubsection{Diversity and correlations among spectral and
  environmental observables}
\label{sec:correlations}

Observers have long sought, and sometimes found, relations 
among the observed properties of SN~Ia
events. These include relations among the various spectral and
photometric properties, and between those properties and the
properties of the host galaxies as a whole, or of the specific
locations of SNe~Ia within a galaxy.  An observed correlation,
assuming it is not due to selection effects, may be merely a
consequence of some basic physics of SN~Ia explosions in general,
rather than of a specific progenitor channel leading up to the
explosion. Nevertheless, in such relations there is a potential for
clues to the progenitor problem.

{\bf The width-luminosity relation:}
The strongest and clearest correlation seen among SN~Ia properties is
the so-called \cite{Phillips93} relation, connecting the maximum-light
luminosity of an event to its light-curve evolution (as parametrized
in a variety of ways), and to the observed color at maximum light
(e.g. \citealt{Kattner12,Hillebrandt13}). Less-luminous SNe
Ia, which have synthesized less $^{56}$Ni, evolve more quickly.
Less-luminous SNe~Ia are also redder, due to some combination of
intrinsically redder color, and reddening by dust,
a combination that has proved hard to disentangle. The problem is
compounded by evidence, often contradictory, that the dust involved in
the effect might have properties distinct from Milky Way dust, with a low
ratio of extinction to reddening, $R_V\equiv A_V/E(B-V)\approx 1.5-2$,
as opposed to the typical Galactic $R_V\approx 3.1$ (see
\citealt{Howell11}, but \citealt{Scolnic14} for a different view).

The Phillips relation itself may be a consequence of SN~Ia physics,
rather than of a particular progenitor scenario. The mass of
synthesized $^{56}$Ni largely determines the total energy of the
explosion.  The larger the $^{56}$Ni mass, the higher also the peak
luminosity due to the radioactive decay of $^{56}$Ni. In parallel,
iron-group elements also cause the bulk of the opacity, slowing the
evolution of the light curve, i.e., leading to a higher light-curve 
``stretch'' factor (e.g. \citealt{Hoeflich96,Kasen07}; 
see \citealt{Kasen09}
for additional factors, such as viewing angle,
that can affect the relation).  Nevertheless, progenitor models invoking
explosions always near $M_{\rm Ch}$ must explain the diversity, of
factor $\sim 4$, in explosion energy among ``normal'' SNe~Ia (\citealt{Branch93}, i.e.  excluding
under-luminous, SN1991bg-like, and over-luminous, SN1991T-like, events).

\cite{Timmes03} have proposed progenitor metallicity as the main
driver for explosion energy. Main-sequence stars with a higher
abundances of CNO produce, during helium burning, higher abundances of
neutron-rich nuclei, particularly $^{22}$Ne, that end up in the WD.
In a WD undergoing combustion as a SN~Ia, neutron richness leads to
more synthesis of stable elements such as $^{54}$Fe and $^{58}$Ni, at
the expense of $^{56}$Ni, and hence to a lower luminosity. Although
\cite{Timmes03} show that a large-enough range in progenitor metallicity
could reproduce the observed range in normal SN~Ia explosion energies,
\cite{Piro08a} and \cite{Howell09a} have shown that the
observed range of SN~Ia host galaxy metallicities does not reach the
extreme values required for this mechanism to have a significant
effect on SN~Ia explosion energy. The observed metallicity range of
SN~Ia host galaxies, even accounting for local extremes within a
galaxy, would lead to $<10\%$ variation in explosion energy.
\cite{Piro08a} further showed that pre-explosion neutron richness is
not controlled solely by metallicity, but rather can be dominated by
neutronization through carbon burning during the $\sim 1000$~yr 
``simmering'' phase that $M_{\rm Ch}$-explosion models generally undergo 
before explosion. 
\cite{Mazzali06} showed that, even if metallicity
variations produce the range in peak luminosity, they will not
 affect light curve shape, i.e. the Phillips relation is not reproduced.
\cite{Foley13a} have analyzed two ``twin'' SNe Ia (2011fe and
2011by) that have identical light curve shapes and optical spectra,
but differ in their UV spectra and in their peak
luminosity. Modeling the difference in UV spectral opacity by means
of progenitor metallicity differences, the \cite{Timmes03} model correctly
predicts the difference in peak luminosity. Metallicity effects may thus be at
work, even if they do not explain the explosion energy diversity or
the Phillips relation. 

Another way to explain the observed diversity in explosions of 
$M_{\rm Ch}$ models is through the so-called ``deflagration-to-detonation
transition''. It has long been known that the
energetics and spectra of 
$M_{\rm Ch}$ models do not match observations,
unless finely (and artificially) tuned as an initial subsonic
deflagration that allows the WD to expand and, at the right 
time, spontaneously to evolve into a supersonic detonation 
\citep{Khokhlov91,Hillebrandt00,Blondin13,Ma13b}.
  Different progenitor metallicities and
masses, and different accretion histories, could affect the number and
location of ignition ``kernels'', and thus the transition to
detonation
(e.g. \citealt{Umeda99,Kasen09,Blondin11b,Seitenzahl13a}).
Different transition times lead, in the final ejecta composition, to
different proportions of unburnt carbon and oxygen, intermediate-mass
elements, stable iron-peak elements, and light-curve-driving
radioactive $^{56}$Ni.  This, in turn, could perhaps produce the
observed diversity in SN~Ia explosion energy (see, in this context,
also ``pulsational delayed detonation'' models, e.g. \citealt{Dessart13b}). 

DD models can also readily produce a range in explosion
energies. \citet{Ruiter13} find, for ``violent'' mergers that include
detonation of small helium layers on top of the C/O WDs
\citep{Pakmor13}, that the explosion energy varies with the mass of
the primary WD in the binary, and can reproduce the full observed
range of SN~Ia luminosities by means of primary masses between 0.9
  and 1.3 \Msun. Combining this with a BPS calculation, they find
that the relative frequency of SN Ia luminosities can also be
reproduced, if the primary WD accretes a substantial amount
of helium from the companion star when it is a helium giant, prior to
becoming the secondary WD. In their own violent-merger simulations,
\cite{Moll13}, also reproduce the range of SN~Ia luminosities and the
Phillips relation through a combination of primary WD mass and viewing
angle of the highly asymmetric explosion.
\cite{Taubenberger13a} analyze the late-time
``nebular''-epoch spectra of several ``super-Chandrasekhar'' SNe~Ia to
estimate $\sim 2M_\odot$ of ejecta, about half of it in $^{56}$Ni,
which they interpret as possibly arising from the merger of two
massive WDs. In nebular-phase spectra of the sub-luminous SN~Ia
2010lp, \cite{Taubenberger13b} find oxygen emission lines which, they
argue, supports a violent merger scenario.  In their collisional
DD model, \cite{Kushnir13a} also reproduce the observed range of
$^{56}$Ni masses via different combinations of WD masses. The typical
observed $^{56}$Ni mass of $0.6 M_\odot$ requires collisions of 
WDs of at least $ 0.7 M_\odot$ each (see also \citealt{Piro13d}).

{\bf Explosion energy, colors, spectral features, and velocities:}
A second, potentially strong, correlation between SN~Ia properties 
that has been claimed is
between luminous energy (as measured by the light curve ``stretch'')
and the kinematic width of iron emission features, particularly
[Fe~III]~$\lambda 4701$, in late-time, nebular-phase, spectra 
\citep{Mazzali98}.  \cite{Blondin12} and \cite{Silverman13c}
have argued that the
relation is driven only by very under-luminous events, and no
correlation remains once they are excluded. \cite{Kushnir13a}, however,
have re-measured the velocities in the late-time spectra by fitting
broadened narrow-line templates, rather than measuring the width of
individual lines. With this procedure, they recover a correlation
between $^{56}$Ni mass and late-time velocity width.  Physically, such
a correlation may simply mean that more energetic explosions produce
larger ejecta velocity distributions, without particular relevance for
the progenitor issue.  

Other correlations that have been reported between observed SN~Ia
parameters tend to be rather weak trends, whose details and strengths
sometimes vary among different samples. The blueshifted velocity of
the Si II $\lambda 6150$ and $\lambda 6355$ absorptions at maximum
light, $v_{\rm Si}$, tracing the photospheric expansion velocity of
the ejecta, has been crossed with various other observables.
\cite{Foley11a} found that $v_{\rm Si}$ correlates with stretch, but
that this is driven solely by the least-luminous events with
$\Delta_{m15,B}>1.5$ (where $\Delta_{m15,B}$ is the decrease in $B$
magnitudes at 15 days after maximum light, and is inversely related to
stretch). Without those events, there is no significant trend between
stretch and ejecta velocity.  \cite{Wang09i},
\cite{Foley11a} and \cite{Foley11b} do
find a weak trend for events with faster ejecta to have redder
intrinsic colors, and \cite{Foley12a} finds a weak anti-correlation
between $v_{\rm Si}$ and host-galaxy mass.  \cite{Maguire12} show a
weak trend of increasing near-UV Ca~II H \& K absorption velocity and
stretch (but \citealt{Foley13d} argue that this is an artifact).
They also find a rather significant dependence
on stretch of the blueshift of the ``$\lambda_2$'' pseudo-emission
feature at 3180~\AA. The ``$\lambda_2$'' feature is thought to arise
from a spectral interval of slightly lower relative opacity in the
heavily line-blanketed near-UV region \citep{Walker12}.
\cite{Wang13b} have shown that $v_{\rm Si}$ correlates with a number
of environmental parameters. High-$v_{\rm Si}$ events tend to occur at
relatively small normalized galactic radii, while the lower-$v_{\rm
  Si}$ events ($<12,000 {\rm km~s}^{-1}$) occur at all
radii. Furthermore, the stellar surface brightness distribution at the
locations of high-$v_{\rm Si}$ events is similar to that of CC SNe,
suggestive of a young progenitor population, while lower-$v_{\rm Si}$
events tend to come from regions of relatively lower surface
brightness. Finally, high-$v_{\rm Si}$ events tend to be in
larger-diameter and more luminous galaxies (in apparent contradiction
to the inverse relation with galaxy mass observed by
\citealt{Foley12a}).  \cite{Wang13b} interpret these results in terms
of two different SN~Ia progenitor populations, distinguished by age
and metallicity. The observed correlations are intriguing, but the
conclusion may be premature, given the limited direct evidence for a
dependence on age and metallicity, and given the different
phenomenology that is observed regarding metallicity
(\citealt{Howell09a}, see above) and regarding progenitor age (see
below). 
We note here, in passing, that \cite{Foley12c} find a
correlation between $v_{\rm Si}$ and the probability for the presence
of a blueshifted absorbing Na~I~D system in the SN spectrum, a result to
which we return in Section~\ref{sec:absorption}, below.

\cite{Benetti05} have separated SNe Ia into two classes -- those
with ``low velocity gradients'' and those with ``high velocity
gradients'' -- based on the decrease with time in the velocity of the
Si~II $\lambda 6355$ absorptions, $\dot v_{\rm Si}$, with the border
at 70~km~s$^{-1}$~d$^{-1}$. Events with high velocity gradients
tend to also have high velocities at maximum light \citep{Foley11b}.
\cite{Maeda10} showed, for a sample
of 20 SNe~Ia, that high-velocity-gradient objects, in their late
nebular phase, have nebular lines of [Fe~II]~$\lambda 7155$ and
[Ni~II]~$\lambda 7378$ that are redshifted relative to the
higher-excitation line of [Fe~III]~$\lambda 4701$, while
low-velocity-gradient objects tend to have velocities that are
blueshifted.  \cite{Maeda10} interpreted this result, in the
context of the deflagration-to-detonation transition, by means of a
combination of an off-center-ignited explosion and a viewing angle
effect. In their picture, the nebular [Fe~II] and [Ni~II] lines come
from the ashes of the initial deflagration phase, while the [Fe~III]
line, which requires low density and heating by $^{56}$Ni decay,
traces the detonation ashes.  If the explosion is ignited off-center,
e.g. within the WD hemisphere on the observer's side, the
deflagration-ash lines will be blueshifted with respect to the
detonation-ash lines. The observer will see a low velocity gradient
in the photospheric lines because of the density and the velocity
structure encountered by the photosphere as it moves inward. 
\cite{Maeda11} and \cite{Cartier11} both find that the nebular line
shifts also correlate with the SN colors near maximum light,
strengthening the same idea of a dependence of observed properties on
viewing angle.

Another spectral signature in SNe Ia, that tends to appear in early-time
spectra of some events, is the ``high-velocity features'' 
(e.g. \citealt{Gerardy04,Mazzali05}), absorptions in 
Si~II $\lambda 6355$ or the Ca~II NIR triplet, whose velocities are
higher by several thousands of km~s$^{-1}$ than those of the normal
photospheric absorptions in these and other lines. The origin of these
features is unclear. Recently, \cite{Childress14}
have studied them in a sample of 58 SNe. They find that the existence
and strength of high-velocity features correlates with SN Ia stretch,
and anti-correlates with $v_{\rm Si}$. In a subset of SNe with weak
high-velocity features, stretch and $v_{\rm Si}$ are correlated.
 
{\bf SN stretch/luminosity and host-galaxy age/star-formation rate:} 
A correlation between peak SN Ia luminosity, or light-curve stretch,
and the age of the host galaxy has been noted for some time: the
oldest hosts, with little star formation, tend to host faint,
low-stretch, SNe~Ia, while star-forming galaxies are more likely to
host bright-and-slow SNe~Ia
(e.g. \citealt{Hamuy00,Howell09b,Hicken09b, Lampeitl10,Smith12,Xavier13,Pan13}). 
\cite{Rigault13} 
have shown this trend
at the local level in SN~Ia host galaxies, using  local H$\alpha$
surface brightness as a star-formation tracer.
While all of these studies see the same general trend,
 the scatter in 
the relation is large, and the various samples give different
pictures of how the trend arises. For example, in \cite{Hamuy00}, the
trend is driven both by a lack of bright SNe in early-type galaxies,
and a lack of faint SNe in late-type galaxies.  In \cite{Howell09b},
galaxies with luminosity-weighted ages $\lesssim 4$~Gyr are seen to
host SNe~Ia of all explosion energies, but galaxies with ages $\gtrsim
4$~Gyr host almost only faint SNe~Ia.  Conversely, in
\cite{Hicken09b} and \cite{Smith12}, 
early-type galaxies host the full range of
light-curve widths, but late-type galaxies host only SNe~Ia
with broad light-curve shapes. This situation likely arises
from a combination of effects. First, the trend, as noted, has a large
scatter to start with. Second, different samples may have different
selection effects that could bias for or against the inclusion of more
or less luminous SNe~Ia hosted by different types of galaxies. This
can become an issue particularly when SN samples are small, such that
regions of the age-luminosity plane that have low SN rates are
sparsely populated.  The appearance of these scatter plots is
likely affected by the different ways of defining light-curve
parameters (this is evident from comparing the plot in
\cite{Hicken09b}, that uses the $x_1$ light-curve stretch parameter,
with their plot that uses the $\Delta$ light-curve-width parameter), or
by the surrogates used for the explosion energies (as in
\citealt{Howell09b}).  Similarly, there are different ways of defining
galaxy morphological type, or of considering a galaxy ``age'', which
is a poorly-defined concept (galaxies have composite populations, and
their full star-formation histories are the more relevant observable).
Finally, galaxies are characterized by multiple correlations between
morphology, age, size, metallicity, star-formation rate, extinction,
and more (e.g., \citealt{Mannucci10}), and it is not clear what is 
the main driver of the
correlation with SN~Ia luminosity.  We revisit this issue when
discussing SN~Ia rates and the delay-time distributions for SNe~Ia of
different stretches, in Section~\ref{sec:rates}, below.  In the
meantime, we note that this trend between explosion energy and
host-population youth could well be an important clue to the
progenitor question.

{\bf Synopsis of correlations:}
  From all of the above, there are relations among
 the observed properties of SNe~Ia,
 and one can try to sketch a rough trend of properties that go together (see
 e.g. \citealt{Maguire13}). At one extreme are events that are luminous,
 have broad light curves, red intrinsic colors, high-velocity ejecta, 
large velocity gradients, redshifted and broad nebular lines, 
no carbon signatures, 
and star-forming hosts. At the other extreme are events with the
 opposite properties. We further discuss the possible source of such a
 trend in Section~\ref{sec:Summary}, below. 

{\bf Peculiar SNe Ia:}
Apart from the diversity among normal SNe Ia, there is a growing ``zoo''
of abnormal events that may nonetheless belong to the SN Ia class, though
with varying degrees of certainty. The numbers of known peculiar
SNe~Ia have grown as a result of recent untargeted field transient surveys
with large SN number yields, 
fainter flux limits, and faster cadences than previous surveys.
Their pecularities could shed light on the
progenitor question if, e.g., the pecularities can be  identified with
some particular progenitor channels or with 
specific deviations from the progenitor parameters of normal events. 

 Of particular recent interest is the
class of explosions, sometimes
called ``SNe Iax'' (e.g. \citealt{Foley13b}), 
that are similar to the prototype event SN~2002cx. 
Although spectroscopically similar to normal SNe Ia,
SNe Iax are characterized by low photospheric velocities at maximum light, of 
2000-8000~km~s$^{-1}$, hot photospheres, based on the presence of 
high-ionization
lines, and peak luminosities that are typically several 
magnitudes faint for the observed
light-curve width. Their light curves lack the ``second IR bump''of
normal events.
At late times, instead of the nebular-phase
broad forbidden lines seen in normal SNe Ia, the spectrum  shows narrow 
permitted lines, indicating a high gas density, $n_e> 10^{9}~{\rm
  cm}^{-3}$, even over a year after explosion
(e.g. \citealt{McCully13}). 
 Their spectra always show carbon features, and sometimes helium. 
Only a few tens of SNe Iax are known, but accounting for their
detectability, \cite{Foley13b}
estimate that there are $\sim 20-50$ such events for every 100 normal SNe Ia,
which would make SNe Iax the most common type of peculiar SNe Ia. 

\cite{Jordan12,Kromer13} and \cite{Fink13}
have modeled SNe Iax as ``failed deflagrations'' -- SN Ia
explosions in which a transition from deflagration to detonation
fails to occur, and furthermore the explosion fails to
completely unbind the WD, leaving behind a $\sim 1$~\Msun~ bound remnant.
 This could explain the
low $^{56}$Ni yields, the low velocities, the unburnt carbon and helium
 in the ejecta, the high degree of mixing, and the clumps of high-density
 material. On the other hand, SNe Iax 
occur predominantly in star-forming galaxies 
(but there is one case, SN2008ge, occurring in a S0 galaxy, 
with no signs of star formation or pre-explosion massive
stars at the explosion site, \citealt{Foley10b}), 
and their
locations within these galaxies track the star-formation rate
similarly to the common Type-IIP CC-SNe \citep{Lyman13}.
Indeed, \cite{Valenti09}
have argued that SNe Iax
are actually CC-SNe with low ejecta velocities, derived from
$7-9$~\Msun~ or $25-30$~\Msun~ progenitors, with cores collapsing into
black holes. 
On the other hand, \cite{Foley10c}
have pointed to the presence of sulfur in the spectra of some SNe Iax,
as evidence for thermonuclear burning in a C/O WD, and against a CC-SN.
It thus remains to be seen whether or not  SNe Iax are the manifestations of
incomplete pure deflagrations of $M_{\rm Ch}$ WDs.


\subsubsection{Polarization and symmetry}
\label{sec:polar}

Polarization can reveal deviations from circular symmetry of a source,
as projected on the sky. In the context of SNe, linear polarization
arises from electron and line scattering of photons emerging from an
asymmetric source. The percentage of total flux that is polarized is
proportional to the projected geometrical axis ratio, and the angle of
polarization shows the asymmetry's orientation on the sky. The
bandpass dependence of continuum polarization, and the polarization
across individual line profiles, can provide further information on
asymmetry and orientation for specific emitting elements and
velocities, and can show how they change with time as a SN
evolves. \cite{Wang08b} have reviewed spectropolarimetry of
SNe, including SNe Ia. Spectropolarimetry can guide the SN Ia
progenitor problem, since most SN Ia models have some built-in
asymmetry, such as the presence of a donor or an accretion disk, a
recently merged or collided secondary WD, rotational flattening, or an
off-center ignition (e.g. \citealt{Ropke12}).

As summarized by \cite{Wang08b}, SNe Ia tend to have low or zero
continuum polarizations, typically fractions of a percent, and
polarization is generally detected only pre- or near maximum
light. This indicates that the expanding photospheres are slightly
asymmetric on the outside, at levels of up to of order 10\%, but quite
round in the inner layers of the explosion. On the other hand, line
polarizations of up to a few percent are sometimes seen in some line
transitions, sometimes with different orientations than those of the
continuum. While, again, line polarization tends to be seen at early
times, there are exceptions (e.g. \citealt{Zelaya13}).  The line
polarizations have been interpreted in terms of clumpy distributions
of the ejecta of particular elements \citep{Kasen03,Kasen09,Hole10}.
\cite{Smith11c} have presented spectropolarimetry of the nearby
SN~2011fe. Like its other properties, its polarization behavior is
typical of normal SNe Ia.

\cite{Wang07} have shown a correlation between 
Si~II~$\lambda 6355$ polarization and $\Delta m_{15}$,
suggesting that more luminous SNe Ia tend to be more symmetric, but
this relation is driven by one or two objects, after excluding some outliers.
\cite{Tanaka10} have extended this result to the case of the
proposed super-$M_{\rm Ch}$ event SN 2009dc. Although this SN's line
polarization is somewhat above that expected from the \cite{Wang07}
trend, the low continuum polarization argues for little
asphericity. \cite{Tanaka10} use this to argue against the
explanation by \cite{Hillebrandt07} that this was an off-center
$M_{\rm Ch}$ explosion, and in favor of a super-$M_{\rm Ch}$
progenitor. However, asymmetry would be expected also from a
super-$M_{\rm Ch}$ rotation-supported configuration.  In 
the normal 
 SNe~2012fr and 2011fe,
\cite{Maund13} and \cite{Soker13c}, respectively,
have used the very low continuum
polarization level ($<0.1\%$), and the implied circular symmetry, to
argue against the violent DD-merger scenario \citep{Pakmor12}, in
which an asymmetric $^{56}$Ni distribution is predicted. At the other
extreme of the same trend, \cite{Howell01} and \cite{Patat12}
have used spectropolarimetry to deduce significant,
co-aligned asphericity in both lines and continuum in the sub-luminous
SNe Ia 199by and 2005ke, respectively, supporting fast rotation or a
merger origin in these cases.

\cite{Leonard05,Wang06b}, and \cite{Patat09a}
show that high Si~II~$\lambda 6355$ line polarizations and line
velocities are correlated.  \cite{Maund10} find, for a sample of
normal SNe Ia, that the polarization of the Si~II~$\lambda 6355$ line
at 5 days pre-maximum is correlated with the line velocity's time
gradient $\dot v_{\rm Si}$ for a given SN. They suggest that
normal SNe Ia have a single, asymmetric distribution of intermediate
mass elements, with the diversity of observed properties arising
partly from orientations effects.

At the two extremes in trends in SN~Ia properties, discussed in
Section~\ref{sec:correlations}, above, it is not clear in which
``family'' fall SNe Ia with relatively high polarization. Based on
\cite{Maund10} it would be in the high-velocity, high-stretch
family. Based on \cite{Wang07}, it would be in the opposite family.


\begin{figure}[ht!]
\centerline{
  \includegraphics[trim= 0mm 10mm 0mm 152mm, clip, angle=0,width=0.6\textwidth]{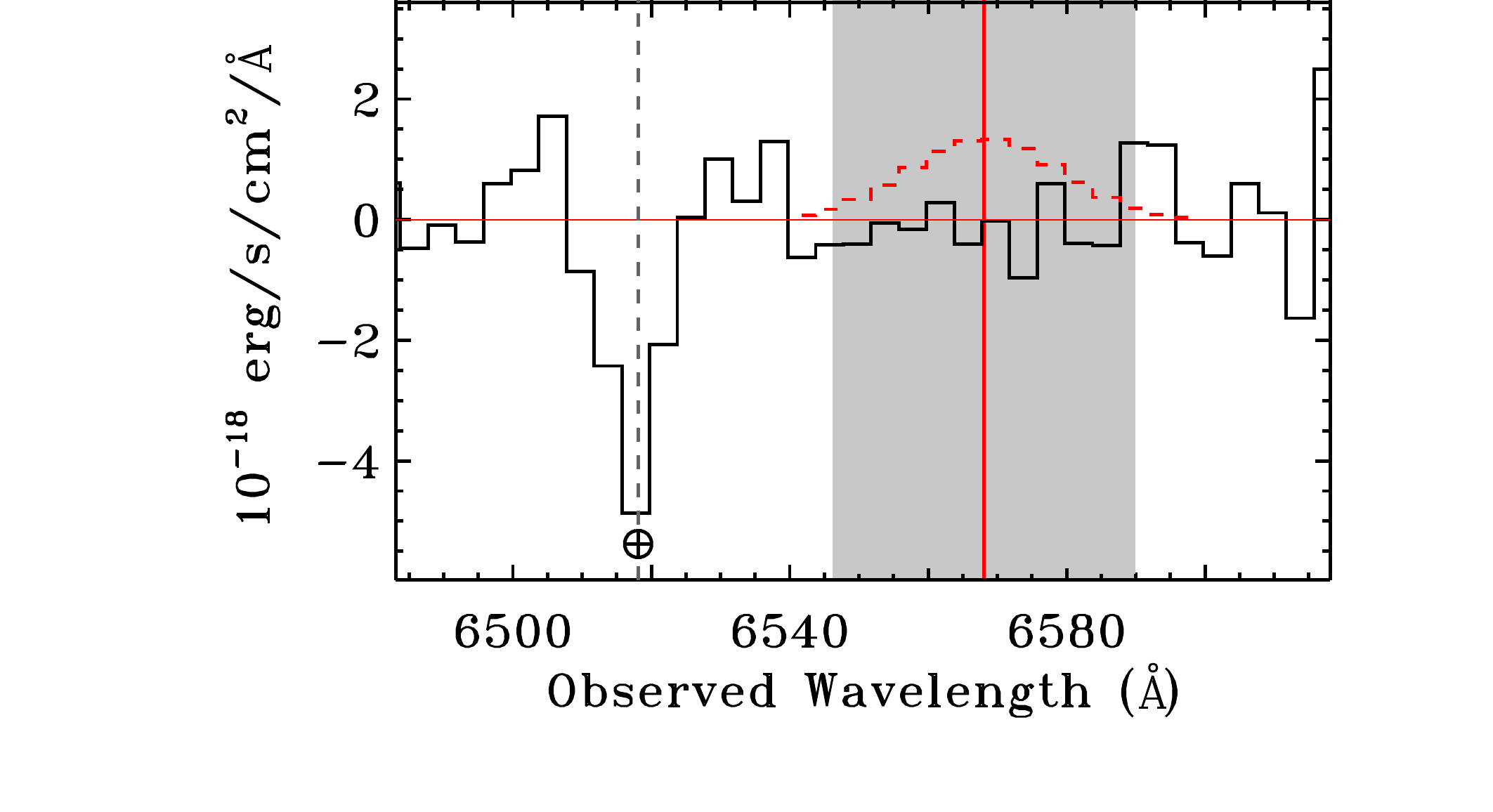}
}
\caption{
Nebular-phase (day 274 after maximum) H$\alpha$ region of the spectrum
of SN~2011fe, showing upper limits on H$\alpha$ emission from
any hydrogen stripped from a companion and entrained in the
ejecta.  The vertical red line and gray band 
show, respectively, the
expected wavelength of H$\alpha$ emission and $\pm 1000$~km~s$^{-1}$
velocity width. Black solid histogram is the observed continuum-subtracted 
spectrum and red
dashed curve is the $3\sigma$ upper limit on line emission, which implies 
a hydrogen mass $<0.001 M_\odot$. Reproduced 
from \cite{Shappee13a}.
}
\label{fig:Shappee13a}
\end{figure}

\subsubsection{Searches for emission from hydrogen}
\label{sec:hydrogen}

In the SD scenario, almost by definition, some signs of hydrogen or
helium from the non-degenerate companion should be visible at some
point.  Apart from the shock from the ejecta impacting a SD companion
(Section~\ref{sec:shock}, above), the ejecta are expected to strip and
entrain material from the companion.  This stripped material, of mass
$\sim 0.1-0.2 ~M_\odot$, which is excited by energy from the
radioactive decay of $^{56}$Co, is then expected to become visible in
the form of emission lines of width $\sim 1000$~km~s$^{-1}$ during the
nebular phase of the explosion, when the ejecta are optically thin
\citep{Marietta00,Mattila05,Liu12b}.  \cite{Mattila05},
\cite{Leonard07}, and \cite{Lundqvist13} have 
 used high signal-to-noise nebular spectra of five
SNe~Ia to set upper limits on the presence of any H$\alpha$
emission. These limits translate to upper limits on hydrogen mass of
0.01 to 0.03 $M_\odot$. The most stringent application of this test
has been, yet again, in the case of the nearby SN 2011fe, where
\cite{Shappee13a} set a limit of 0.001 $M_\odot$ on the stripped
hydrogen mass (see Figure~\ref{fig:Shappee13a}) 
These results
strongly suggest that, at least in the six cases examined, there was
no main-sequence or red-giant SD companion in the system at the time
of explosion. 

While the absence of H and He is a defining characteristic of SNe~Ia,
there are exceptions.  Until recently, H and He in emission was
reported in only two SNe~Ia: SN 2002ic \citep{Hamuy03}  
and SN 2005gj \citep{Aldering06,Prieto07}. 
These events
displayed a SN-Ia-like spectrum, but topped by strong variable
H$\alpha$ line emission, visible during most epochs, and reminiscent
of a core-collapse Type-IIn spectrum. The spectra were interpreted as
showing the interaction of the SN ejecta with the circumstellar medium
(CSM) of an evolved SD companion \citep{Wood-Vasey06,Han06}.
Another possibility that was raised was a prompt
DD merger that encountered a recently ejected common envelope
\citep{Livio03}. \cite{Chugai04} speculated that such
events are examples of ``SNe 1.5'' \citep{Iben83}, i.e. thermonuclear
runaways in the CO cores of single AGB stars that have not yet  lost their
envelopes. Alternatively, they proposed that such ``strongly
CSM-interacting'' SNe Ia are SD systems in which the donor is a
massive red supergiant.  In either case, the CSM envelope has a mass
of at least several $M_\odot$.  Some doubts were initially raised as
to whether these rare events were, in fact, true SNe~Ia.  
\cite{Benetti06} argued that these explosions are actually a subclass of
Type-Ic core-collapse SNe, in which the ejecta are encountering the
collapsed star's previously ejected envelope, in a variant of what is
thought to be the case for Type IIn SNe. 
\cite{Trundle08} pointed out the resemblance of the absorption features in
such events to those of the class of ``luminous blue variables'',
which are thought to be the progenitors of type Ib/c SNe.
 
The most recent and well-studied object of this type is PTF11kx
\citep{Dilday12,Silverman13a}, further discussed below
(Section~\ref{sec:absorption}) in the context also of SNe~Ia that have
intervening Na~I~D absorption. In PTF11kx, the H$\alpha$ emission is
somewhat weaker and develops somewhat later than in the above two
cases of CSM-interacting SNe~Ia, and the light curve is more similar
to those of normal SNe~Ia. From analysis of the evolving spectra,
\cite{Dilday12} deduce the progenitor was a SD system with a red-giant
donor and multiple shells of equatorially distributed circumstellar
material, swept up by ejecta from pre-explosion nova eruptions, similar
to those in the symbiotic recurrent nova RS Ophiuchi, in which a WD
accretes through a wind from a red giant (see
Section~\ref{sec:recurrent}, above).  \cite{Soker13b}, however,
have estimated that the circumstellar mass in PTF11kx is at least $\sim
0.1-0.6~M_\odot$, which is higher than the CSM expected in SD
models. Instead, they advocate a variant of the \cite{Livio03} picture
(ejecta from a DD merger encountering the previously ejected common
envelope). In the context of the \cite{Soker13a} core-degenerate
scenario, they propose that during the common envelope phase, the WD
disrupts and accretes the less-dense AGB core (rather than vice versa),
resulting in a prompt (rather than long-delayed) explosion.
 
\cite{Silverman13b} have compiled a sample of 16 such
strongly CSM-interacting events, some by re-classifying old events
(one of them, SN~2008J, was concurrently studied by
\citealt{Taddia12}) 
and some from new discoveries from the PTF survey. They show that such
events tend to have higher peak luminosities and slower light-curve
evolution than normal SNe~Ia, and that their hosts are always galaxies
with ongoing star formation.  \cite{Dilday12}
estimate that the fraction of strongly CSM-interacting SNe~Ia among
all SNe~Ia could be 0.1\% (based on the discovery rate in the PTF
survey) to 1\% (based on the SDSS-II survey), but potentially higher,
if cases with weaker H signatures have been overlooked, or if some cases
with stronger signatures have been misclassified as type-IIn SNe.

For the two prototype CSM-interacting cases, SN 2002ic and SN 2005gj,
late-time (2.2--3.8~yr and 0.4--1.4~yr post-explosion, respectively)
mid-IR {\it Spitzer} photometry analyzed by \cite{Fox13a} shows
variable emission by warm (500-800~K) dust at these late times. In the
case of SN~2005gj, the dust luminosity increases monotonically by a
factor $\sim 2$ over the observed period. The inferred dust masses are
of order $M_{\rm d}\sim 10^{-2}M_\odot$, at distances of $r\sim
10^{17}$~cm. It is unclear if this is newly formed dust condensing in
the SN ejecta, pre-existing CSM dust that is being heated
by UV and X-ray radiation from CSM interaction (see
Section~\ref{sec:csminteractions}), or
 light echoes by dust that is far from, and
unrelated to a CSM (\citealt{Patat05}, see Section~\ref{sec:echoes}).
In any event, similar late-time IR emission, implying similar
dust parameters, is seen in many Type-IIn SNe studied by 
\cite{Fox13b}.

Thus, in several respects, CSM-interacting SNe Ia bear a strong
resemblance to Type-IIn SNe, in which ejecta impact a large mass of
CSM, of order $1~M_\odot$, thus converting kinetic energy to
optical-band energy that accrues significantly to the radioactively
powered part of the light curve. This suggests that CSM-interacting
SNe Ia are a rare subclass in which the ejecta interact with the CSM
of a relatively massive star, rather than the CSM of a donor star in a
traditional SD scenario. The presence of a massive star
would explain why all such events identified to date have
occurred in star-forming galaxies \citep{Silverman13b}.  

Returning to SN~2002cx-like events (``SNe Iax'', 
see Section~\ref{sec:correlations}, above), \cite{Liu13c}
predict that if such events are indeed failed pure deflagrations in 
$M_{\rm Ch}$ SD systems, the low kinetic energy of the explosion means that  
only small amounts of HI, $\lesssim 0.01$~\Msun, are expected to be
seen in the
nebular-phase spectra of such objects.

\subsubsection{Radio and X-ray emission from CSM interactions}
\label{sec:csminteractions}

Interaction of SN ejecta with a pre-explosion CSM or ISM is expected
to produce radio synchrotron emission, from shock-accelerated electrons
in an amplified magnetic field, and X-ray emission through inverse
Compton upscattering, by those electrons, of the optical emission of the
SN. The CSM could be from previous mass loss from a donor star, or from
losses from the accretion flow onto the WD.  The physics and
signatures of such interaction have been computed by 
\cite{Chevalier82,Chevalier98} and \cite{Chevalier06}; 
see \cite{Chomiuk12a} and \cite{Margutti12}, 
for recent generalizations. 
Briefly, mass
conservation dictates that a CSM produced by a constant pre-explosion
mass-loss rate, $\dot M$, driving a wind of constant velocity, $v_w$,
will have a radial density profile $\rho_{\rm csm}(r)=(\dot M/4\pi
v_w)~r^{-2}$. As the SN shock advances through the CSM with speed
$v_s$, it accelerates particles to relativistic energies, and
amplifies the ambient magnetic field. From the post-shock energy
density, $\rho_{\rm csm} v_s^2$, the fractions in relativistic
electrons and in magnetic fields are usually parametrized as
$\epsilon_e$ and $\epsilon_B$, respectively. Post-shock, the electrons
assume a power-law energy distribution, $N(E)\propto E^{-p}$, above a
minimum energy $E_{\rm min}$ which, in turn, depends on $p$, $v_s$,
and $\epsilon_e$.

The electrons gyrate along magnetic field lines, emitting a
synchrotron spectrum with specific luminosity $L_\nu \propto
\nu^{5/2}$, shaped by synchrotron self-absorption, up to a 
frequency, $\nu_S$ (generally in cm-wave radio bands), 
and an optically thin synchrotron spectrum, $L_\nu
\propto\nu^{-(p-1)/2}$, at higher frequencies. As the shock advances
through progressively less-dense gas, the peak emission shifts to
lower frequencies. Measurements at a fixed frequency will therefore
first see an intensity rising with time, $t$, when that frequency is
in the self-absorbed part of the spectrum, up to a maximum, followed
by a declining signal when observing at optically thin frequencies.
In those Type Ib/c SNe in which synchrotron emission has been detected,
the electron energy distribution has an index $p\approx 3$, producing an
optically-thin synchrotron power-law index 
of $\approx -1$ \citep{Chevalier06,Soderberg12}. 
For these SNe,
$\epsilon_e\approx 0.1$, and $\epsilon_B$ values in the range 0.01 to
0.1 have been inferred. The shock velocity, $v_s$, which is set by the
density profiles of the CSM and of the ejecta, has a weak time
dependence.  In broad terms, the optically thin synchrotron luminosity
then behaves as
\begin{equation}
L_\nu \propto v_s^3~ E_{\rm min}~ \nu^{-1}~ t^{-1} ~\epsilon_e~ \epsilon_B~
\dot M^2 ~ v_w^{-2}.
\end{equation}
Given an estimate of $v_s$, a radio measurement of $L_\nu$ (or an
upper limit on it) can therefore constrain the product 
$\epsilon_e \epsilon_B\dot M^2/ v_w^2$.  
A value of $\epsilon_e=0.1$ is
generally assumed. Values of $\epsilon_B$, as noted, span one order of
magnitude, and stellar wind velocities, $v_w$, are generally in the
range of tens to a few hundred km~s$^{-1}$. Thus some useful
information can be derived regarding the interesting parameter $\dot M$.
In the case of a uniform ISM ($\rho_{\rm csm}(r)={\rm const.}$),
$L_\nu$ grows slowly with time roughly as $t^{0.35}$ 
\citep{Chomiuk12a}
and the observations will constrain the product 
$\rho_{\rm csm}~\epsilon_B^{0.9}$.

In X-rays, the inverse-Compton emission for an
 electron population with $p=3$ is 
\begin{equation}
L_\nu \propto v_s ~E_{\rm min}~ \nu^{-1}~ t^{-1}~ \epsilon_e ~\dot M~
v_w^{-1}~ L_{\rm SN}(t),
\end{equation}
 where $L_{\rm SN}(t)$ is the bolometric optical-UV luminosity from
 the SN photosphere \citep{Chevalier06,Horesh12,Margutti12}.  The
 inverse-Compton emission, in contrast to the synchrotron emission,
 does not depend on the uncertain $\epsilon_B$ parameter.  In the case
 of a shock expanding into a constant-density ISM, rather than a CSM,
 the ratio of the inverse-Compton luminosity to the optical-UV 
 luminosity is roughly constant with time.  

No radio 
or X-ray emission from a SN Ia has ever been detected.  In
X-rays, \cite{Hughes07} set upper limits on flux from four SNe~Ia,
including two strongly CSM-interacting cases, SN2002ic, and SN2005gj
(see Section~\ref{sec:hydrogen}, above).  \cite{Russell12} set
X-ray flux upper limits on 53 individual SNe~Ia, and on their stacked
images, and concluded that winds from evolved donor stars can be
excluded.  In the radio, \cite{Panagia06} obtained VLA observations at
wavelengths of 0.7 to 20~cm of 27 nearby SNe Ia at 46 epochs, roughly
$\sim 10-100$~d after explosion, and used them to set limits on CSM
interaction in each case.  \cite{Hancock11} performed a stacking
analysis of these data, to obtain deeper limits.  Assuming typical
values for the other parameters involved, they set a limit of 
$\dot M \lesssim 10^{-7}~M_\odot~{\rm yr}^{-1}$.  In radio, for SN~2006X,
which showed variable Na~I~D lines (see Section~\ref{sec:absorption},
below), VLA non-detection constrains any CSM at
$r\sim 10^{17}$~cm to arise from a wind with $\dot M \lesssim 3\times
10^{-7}~M_\odot~{\rm yr}^{-1}$, assuming $v_w=100$~km~s$^{-1}$
\citep{Patat07}.  This suggests a limit on the mass in the region of
$M\lesssim \dot M~ r~/~v_w=10^{-4}~M_\odot$.

\begin{figure}[ht!]
\centerline{
  \includegraphics[angle=0,width=0.8\textwidth]{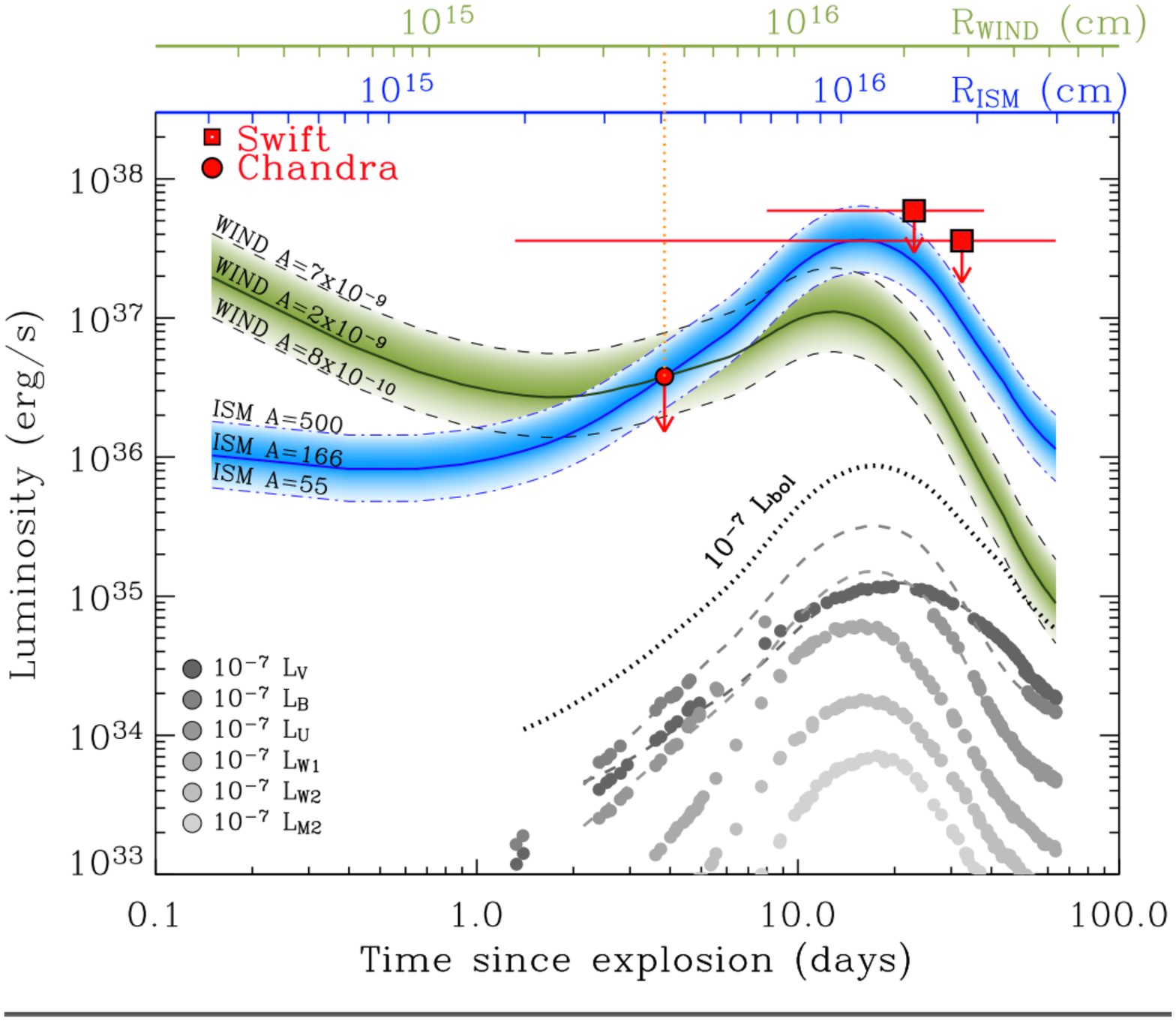}
}
\caption{X-ray $3\sigma$ upper limits (points) 
from {\it Swift} and {\it Chandra} 
on the luminosity of SN~2011fe, compared to the expected X-ray evolution
from inverse Comptonization of the UV-optical light curve (shown by gray
filled circles for the various UV-optical bands, and with a black dotted line
for the estimated bolometric light curve.)  Green band is for
comptonization by a CSM wind, with $\dot M$ values as marked in units of 
$M_\odot ~{\rm yr}^{-1}$, for a wind
speed of 100~km~s$^{-1}$. Blue band is for comptonization by a
uniform-density ISM, with densities in units of cm$^{-3}$, as marked. 
The data constrain the progenitor system's mass loss through a wind to
$\dot M \lesssim 2\times 10^{-9}~ M_\odot ~{\rm yr}^{-1}$, and a  
 uniform-density CSM to have 
$\lesssim 150$~cm$^{-3}$.
 Reproduced from \cite{Margutti12}.
}
\label{fig:Margutti12}
\end{figure}

\begin{figure}[ht!]
\centerline{
  \includegraphics[angle=0,width=0.6\textwidth]{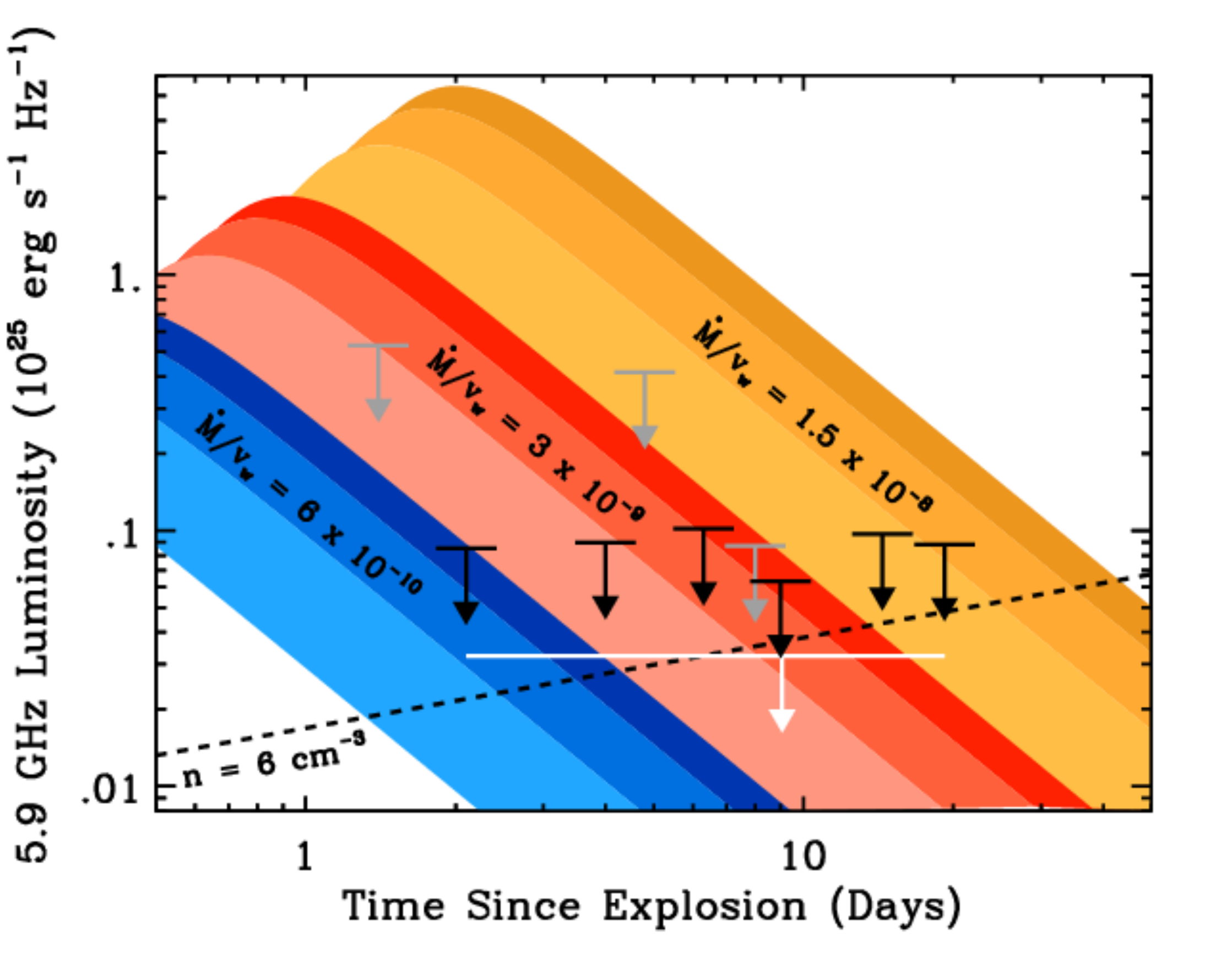}
}
\caption{
EVLA $3\sigma$ upper limits (black arrows) 
on the 5.9~GHz radio luminosity of SN~2011fe,
compared to expectations from synchrotron emission due to interaction
of its ejecta with a CSM. Also shown (gray arrows) are upper limits from
\cite{Horesh12}, 
scaled to 5.9~GHz, and a time-stacked limit (white arrow). Colored
swaths are the light curves expected from interaction with a wind, 
for the marked values of $M_\odot ~{\rm yr}^{-1}/v$, in units of   
$M_\odot ~{\rm yr}^{-1}/100~{\rm km~s}^{-1}$, with shades
corresponding to $\epsilon_B$ in the range $0.01-0.1$, and
$\epsilon_e=0.1$. 
The dashed black line  is the expected evolution for a uniform-density ISM
of density $n=6~{\rm cm}^{-3}$. The mass loss limit of 
$\dot M \lesssim 6\times 10^{-10}~ M_\odot~ {\rm yr}^{-1}$ argues
against any Roche-lobe-overflow SD scenario with $\gtrsim 1$\% mass
loss in the system. Reproduced 
  from \cite{Chomiuk12a}.
}
\label{fig:Chomiuk12}
\end{figure}

In the case of the nearby and well-studied SN~2011fe, \cite{Horesh12}
obtained radio data from 0.3 to 20~cm, using the CARMA,
EVLA, and WSRT telescopes, as early as 1.3~d after the explosion time
deduced by \cite{Nugent11} (see Section~\ref{sec:earlylc}, above).
{\it Swift} and {\it Chandra} observations were secured at 1.2~d and
4~d, respectively. Comparing the upper limits on radio and X-ray flux
to the expected light curves, \cite{Horesh12} set a limit of $\dot M
\lesssim 10^{-8}~M_\odot~{\rm yr}^{-1}$, assuming $v_s=4\times
10^4$~km~s$^{-1}$, $v_w=100$~km~s$^{-1}$, $\epsilon_e=0.1$, and also
assuming $\epsilon_B=0.1$ in the radio case. This excludes the
presence of a circumstellar wind from a giant donor.  
\cite{Margutti12} used additional {\it Swift} and {\it Chandra} epochs,
including stacks of epochs that are not far from maximum optical light
(which corresponds also to inverse-Compton X-ray maximum light) to obtain
deeper and later-time X-ray limits. They use them to refine the
mass-loss limit to 
$\dot M \lesssim 2\times 10^{-9}~ M_\odot ~{\rm yr}^{-1}$. 
These stronger limits are used to argue, beyond giant
donors, also against any stable hydrogen-burning Roche-lobe overflow
scenario (i.e. with 
$\dot M > 3\times 10^{-7}~ M_\odot ~{\rm yr}^{-1}$; 
see Section~\ref{sec:supersoft}) in which more
than 1\% of the accreted material is lost, e.g. through the outer
Lagrange points (i.e. other than L1).  A uniform-density CSM is
constrained by \cite{Margutti12} to have a particle density
$<150$~cm$^{-3}$. These constraints are shown in Figure~\ref{fig:Margutti12}.
Finally, deeper limits have been set also in the radio
by \cite{Chomiuk12a}, using additional EVLA epochs between 2 and
19 days past explosion, mostly about 5 times deeper than those of
\cite{Horesh12}. Using also a higher assumed $v_s=0.35c$, they
interpret their radio flux limits as a stringent limit on CSM mass
loss, $\dot M \lesssim 6\times 10^{-10}~ M_\odot~ {\rm yr}^{-1}$,
during the last 100--1000~yr leading up to the explosion. With this
limit, a 1\% mass loss through outer Lagrange points can be excluded
even for recurrent novae, unless the wind speed $v_w$ is particularly
high, with a correspondingly rarified CSM.  A uniform ISM density 
$n_{\rm ism}\gtrsim 6$~cm$^{-1}$ can be excluded (for $\epsilon_B=0.1$), or
$n_{\rm ism}\gtrsim 44$~cm$^{-1}$ can be excluded (for
$\epsilon_B=0.01$), at radii $r\sim 10^{15}-10^{16}$~cm from the
explosion.  Figure~\ref{fig:Chomiuk12} shows these limits.  We
note that all of the above limits are slightly weakened if the
explosion was in fact earlier than deduced by \cite{Nugent11}, by
$\sim 0.5-1.5$~d (\citealt{Piro13b,Mazzali13}; see
Section~\ref{sec:earlylc}, above).

In summary, sensitive X-ray and radio observations of a good number of
SNe Ia have so far failed to detect a CSM, including in several SNe Ia
that show ejecta-CSM interactions and variable Na~I~D absorptions. The
limits make giant SD donors unlikely. For SN~2011fe, these data show a
particularly ``clean'' environment around a SN Ia, and argue against
most of the traditional SD scenarios.

\subsubsection{Intervening absorption}
\label{sec:absorption}

In the SD scenario, SN~Ia spectra may be expected to show variable
narrow blueshifted absorption lines from a circumstellar wind from the
companion, from accretion overflows, or from pre-explosion nova-like
outbursts (e.g. \citealt{Patat07}).  As detailed below, absorptions
interpreted as such signatures have been detected in a few SNe~Ia.
For the DD scenario, in turn, as already discussed in
Section~\ref{sec:shock} above, DD merger simulations differ on
the amount and radius of merger-related debris at the time of
explosion. Furthermore,
\cite{Shen13b} have simulated a CSM that is shaped prior to DD
mergers that involve a CO primary WD and a He WD secondary 
(a combination raised in the context of sub-Chandrasekhar, or
double-detonation, DD models, e.g. \citealt{Pakmor13,Shen13a}).
A thin hydrogen layer, expected on the He WD's
surface, accretes onto the CO WD and leads to nova-like eruptions on
timescales of $\sim 10^2- 10^3$~yr prior to the SN~Ia explosion.  In
the ISM densities present in spiral galaxies, these pre-SN-Ia
eruptions can shape the CSM so as to produce circumstellar
absorptions, similar to SD model expectations and to those sometimes
observed in SNe~Ia. \cite{Raskin13a} likewise use simulations to show
that, under some circumstances, tidal debris from DD mergers could
produce observable absorptions in SN Ia spectra.

\begin{figure}[ht!]
\centerline{
 \includegraphics[
angle=0,width=0.6\textwidth]{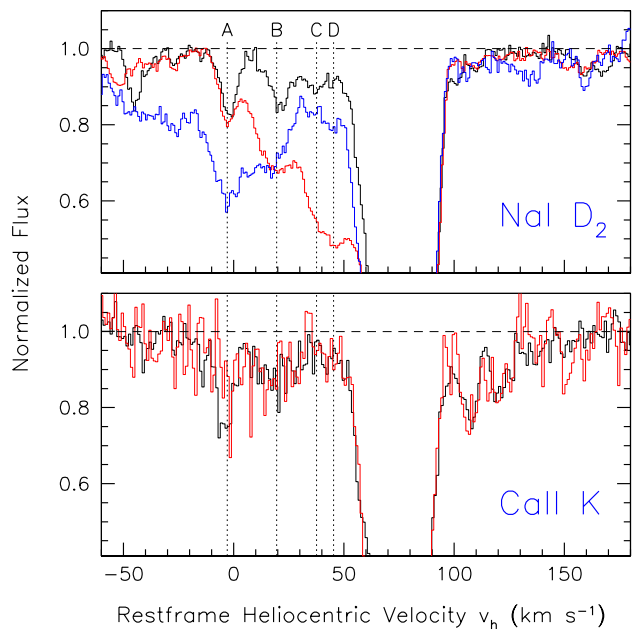}
}
\caption{Spectra of SN~2006X, showing time evolution of the
  Na~I~D$_2$ absorption feature (top panel) 
between days $-2$ (black), $+14$ (red), and
  $+61$ (blue). Individual variable absorption components are labeled
  A-D. For comparison, the non-variable absorptions in the Ca~II~K line
  (bottom panel, only first two epochs) 
are also shown. The growing absorption strength is thought 
to be due to recombining Na~II ions in a CSM, ionized earlier by UV photons
  from the explosion. The strongly
  saturated Na~I and Ca~II absorptions 
come from unrelated interstellar gas clouds
  in the disk of the host galaxy.  
>From \cite{Patat07}.
}
\label{fig:Patat07}
\end{figure}

In terms of the observations, variable, blue-shifted 
($\sim 10-100$~km~s$^{-1}$) absorption in the 
Na~I~D doublet ($\lambda\lambda 5890, 5896$) 
has been detected in the high-resolution, multi-epoch
spectra of three SNe~Ia (SN~2006X, \citealt{Patat07}; SN~2007le,
\citealt{Simon09}; and the CSM-interacting case PTF11kx,
\citealt{Dilday12}, see Section~\ref{sec:hydrogen}, above). In the
case of PTF11kx, the absorption is also seen in narrow lines of Fe~II,
Ti~II, and He~I.  A fourth variable Na~I~D case was found in
low-resolution, but high-signal-to-noise, spectra of SN~1999cl by
\cite{Blondin09}. Because of the low resolution, in this case one
cannot determine whether the absorption is blueshifted or
otherwise. However, \cite{Blondin09} noted that absorption
variations of the observed magnitude exist for only two out of 31
SNe~Ia with data that could have revealed them (SN~1999cl, and
SN~2006X, whose variations \citeauthor{Blondin09} could ``rediscover'' at low
resolution). \cite{Sternberg13} 
give statistics for the detection of variable Na~I~D at high spectral
resolution, of 3/17 (or 2/16, if excluding the unusual PTF11kx).  
In all four known cases, the Na~I~D absorption generally grows
in strength on a timescale of about 10~d.  In the three
high-resolution cases, it is seen that the growth is in individual
blueshifted velocity components, and that the absorption in the Ca~II
H\&K lines ($\lambda\lambda 3968, 3933$) does not change in strength
during the same period. Figure~\ref{fig:Patat07} shows SN~2006X, as an
example. At least three of the cases of SNe~Ia with
time-variable NaI~D absorption belong to the subclass of SNe~Ia with
``high velocity gradients'' \citep{Benetti05} (see
Section~\ref{sec:correlations}, above).
In all four cases the hosts are late-type spirals. 

Using order-of-magnitude arguments \citep{Patat07} and detailed
photoionization modeling \citep{Simon09}, the growing Na~I~D absorption
has been interpreted as due to recombining Na~II ions in circumstellar
material of density $n_{\rm csm}\sim 10^7$~cm$^{-1}$, at distances of
$r\sim 10^{16}-10^{17}$~cm from the explosion, with a total mass of
$10^{-5}$ to $10^{-2}$~$M_\odot$. Estimates of the UV spectrum and
light curve of a SN~Ia, based on model templates and extrapolations of
the few SNe~Ia with UV data, suggest that the SN flux can ionize Na~I
atoms (ionization potential 5.1~eV) out to such distances, but not
Ca~II ions (ionization potential 11.9~eV) for which the required
ionizing photons are orders of magnitude rarer in the SN spectrum.
The short recombination timescale of 10~d further requires a high
electron density, which implies a significant hydrogen ionization
fraction, and hence limits the distance to the absorbing clouds.  In
SN~2006X, part of the Na~I~D absorption actually weakens again 2 months after
maximum light, which \cite{Patat07} interpret as the SN ejecta
reaching and reionizing some components of the CSM (the 
corresponding
weakening expected also in the Ca~II absorption, due to
 such a process, could not be tested at that epoch because of low signal-to-noise ratio).

The absorbing gas may be a clumpy wind from a red-giant donor, or
an equatorial outflow from pre-explosion recurrent nova eruptions
during the decades preceding the SN~Ia explosion. The asymmetric
nature of such outflows could introduce a viewing angle dependence
that would explain the apparent rarity of SNe~Ia with variable
Na~I~D. An alternative explanation, invoking variable line-of-sight
absorption by ISM material far from the SN \citep{Chugai08}, explains the
non-variability of the Ca~II lines through abundance effects, 
by means of differing fractions of Ca and Na locked in dust grains in
intervening clouds.
\cite{Patat11} have noted the similarities of these variable
absorptions with post-outburst features in the recurrent nova RS
Ophiuchi (see Section~\ref{sec:recurrent}, above).  On the other hand,
U Sco did not show any variable absorption during its outburst in 2010
\citep{Kafka11}.   In the well-studied normal SN~Ia 2011fe, despite the high
signal-to-noise of the available spectra, \cite{Patat13} find no
evidence for variable intervening absorption that could be associated with the
CSM of the progenitor system.

\cite{Sternberg11} compared the incidence of narrow blueshifted and
redshifted Na~I~D absorptions in the single-epoch high-resolution
spectra of 35 SNe~Ia. They assigned a zero velocity to the strongest
Na~I~D absorption, associating it with the host galaxy's ISM, and then
identified additional absorption lines. They found 12 cases with
blueshifted absorptions, 5 with redshifted absorption, 5 with either
single or symmetric blue/red Na~I~D absorption, and 13 cases with no
detected Na~I~D absorption (most of them in early-type galaxies).
\cite{Sternberg11} interpreted the 12-over-5 excess of blueshifted
vs. redshifted absorptions as signatures of the CSM, and concluded
that $>20-25$\% of SNe~Ia in spirals derive from SD progenitors
(although, as noted, such absorptions may also arise in a DD scenario;
\citealt{Shen13b}).  

\cite{Foley12c} have analyzed a somewhat-modified sample of 23 SNe~Ia with
high-resolution spectra, finding 10 cases with blueshifts and 5 with
redshifts. They further find an association between ejecta velocity,
as measured by Si~II $\lambda 6355$ near maximum, and the presence of
blueshifted Na~I~D; SNe~Ia with ejecta velocities $v_{\rm Si}\gtrsim
12,000$~km$^{-1}$~s$^{-1}$ are more likely to display blueshifted
Na~I~D absorption.   \cite{Foley12c} point out that the run of
ejecta velocities and their possible association with the presence of a CSM
outflow suggest asymmetric progenitors and explosions, such that
higher velocity ejecta are aligned with higher density
CSM. Alternatively, the result could imply a variety of progenitor
systems, such that progenitors with denser or closer CSMs tend to
produce more energetic explosions. However, 
as already discussed, explosion energy and host galaxy type are correlated,
which could provide an indirect link between explosion energy and ISM 
(rather than CSM) absorption.

\cite{Maguire13} have re-analyzed spectra for 16 of the
 \cite{Sternberg11} events, together with new 
data for 16 additional SN~Ia, while defining differently the zero absorption
 velocity.
 In this combined sample, the excess of
blueshifted over redshifted Na~I~D absorptions is 11 versus 4, again
 suggesting $\sim 20\%$ CSM-related absorption. \cite{Maguire13}
confirm the tendency for $v_{\rm Si}\gtrsim
12,000$~km$^{-1}$~s$^{-1}$ to have blueshifted Na~I~D, and also find a
 correlation between $B-V$ color at maximum and the equivalent width of
 the Na~I~D absorption (see also \citealt{Forster13}).

\cite{Phillips13}
have studied a sample of 32 SNe Ia at high spectral resolution. They
find that a quarter of the objects show anomalously large Na~I~D
absorption, relative to the dust extinction implied by the SN colors.
The same extinction is indicated by the SN colors and 
by the diffuse interstellar band at 5780~\AA, suggesting that the
dust, despite its usual low values of $R_V$, is in the host's ISM,
rather than in the CSM of the progenitor. However, all of the
objects with excess Na~I-absorption 
have Na~I~D blueshifts, hinting that the Na is indeed in the CSM,
which was perhaps pre-enriched with this element (\citealt{Phillips13}
outline several scenarios for this). On the other hand, three SNe Ia
in their sample having variable Na~I~D, and which are therefore strong
candidates for having a CSM, do not show unusually strong Na~I~D
lines.

\subsubsection{Circumstellar dust and light echoes }
\label{sec:echoes}

Spectra of light echoing off interstellar dust clouds has permitted
spectacular classification of several historical Galactic SNe,
centuries after their explosions, including two SNe Ia, SNR 0509-67.5
\citep{Rest08a} and Tycho's SN \citep{Rest08b,Krause08}.  In
principle, however, monitoring of SN light echoes at much earlier
times, whether by spatially resolving the echo or by inferring its
integrated effect on spectra and light curves, can provide a
three-dimensional mapping of the CSM and the ISM around the SN
(\citealt{Patat05,Patat06}; see \citealt{Rest12} for a review). 
To date, there have been only four clear detections of SN Ia light echoes: 
SN~1991T \citep{Schmidt94,Sparks99};
SN~1998bu \citep{Cappellaro01,Garnavich01}; 
SN~1995E \citep{Quinn06}; 
and SN~2006X \citep{Wang08c,Crotts08}.
In each of these cases, the optical light curve's decline flattened
abruptly after a few hundred days, and the late-time spectra showed
features from earlier phases. The late-time spectra could be modeled well
as luminosity weighted averages of the past spectral evolution of the
SNe. Finally, HST imaging showed the actual resolved light-echo rings
expanding around the SN site. In several of these cases, polarimetry
provided further constraints on the scattering geometry.  Modeling of
these data showed, in all cases, that the scattering dust is at
distances of a few tens to a few hundred pc, generally in front of the
SNe. \cite{Wang08c} raised the possibility that there exists an
additional inner circumstellar echo component, but analysis of the
same data by \cite{Crotts08} argues against this option. The
presently known cases of light echoes most likely probe dusty regions
far from the event, with no direct bearing on the progenitor problem.
 
As noted above (Section~\ref{sec:hydrogen}), 
signatures of possibly circumstellar dust have been
seen in several CSM-interacting SNe Ia \citep{Fox13a}. In three normal
SNe Ia (including SN~2011fe),
\cite{Johansson13} have used {\it Herschel} 
far-IR non-detections to set upper limits on the circumstellar dust
mass, of $\lesssim 10^{-2} - 10^{-1} M_\odot$.

\subsection{Post-explosion evidence in SN remnants}
\label{sec:remnants}

\subsubsection{Searches for surviving companions}
\label{sec:remnants.donors}

The donor star, in a SD scenario, will survive the explosion, and is
likely to be identifiable by virtue of its anomalous velocity,
rotation, composition, temperature, or luminosity 
(e.g., \citealt{Ruiz-Lapuente97,
Marietta00,Canal01,
Wang10a,Pan13a,Liu13a,Shappee13b}).

Searches for a surviving donor star in Tycho's SN of 1572 (as
noted, a SN~Ia confirmed with a light echo spectrum, 
\citealt{Rest08b,Krause08}), 
based on
chemical abundances, radial velocities, proper motions, and rotation
velocities, have not been able to reach a consensus
\citep{Ruiz-Lapuente04,Fuhrmann05,Ihara07,Gonzalez-Hernandez09,
Kerzendorf09}.
The individual studies have
pointed out distinct preferred candidates and argued against those of
the other studies.  Most recently, \cite{Kerzendorf12b}, based on Keck
spectra
and on proper motions from HST imaging, have
concluded that there are no good candidates for SD survivors of any
type (giant, sub-giant, or main-sequence) in this remnant. Conversely,
based largely on the same data, \cite{Bedin13}
continue to advocate a star labeled Tycho-G
as the likely surviving SD-scenario donor star. While the two groups
broadly agree regarding the measured abundances, proper motion, and
radial velocity of Tycho-G, they differ regarding the degree that these
parameters are unusual or indicative of a donor-star origin for the star.
 
SN 1006 is widely considered a likely SN~Ia, based on its 500~pc
height above the Galactic plane, its symmetry, and the absence of a
neutron star \citep{Stephenson02}.  However, the total iron mass in
its remnant is apparently very low for a SN~Ia, $\sim 0.06~M_\odot$,
based on X-ray emission lines of Fe-L \citep{Vink03} and Fe-K
\citep{Yamaguchi08}, and based on UV and optical absorption lines of
cold Fe II along multiple lines of sight to background sources
\citep{Wu93,Winkler05}.  
On the other hand, the strong
X-ray emission seen in Si, S, Ar, C, and O, seems consistent with
expectations from the layered outer parts of a dynamically young SN~Ia
remnant, in which the reverse shock has yet to reach and heat the
inner, iron-rich, regions \citep{Badenes07}. Those regions might
nonetheless be photoionized by the X-ray emission from the reverse
shock \citep{Hamilton84}, and hence invisible to the UV-optical
low-ionization-Fe absorption studies.  Be that as it may,
\cite{Kerzendorf12a} and \cite{Gonzalez-Hernandez12}, for this SN
remnant, agree on the absence of any apparent surviving companions
with luminosities greater than $L_V \sim 0.5~ L_{\odot,V}$. Thus, if
SN1006 was a SN~Ia, all evolved SD donors are ruled out. Normal
main-sequence donors are still allowed, unless their luminosities are
significantly increased by the interaction with the ejecta, as
predicted by the models for the post-explosion appearance of donor
stars, cited above.

\begin{figure}[ht!]
\centerline{
  \includegraphics[angle=0,width=0.58\textwidth]{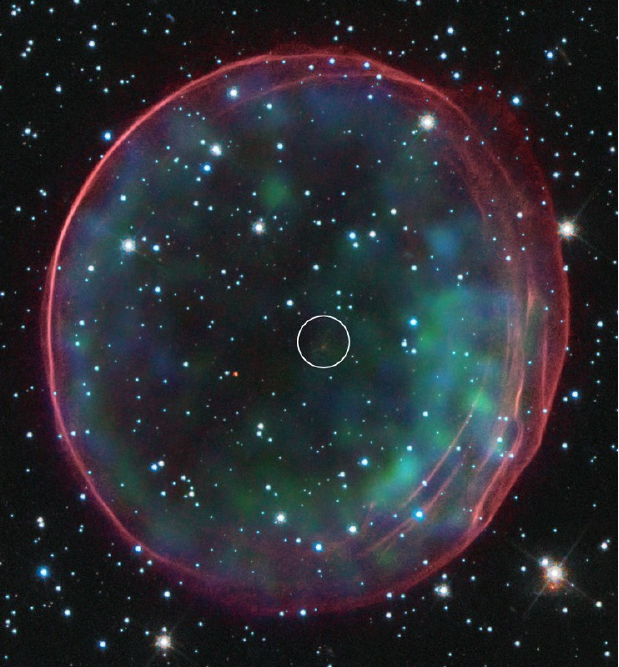}
}
\caption{ 
Composite HST/{\it Chandra} image, in $B,V,I$, H$\alpha$, and X-rays,
of the $\sim 400$-year-old SN~Ia remnant 0509-67.5 in the LMC.
The central 1.4-arcsec radius circle
is the $3\sigma$-confidence region that could host a surviving donor
star, considering uncertainties in the geometrical center and in the
remnant age, and assuming the maximum plausible proper motion.
As argued by \cite{Schaefer12}, the non-detection
of any potential surviving stars within this region, down  to
$L_V=0.04~L_{\odot,V}$,
essentially rules out the possibility of a surviving SD companion.
Credit: NASA/ESA/Hubble Heritage Team (STScI/AURA).
}
\label{fig:Schaefer12}
\end{figure}

SNR~0509-67.5 in the Large Magellanic Cloud (LMC) is the remnant of an
over-luminous SN~Ia from {\it circa} 1600 \citep{Badenes08}, confirmed
as such with a light-echo spectrum by \cite{Rest08a}.  Fortuitously,
the remnant is far from the center of the LMC, in a region with low
stellar density, and it has been imaged to great depth with HST.
\cite{Schaefer12} have shown (Figure~\ref{fig:Schaefer12}) 
that there are no
stars, down to V-band luminosities as low as $L_V=0.04 L_{\odot,V}$,
in the area around the remnant's geometrical center that could be
populated by a runaway donor star.  This luminosity corresponds to
late-K-type main sequence stars of mass $\sim 0.5 M_\odot$, and
essentially rules out all traditional SD companions.
Some 
diffuse emission seen in the center of the remnant 
is an unrelated background galaxy (Pagnotta, Walker, \& Schaefer, in prep.).

Another LMC remnant, SNR~0519-67.5, is projected on a denser stellar
region, and therefore provides weaker limits on a surviving
companion. A light echo found by \cite{Rest05} indicates an explosion
that is $600\pm200$ years old, and \cite{Rest12} indicate that the
echo's spectrum confirms a SN~Ia classification (but the spectrum is
still unpublished). In this case, \cite{Edwards12} again using HST
imaging, are able to rule out the presence of surviving
post-main-sequence donor stars, but 27 main-sequence stars, that are
close enough to the explosion center, cannot be excluded as surviving
companions. However, as already noted, the interaction with the ejecta
would likely brighten such stars significantly, and hence the main
sequence donor option is also in doubt.

Kepler's SN of 1604 is thought to have been a SN~Ia, but this has not
been established beyond doubt (see further discussion in
Section~\ref{sec:remnants.csm}, below). \cite{Kerzendorf13} have
obtained spectroscopy for the 24 stars with $L>10 L_\odot$ in the central 38
arcsec of the remnant, none of which show any signatures expected of a former donor
star. There are also no potential red giant, 
AGB (as has been 
  proposed from hydrodynamical models, see below), or post-AGB  
surviving donor stars in this region.

The SD scenario can perhaps circumvent the above constraints via the
spin-up/spin-down ideas (\citealt{Di-Stefano11,Justham11,Hachisu12b};
see Section~\ref{sec:doubledet}), if the delay between the end of
accretion and the explosion is long enough for the donor star
to evolve to a dim, undetectable, WD itself.
However, as discussed above in Sections~\ref{sec:models} and
\ref{sec:spinningWDs}, the viability of fast WD rotation in general,
and differential rotation in particular, faces theoretical and
observational challenges.

\subsubsection{Searches for CSM interaction in remnants}
\label{sec:remnants.csm}

Another approach to search for clues of the progenitor systems in SN
remnants is through hydrodynamical models for specific remnants,
thought to have been SN~Ia events.  The modeling elements in such
analyses have included combinations of a pre-explosion environment,
possibly shaped by outflows from the system, evolution of the
temperature, density, and ionization of the ejecta and the CSM, and
attempts to reproduce, the observed
geometry, dynamics, and X-ray spectrum of a remnant (e.g.,
\citealt{Sorokina04,Badenes06,Badenes08,Kosenko08,Kosenko11,Patnaude12};
see \citealt{Vink12} for a recent review).

\cite{Badenes07} have modeled the X-ray emission in seven young SN~Ia
remnants to test for the presence of large ($3-30$~pc) wind-blown
cavities in the ISM, with densities $n_{\rm csm}\lesssim 10^{-3}~{\rm
  cm}^{-3}$, centered on the explosion sites. They show that such
swept-out cavities are expected in the context of the optically thick,
$\sim 1000$~km~s$^{-1}$, outflows from rapidly accreting WDs in the
wind-regulated accretion picture (\citealt{Hachisu96}, see
Section~\ref{sec:rapidaccretion}).  In all of the seven cases
examined, the observations do not support this scenario.  In fact,
\cite{Badenes07} find that line fluxes and energy centroids in the
X-ray spectra, and the observed radii of the forward shocks, at the
known ages of the remnants, are always as expected for ejecta
advancing through a uniform-density ISM of $n_{\rm csm}\approx 1~{\rm
  cm}^{-3}$.  The fast pre-explosion outflows, leading to rarified
CSMs, would have resulted in larger forward-shock radii and weaker
lines, by factors of a few.  This result shows that, if these SNe had
SD progenitors, the growth of the WDs proceeded in some way other than
wind-regulated accretion, or that such growth ceased long before the
explosion, as in spin-up/spin-down models.  A CSM with a
$\rho\propto r^{-2}$ radial density profile is expected from any
constant-velocity outflow, fast or slow, be it from an evolved donor
star or from the accreting WD itself. \cite{Badenes07} found that such
density profiles also gave poor correspondence with the data, but only
a limited range of these $\rho\propto r^{-2}$ models was studied.
Support for a uniform-density CSM comes also from modeling by
\cite{Raymond07} of the H$\alpha$ filaments in the remnant of SN~1006,
as seen in HST imaging. They conclude that $n_{\rm csm}\approx
0.25-0.4~{\rm cm}^{-3}$, with variations of only $\sim 20\%$ on pc
scales.

An exception to the absence of cavities in SN~Ia remnants could be
RCW~86, a proposed remnant for the SN of 185 C.E. \citep{Vink06},
possibly a SN~Ia, based on the large
observed mass of Fe and the absence of a neutron star. 
\cite{Vink97}, \cite{Badenes07}, and most recently \cite{Williams11b} do find
evidence for a 12-pc-radius central cavity in this case.  On the other
hand, this remnant, which is in an OB association, has traditionally
been considered a CC~SN (e.g. \citealt{Ghavamian01}).

Among potential SN~Ia remnants, the remnant of Kepler's SN of 1604 has
been studied perhaps in the most detail, but has also led to some
confusing results.  It was once thought to have been a
CC SN, based on the asymmetrically bright, dense, and
nitrogen-rich shell of material in the remnant's northern side.  
More recently, however, it is generally assumed to have been a
SN~Ia, based on its high altitude above the Galactic plane (at least
350 pc, but possibly $>800$~pc, given persistent incompatible
estimates of its distance, see \citealt{Patnaude12}), its high iron and
low oxygen content, and the lack of a detected neutron star
\citep{Reynolds07}.  \cite{Chiotellis12} modeled the remnant
assuming a SN~Ia explosion in a SD configuration with an AGB-star
donor of initial mass $4-5 ~M_\odot$.  The binary system likely had a
high velocity of 250~km~s$^{-1}$ away from the Galactic disk
\citep{Borkowski94}, a velocity which may be difficult to explain for
a runaway binary \citep{Vink12}. In this picture, the northern shell
arises from the forward shock encountering a pre-explosion CSM of mass
$\sim 1~M_\odot$, with the CSM previously shaped by its movement
through the ISM at the said high velocity
\citep{Borkowski92,Vink08}. \cite{Burkey13} find further evidence for
an AGB progenitor by using {\it Spitzer} 24~$\mu$m data to distinguish
shocked CSM from shocked ejecta emission. They identify an equatorial
CSM component that they associate with an edge-on disk-like outflow
from an AGB donor. 

However, as noted in Section~\ref{sec:remnants.donors}, above, there
is no surviving AGB star in the remnant within the central $38''$
\citep{Kerzendorf13}.  A
post-AGB star or young WD that is only a few hundred years old would
have a luminosity of $\sim 10^4~L_\odot$ (Bloecker 1995), 
but the brightest star, even within 
a conservative 60~arcsec radius of the geometrical center, 
has a $V$-band luminosity of only
$330~L_\odot$ \citep{Kerzendorf13}. 
Furthermore, \cite{Patnaude12}, modeling
the south side of the remnant, report that a simple $\rho\propto
r^{-2}$ wind-shaped CSM, as favored by \cite{Chiotellis12}, cannot
simultaneously fit the observed dynamics and the X-ray spectrum.  To
fit all the data, \cite{Patnaude12} find that a 0.1~pc low-density
central cavity needs to be included in the CSM model.  Alternatively,
a model that also works is one with no wind-shaped CSM at all, but
rather with a remnant that expands into a more-or-less uniform-density
ISM. A north-south gradient in this ISM (a gradient which has some
support from direct 160~$\mu$m emission maps of the region by
\citealt{Blair07}) could explain the northern bow shock.  Each of
these solutions (small central cavity or uniform ISM) corresponds to a
different and incompatible distance to the object, which in turn
implies a sub-energetic or super-energetic SN~Ia event (if, indeed,
Kepler's SN was a SN~Ia). With its multiple puzzles and
contradictions, Kepler's SN remnant probably does not illuminate much
the SN~Ia progenitor issue at this point in time, but it could
  if its type and distance became known.

\cite{Borkowski06} have modeled two remnants in the Large Magellanic
Cloud with strong Fe-L line emission in their interiors, which they
conclude requires a large interior density, as would be expected from
a slow pre-explosion wind. While distinct from the fast
accretion-driven winds of the \cite{Hachisu96} scenario, this result
may be considered as evidence for the presence of pre-explosion SD
companions. On the other hand, without good X-ray spectra and
well-determined ages, the typing of these two remnants as SNe~Ia is
not secure.

Another interesting constraint comes from the
existence of ``Balmer-dominated shocks'' primarily in remnants
thought to be of SN~Ia origin (see \citealt{Vink12}). Optical spectra
of such regions show pure Balmer-line 
emission with a narrow core and a
broad base. The narrow emission is understood to arise by collisional
excitation, by electrons and protons, of cold neutral hydrogen just
being overrun by the forward shock.  The broad component is due to hot
post-shock protons, that undergo charge transfer with atoms, to become
hydrogen atoms at excited levels, or
subsequent collisional excitation of those hydrogen atoms 
\citep{Chevalier80,Heng10}.
Core-collapse remnants generally do not show Balmer-dominated shocks,
which is reasonable given that their SN explosions are bright in the
UV and therefore can photoionize a large surrounding
region. Conversely, \cite{Ghavamian03} have pointed out that the
existence of Balmer-dominated shocks, and hence neutral hydrogen, at
radii of $<30$~pc in SN~Ia remnants (and often below 5 or 10~pc, see
\citealt{Vink12}) limits the pre-explosion photoionizing flux of the
progenitors during the $t\sim 10^{5}/n$ years prior to the explosion
(corresponding to the recombination time at ambient densities of
$n/(1~{\rm cm}^{-3}$). Supersoft X-rays sources in particular, with
their UV/X-ray luminosities of up to $\sim 10^{38}$~erg~s$^{-1}$,
would have carved out ionized Stromgren spheres of radius $\sim 30$~pc
around the remnants \citep{Rappaport94}, in conflict with the observed
Balmer-dominated emission at smaller radii.
    
\subsection{SN Ia rates and the delay-time distribution}
\label{sec:rates}

SN~Ia rates and their dependence on environment and on cosmic time can
provide further clues to the progenitor problem.  In essence, finding
the dependence of the SN rate on the age distribution of the host
stellar population can reveal the age distribution of the actual SN~Ia
progenitors. Different progenitor scenarios involve different
timescales that control the production rate of SN~Ia events, and will
thus predict different SN~Ia age distributions.  For a detailed review
on surveying for SNe for rate purposes, the derivation of SN~Ia rates,
and their application to the progenitor question, see \cite{Maoz12a}.
  
A fundamental function for progenitor-question purposes is the
distribution of times between star formation and SN~Ia explosion,
usually called the delay-time distribution (DTD). The DTD is the
hypothetical SN rate versus time that would follow a brief burst of
star formation, with the burst having one unit of total mass in formed
stars.  It is the ``impulse response'' that embodies the physical
information of the system, free of nuisances -- in the present context
-- the diverse star-formation histories (SFHs) of the galaxies hosting
the SNe. Determining the DTD observationally has
increasingly become the objective of SN~Ia rate measurements.

\subsubsection{Theoretical expectations}

As noted in Section~\ref{sec:models}, theoretical forms for the DTD
can be derived from binary population synthesis (BPS) calculations,
obtained by numerically evolving simulated populations of binaries with chosen
distributions of initial parameters, or alternatively by 
following analytic approximations (e.g. \citealt{Greggio05}).
The DTDs for each progenitor channel 
can be compared to observationally derived DTDs. 

The DD model naturally gives rise to a broad range of
delay times \citep{Yungelson00}.  With some
simplifying assumptions, this DTD form
can be seen to result from the strong dependence of the 
merger time on the post-common-envelope separation of the WD pair
\citep{Greggio05,Totani08}. Suppose the post-common-envelope
separation, $a$, is distributed as a power law, $dN/da\propto
a^\alpha$, over the range of separations that will merge over a
Hubble time. Suppose further that the time until merger depends on the
separation to some power, $t\propto a^\gamma$. Then the form of the DTD
will be given by $dN/dt \propto t^\beta$ with
$\beta=-1+(\alpha+1)/\gamma$. For the angular momentum
loss due to gravitational waves, $\gamma=4$. 
The post-common-envelope distribution of WD separations
is generally found by BPS simulations to be similar in form to the
input distribution of main-sequence binaries, which is about flat in
log separation, i.e., a power law with $\alpha\approx-1$
\citep{Toonen12, Yungelson13}.  A $\sim t^{-1}$ DTD is thus expected
even if the separation distribution is not exactly flat in log $a$, as
long as the value of $\gamma$ is sufficiently large. The ``collisional DD''
model of \cite{Katz12} (see Section~\ref{sec:collisions}), with $\gamma=5/2$
would give a similar DTD.

\begin{figure}[ht!]
\centerline{
   \includegraphics[angle=0,width=0.57\textwidth]{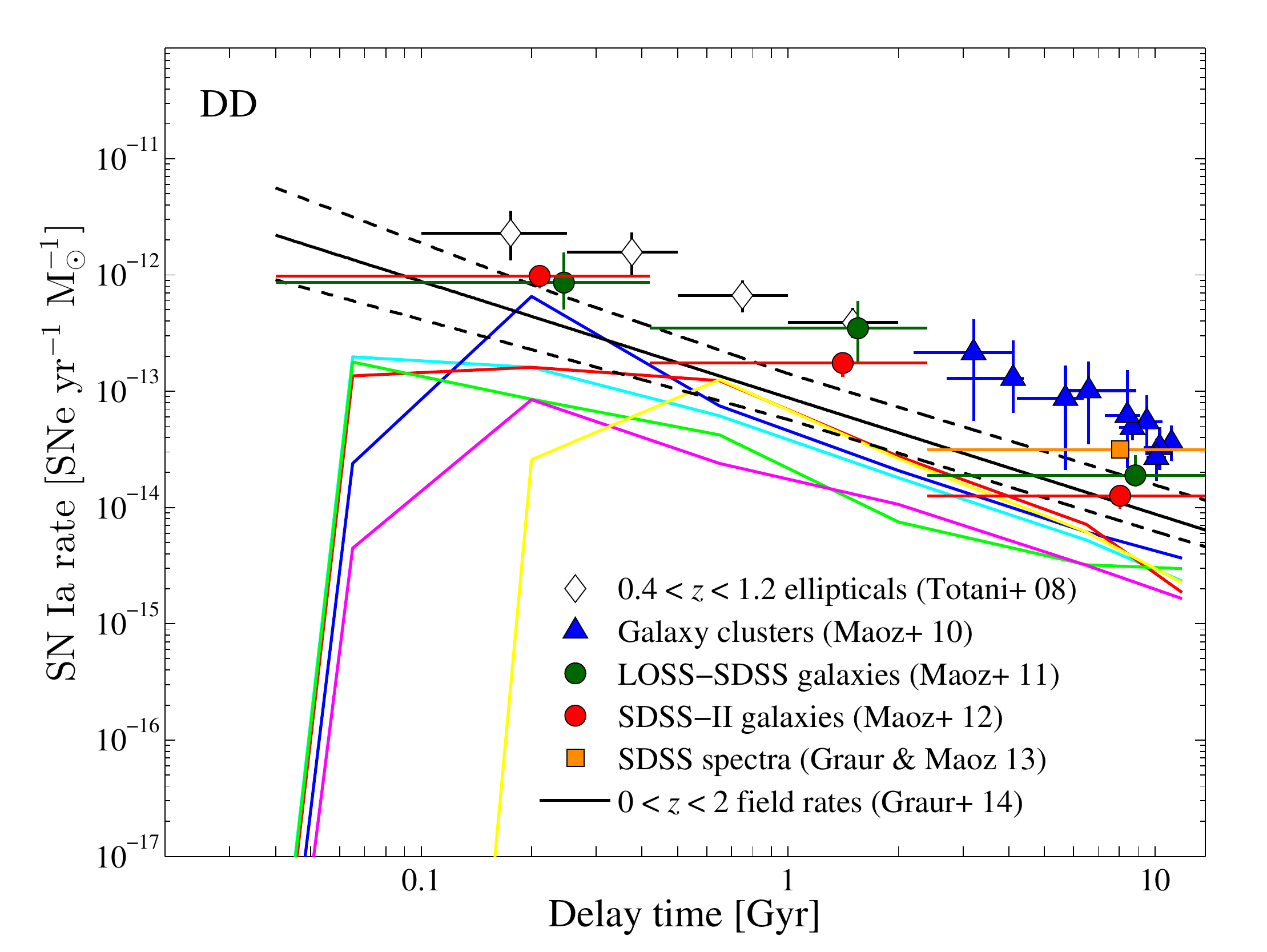}
\includegraphics[angle=0,width=0.57\textwidth]{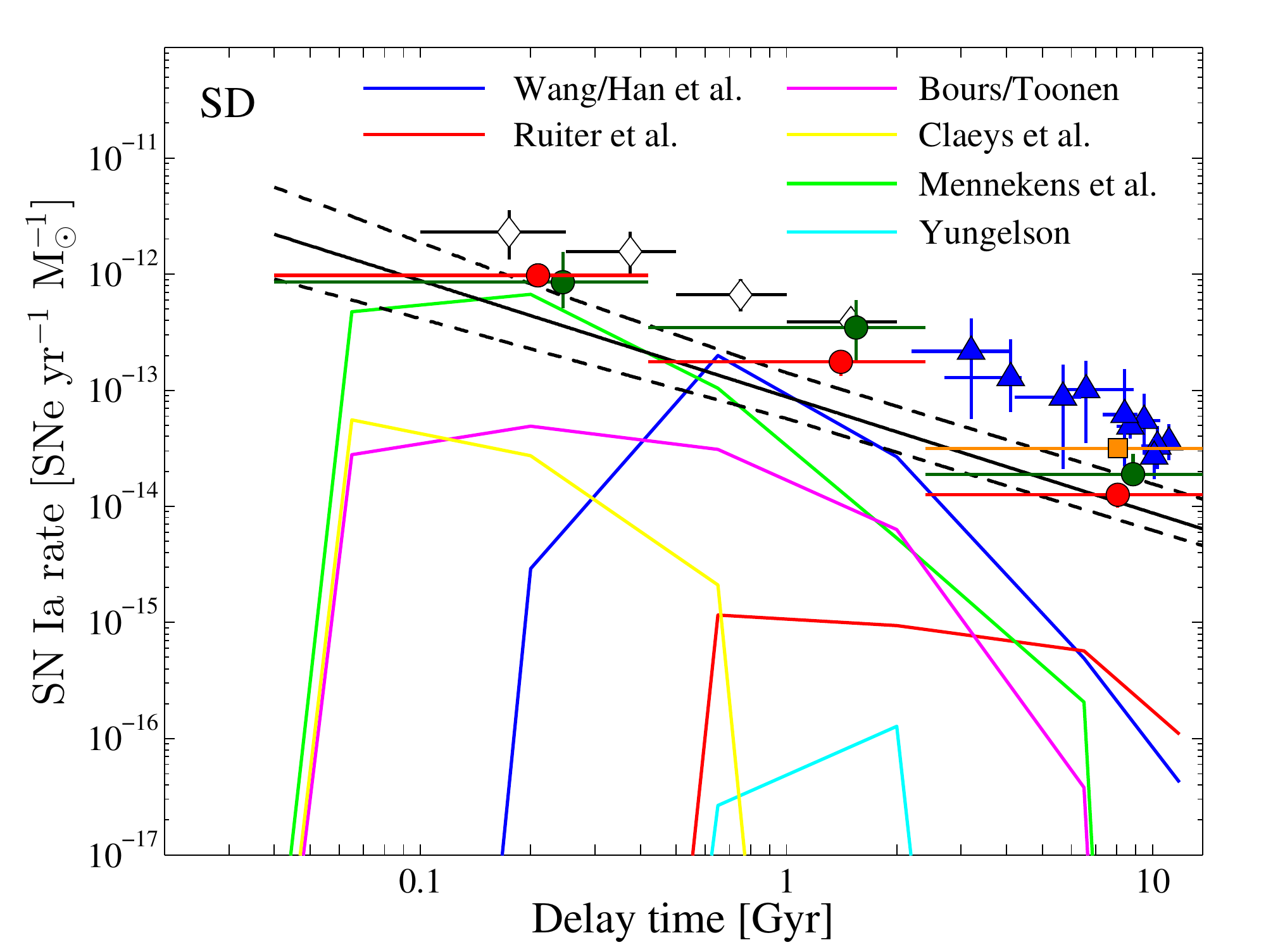}
} 

\caption{
Observed versus theoretical delay-time distributions. 
In both panels, points and the straight black lines are some of 
the observationally derived DTDs discussed here. Colored curves are
theoretical DTDs adapted
from the compilation of BPS predictions  for DD models (left panel)
and for SD models (right panel) by
\cite{Nelemans13},
who adjusted all models to have the same physical input
parameters. All DTDs, observed and theoretical, are shown with
a consistently assumed definition (SNe per year per formed stellar
mass)
 and a consistent IMF. The \cite{Toonen12} curve shown here corrects a
 misprint in the plot shown in \cite{Nelemans13}.
  }
\label{fig:Maoz12}
\end{figure}

The $\sim t^{-1}$ DTD behavior breaks down at short delays. A
``bottleneck'' in the process is the production rate of WDs that can
serve as SN~Ia progenitors.  The production rate is zero for at least
$30-40$~Myr, until main sequence stars with masses below $\sim$8~\Msun~
evolve into the first WDs to emerge from the stellar population.  The
WD production rate then rises to a maximum, and falls as $\sim
t^{-1/2}$ (e.g. \citealt{Pritchet08}), up to a cutoff time, $t_c$,
corresponding to the main-sequence lifetime of the least massive stars
whose descendant WDs contribute to a DD-channel SN~Ia.
 Since this
main-sequence mass is usually considered to be $2-3 ~M_\odot$, $t_c$
is generally $\sim 1$~Gyr.  The DTD will be the convolution of the WD
production rate and the merger rate dependences, resulting in a broken
power law, $t^\beta$, with indices $\beta\sim-1/2$ at $t<t_c$ and
$\beta \sim-1$ at
$t>t_c$. More realistically, the supply of WDs that will eventually
merge is more complex, given that two WDs are required, and that the
time until they are ``ready'' to evolve solely via gravitational decay
depends on the initial binary parameters of their progenitors, which
determine the ensuing interactions among the pair.  BPS simulations
indeed show
a complex behavior (that varies among models, according to the
different physical recipes assumed), for the pre-$t_c$ DTD
shape. However, as expected, the DD models generally do show the generic
$t^{-1}$ behavior after $\sim 1$~Gyr. This can be seen 
in Fig.~\ref{fig:Maoz12}, which includes a compilation by
\cite{Nelemans13} of some recent BPS theoretical DTDs for the DD channel.

For the standard SD model, the defining property is that the accretion
rate onto the WD needs to be in a rather narrow range, near 
$\dot M \sim 3\times 10^{-7} ~M_\odot~{\rm yr}^{-1}$ (see
Section~\ref{sec:models}, above). Such a rate is realized only by
main-sequence, or slightly evolved sub-giant, donor stars that can
transfer of the order of $1~M_\odot$ to the WD on a thermal
(i.e. Kelvin-Helmholtz) timescale, $t_{\rm th}$,
of the donor star (\citealt{van-den-Heuvel92}).  A donor star, upon
losing mass, will contract adiabatically on a dynamical timescale.
Thermal equilibrium is regained on a thermal 
timescale, causing re-expansion and a renewed mass supply on that
timescale (e.g. \citealt{Pols94}). 
The accretion rate is thus
\begin{equation}
\dot M\sim \frac{1~M_\odot}{t_{\rm th}}
\sim \frac{1~M_\odot}{GM^2/RL}\sim 3 \times 10^{-7} M_\odot~{\rm yr}^{-1}
\left(\frac{M}{2~M_\odot}\right)^3, 
\end{equation}
assuming that stellar main-sequence radius scales with mass, as $R\propto
M$, and luminosity as $L\propto M^4$. There is therefore a limited range of 
donor masses, $2 - 3$~\Msun, that can supply material at the rate required
for stable hydrogen burning on the WD surface. This is confirmed 
by detailed
calculations (e.g. \citealt{Langer00,Han04}).
Since these systems explode during, or shortly after, the main
sequence, in many models the DTD is concentrated between a 
few-hundred Myr and 1--2 Gyr, dropping off sharply before and after 
\citep{Yungelson05}.
Fig.~\ref{fig:Maoz12} illustrates this for a set of 
BPS theoretical DTDs for the SD channel.

Similar exponential cutoffs are seen in analytic approximations of  
 SD model DTDs 
(e.g., \citealt{Greggio05}).  
 A limiting factor of such predictions is that the binary evolution
calculations involve many
uncertain assumptions, and therefore the theoretical DTD predictions
vary among different groups. This is apparent, in Fig.~\ref{fig:Maoz12},
in particular for the SD channel.

\citet{Hachisu96} have suggested that an optically thick wind, driven
from the accreting WD, can stabilize the mass transfer in binaries
where
 a low-mass giant fills its Roche lobe. The long
evolutionary time of the low-mass donor can produce a SD model with long
delay times. \citet{Hachisu08a} have proposed that this same optically
thick wind can strip mass from donors in the traditional main-sequence
model, and thereby raise the mass range for donor stars up to
8 \Msun, thereby extending the SD model also to short delay times.
Studies using these extended parameter ranges to calculate the DTD
based on BPS \citep{Mennekens10,Meng10c,Wang10a,Bours13} 
have typically found DTDs that are indeed broader and
(depending on how massive the donors that are allowed) extend to short
delay times, but find very few systems 
with giant donors. The reason
is that the required period range for such systems, of a few hundreds
of days, is sparsely populated in BPS models by post-common-envelope 
binaries (see \citealt{Yungelson00}). There are known 
examples of symbiotic binaries (e.g. T CrB, RS Oph) and post-AGB-star binaries 
(see \citealt{van-Winckel09}) observed in that period range, but their
statistical fraction is unclear. 
\citet{Hachisu99} have proposed that many wide systems can evolve
into this period range by interaction of the binary with the slow wind
of the giant, although the efficiency of this process is
debated. Combining all of these effects, \cite{Hachisu08b}
have suggested that a combination of SD main-sequence
and red-giant channels could together give a  $\sim t^{-1}$-shaped DTD. 

The helium SD channel is expected to contribute at short delay times
($\lesssim 10^8$yr), due to the fact that the helium donors evolve
from rather massive main-sequence stars with little delay (and the WDs
need to have formed even earlier, e.g. \citealt{Wang09a}).

\subsubsection{Observed DTD from SN rates vs. color and Hubble type}
\label{sec:dtdcolors}

The first observational indications for the existence of a range of
SN~Ia delay times (i.e. a DTD) was found by \cite{Mannucci05} and
\cite{Mannucci06} who, analyzing a local SN sample
\citep{Cappellaro99}, demonstrated that the SN~Ia rate per unit stellar
mass changes with host-galaxy color and Hubble type -- parameters that
trace the star-formation rate (SFR).
Nonetheless, early-type galaxies with
no current star formation were seen to have a non-zero SN~Ia rate
(as was already well known).
A similar SN~Ia rate dependence on host
color and specific SFR was found by \cite{Sullivan06} for 
the Supernova Legacy Survey (SNLS) sample of SNe~Ia,
and more recently by \cite{Li11b} for the Lick
Observatory SN survey (LOSS) sample \citep{Leaman11,Li11a},          ,
and by \cite{Smith12} for the SDSS-II SN sample \citep{Frieman08,Sako08}.

Early interpretations of these results argued for the co-existence of two
SN~Ia populations: a ``prompt'' population, with rates proportional to
the CC~SN rates \citep{Mannucci05} or to the star-formation
rate \citep{Sullivan06}, that explodes within $\sim100-500$~Myr; and a
delayed channel, proportional to the stellar mass, that produces
SNe~Ia on timescales of order 5~Gyr. In the similar ``$A+B$''
formulation \citep{Scannapieco05a}, the SN~Ia rate in a
galaxy is proportional to the stellar mass (through the $A$
parameter) and to the galaxy's star-formation rate (through $B$).  In
retrospect, however, the ``prompt-plus-delayed'' interpretation is, in
essence, a DTD with two coarse time bins, and these two ``channels''
appear to represent the two sides of a broad DTD.
More quantitative information on the DTD can be obtained by comparing
the rates with the stellar age distribution of each galaxy type, and
indicates that, indeed,  to
reproduce the SN rate vs. host-color relations, 
a broad DTD is required,
with delays below 0.5~Gyr to trace the SFR, 
and an extended tail to $\gtrsim 5$~Gyr delay times, 
to produce SNe~Ia in galaxies with low current SFR.
\cite{Mannucci06} and \cite{Mannucci08b} showed this
using several SD models, while 
\cite{Maoz12a} showed that a $\sim t^{-1}$ DTD can do so as well.
While The `$A+B$' parametrization is still
sometimes used to analyze SN~Ia rates, it is a poor physical metric.
For example, the $A$ component depends on the mean stellar age of the
monitored population, with younger stellar systems producing more SNe
for any declining DTD.

\subsubsection{DTD from rates in galaxy clusters and in field ellipticals}
\label{sec:dtdclusters}

Measuring the SN rate in galaxy clusters as a function of redshift is
probably the most direct way to recover the DTD.  Optical spectroscopy
and multiwavelength photometry of cluster galaxies has shown that most
of their stars were formed within short episodes ($\sim 100$~Myr) at
high redshift ($z\sim 3$; \citealt{Jimenez07,Rettura11}), 
a good approximation to the $\delta$-function burst defining the DTD.
Thus, an almost-direct measurement of the DTD is
provided by the observed SN~Ia rate per unit stellar mass in galaxy
clusters, as a function of the time since the cluster's star-formation epoch
(i.e. the cosmic epoch at the observed cluster redshift minus the
stellar formation epoch).  Cluster SN~Ia rates have
been measured over the past decade in the redshift range $0<z<1.5$,
corresponding to DTD delays of $2-10$~Gyr  
\citep{Gal-Yam02,Germany04,Maoz04,Sharon07,Mannucci08a,Sand08,
Graham08,Gal-Yam08,Dawson09,Dilday10b,Maoz10c,Barbary12a,Sand12}.
These measurements are still limited by significant
Poisson errors due to small SN numbers. 

\cite{Maoz10c} derive a DTD
based on most of these galaxy-cluster SN~Ia rate measurements, 
which is shown in Figure~\ref{fig:Maoz12}.  They find
consistency with a $\sim t^{-1}$ form
at $t>2$~Gyr,  and show that the
$\sim t^{-1}$ conclusion is not critically dependent on uncertainties in the
input parameters involved, such as the precise redshift of star
formation in clusters, whether it was a brief or extended burst, or
the contribution of ongoing low-level star formation in clusters, as
long as these are at the levels, redshifts, and cluster locations
allowed by direct measurements of star-formation tracers in clusters.

A related approach to DTD recovery was taken by \cite{Totani08}, using
field elliptical galaxies instead of galaxy clusters. Comparing SN~Ia
rates in early-type galaxies of different luminosity-weighted ages,
seen at $z=0.4-1.2$ as part of the Subaru/XMM-Newton Deep Survey, they
were the first to show observationally that the DTD is consistent with
a $t^{-1}$ form. This observed DTD is also shown in
Fig.~\ref{fig:Maoz12}.
 Like the elliptical galaxies in clusters,  field
ellipticals are dominated by old stellar populations created in a 
short burst at
early times, allowing for a direct derivation of the DTD.
 However, in
general, the concept of a ``typical age'' for a host galaxy,
interpreted as a SN~Ia progenitor age, is a risky zeroeth-order
approximation to the full SFH of a galaxy (see \citealt{Mannucci09a}
and \citealt{Maoz12a}).   The SN rate of a galaxy with a wide
distribution of ages is likely to be dominated by the youngest stellar
population, even when those young stars are present only at low levels
that may escape detection \citep{Mannucci08b}. As opposed to
ellipticals in the central regions of clusters, which appear to be
very quiescent,
non-cluster ellipticals  often host non-negligible amounts of late-time
star formation (e.g. \citealt{Zhu10}).

In a possible instance of this, \cite{Della-Valle05} found that
radio-loud early-type galaxies have higher SN~Ia rates than
otherwise-similar radio-quiet galaxies.  A possible confirmation
of the effect was found by \cite{Graham10}.  \cite{Mannucci06}
proposed that quiescent early-type galaxies sometimes acquire new gas
in a minor merger, and that this material leads to both radio activity
(through accretion onto the central black hole) and star formation
which, in turn, produces prompt SNe~Ia. 
\cite{Greggio08} have pointed to the absence of the expected CC-SNe in
the same radio galaxies, or in early-type galaxies in
general (e.g., \citealt{Hakobyan08}.) On the other hand,
CC SNe have been found in some early-type galaxies 
showing clear signs of ongoing star formation, based on UV data \citep{Suh11}.
The observed rate trend in 
radio galaxies still lacks confirmation and, furthermore,
the constraints on the DTD would be quite
model-dependent, while other explanations have also been proposed
\citep{Capetti02,Livio02}.

Other applications of the ``rate vs. age'' approach have been made by
\cite{Aubourg08,Raskin09,Cooper09,Schawinski09c} and
\cite{Thomson11}. They confirm the wide distribution of delays
and, in particular, the existence of ``prompt'' SNe~Ia, though
with a wide range defining the age of that population. 
Uncertainties are again dominated by assumptions regarding the
parent stellar populations.

\subsubsection{DTD from iron abundances in galaxy clusters}

Constraints on the DTD can also come from considering
element abundances in galaxy clusters.
As reviewed in \cite{Maoz10c},
X-ray observations of the intracluster medium (ICM) and optical
observations of the stellar populations determine the ratio of
iron-to-stellar masses. As iron is a main product of SN~Ia explosions
(with a mean of $\sim 0.7 M_\odot$ per event), and nothing escapes the
deep cluster potential, this ratio sets the time-integrated number of
SNe~Ia in clusters, per formed stellar mass (after estimating and
subtracting the iron contributions by CC SNe).
This integral constraint, minus an integral over 
the DTD  at delays $t> 2$~Gyr from direct cluster rate measurements, 
can provide an indirect estimate of the DTD at delays $0<t <2$~Gyr. 
Although the DTD value
thus obtained in this early bin tends to be somewhat high, it is
consistent with the $\sim t^{-1}$ form found directly from the SN~Ia
rates at lower redshifts. Uncertainties on the mass of the iron in the
stellar and ICM components, and on the contribution of CC-SNe to iron
production, propagate into this estimate \citep{Loewenstein13}.
The combined constraint on the DTD power-law
index is $\beta=-1.2\pm 0.3$. 

The time evolution of iron abundance in clusters 
can, in principle, also be used to measure the shape of the DTD.
\cite{Calura07} have shown that a broad DTD that is peaked at short
delays, when combined with assumptions about the winds that transfer
iron from the galaxies to the ICM, can reproduce the observed
evolution of iron abundance in the ICM between $z=0.2$ and $z=1.2$

\subsubsection{DTD from volumetric rate evolution}
\label{sec:dtdvolumetric}

Another observational approach to recovering the DTD has been to
compare the volumetric SN rate from field surveys, as a function of
redshift, to the cosmic SFH. An advantage of this approach is the
averaging-out, over large cosmic volumes, of the uncertainties in SN
rates and SFHs based on few galaxies. 
Furthermore, field surveys
are not biased in favor of SNe in particular types of galaxies, as
opposed to galaxy-targeted surveys.
On the other hand, 
uncertainties in the cosmic SFH have remained a limiting 
factor in deriving a DTD (e.g. \citealt{Graur13b}).
As the DTD is the SN ``response'' to
a short burst of star formation, the volumetric SN rate versus cosmic
time, $R_{Ia}(t)$, will be the convolution of the DTD with the SFH
(i.e. the star formation rate per unit comoving volume versus cosmic
time, $\dot \rho(t)$),
\begin{equation}
R_{Ia}(t)\propto\int_{0}^{t}\dot\rho(t-\tau)\Psi(\tau)d\tau
.
\label{convol}
\end{equation}

Starting with \cite{Ruiz-Lapuente98}
 and \cite{Sadat98}, the volumetric rate evolution has been
compared to cosmic SFH.   
Numerous volumetric rate measurements, progressively more accurate and
at higher-redshift, have accumulated over the past decade (see
\citealt{Graur13b} for a recent compilation).
Early on,
\cite{Dahlen04,Strolger04,Dahlen08} and \cite{Strolger10} argued for a
DTD that is peaked at a delay of $\sim 3$~Gyr, with little power at
either shorter or longer delays. This conclusion was based on their
measurements of the SN~Ia rate at $z>1$ with HST.  However, this
conclusion was contested based both on re-analysis of the same data
\citep{Kuznetsova07,Mannucci07b,Greggio08,Blanc08b,Horiuchi10}, and on
additional $z>1$ SN~Ia rates \citep{Poznanski07b}.
The main problems
pointed out were the small number of SNe, and hence the large
statistical errors in the rates, compounded by 
systematic uncertainties due to
the necessary assumptions regarding SFHs and dust extinction.
\cite{Mannucci06} further pointed out that such a narrow DTD does 
not reproduce the dependence of local SN~Ia
rates on galaxy colors (see Section~\ref{sec:dtdcolors}, above).

\cite{Graur11} discovered a larger sample of 37 likely SNe~Ia at
$1<z<2$ by monitoring the Subaru Deep Field. Additional new high-$z$
SN~Ia rate measurements have emerged also from the recent CLASH 
\citep{Graur13b} and CANDELS \citep{Rodney13} 
surveys with HST, with the highest-redshift SNIa in the latter survey
at $z=2.15\pm 0.1$.  Several
authors \citep{Graur11,Kistler13,Perrett12,Graur13a,Graur13b,Rodney13} 
have analyzed some
or all of these new high-$z$ SN rate measurements, together with
precise new rates at lower redshifts from SDSS-II \citep{Dilday10a}
and SNLS \citep{Perrett12}. They all find that a DTD with a power-law
form, $\sim t^{-1}$, when convolved with a wide range of plausible
SFHs, gives an excellent fit to the observed SN rates. For example,
the best-fit power-law index found by \cite{Graur13b} is
$\beta=-1.00\pm 0.06$ (random error, due to the uncertainties in the
SN rates), $\pm 0.10$ (systematic error, due to the range of possible
SFHs). This power law is shown in Figure~\ref{fig:Maoz12}.
The precise $z<1$ rate measurements are particularly constraining for the DTD.
\cite{Graur13b} show that the rates are incompatible with the functional
forms of DTDs from SD-model BPS calculations, which have cutoffs at
delays beyond several Gyrs. \cite{Rodney13} find that, assuming the
DTD is a $t^{-1}$ power law beyond 500~Myr, 30-65\% of SNe~Ia must
have delays shorter than 500~Myr, consistent with a $t^{-1}$ power law
that extends all the way down to 40~Myr delays.    

\subsubsection{DTD from star-formation histories of individual galaxies}
\label{sec:dtdindividual}

The convolution in Eq.~\ref{convol} between SFH and the DTD, giving
the SN~Ia rate versus time, applies not only to a cosmological volume
but also to individual galaxies. \cite{Brandt10} and \cite{Maoz11}
introduced similar methods for recovering the DTD from SN~Ia samples
for which the monitored galaxies have individual estimated SFHs, based
on spectral synthesis.  In essence, the SFH of every individual galaxy
is convolved with a trial universal DTD, and the resulting current
SN~Ia rate is compared to the number of SNe the galaxy hosted in the
survey (generally none, sometimes one, rarely more).  \cite{Maoz11}
applied the method to a subsample of the galaxies in LOSS, 
and the SNe
that they hosted  with spectral-synthesis-based
SFH reconstructions by \cite{Tojeiro09}, based on spectra from the
SDSS. The limited time-resolution of these SFH reconstructions dictate
that the recovered DTD has only three coarse time bins, 
with a prompt
SN~Ia component, that explodes within $\tau <420$~Myr of star
formation, an intermediate component, at $0.42<\tau<2.4$~Gyr, and a
 delayed component that explodes after $\tau >2.4$~Gyr.
\cite{Brandt10} applied their method to a sample of SNe~Ia from the
SDSS-II survey.  \cite{Maoz12c} applied the
\cite{Maoz11} algorithm to an SDSS-II sample that is larger and more
 complete than the \cite{Brandt10} sample, and corrects several
oversights in the previous analysis. 
Finally, in \cite{Graur13a} the
method was applied to a sample of SNe~Ia discovered in archival SDSS-I
galaxy spectra (the galaxies happened to host a SN within the
spectroscopic aperture during the observations), again using the
spectral-synthesis-based SFHs of the galaxies.  All of these
three-time-bin DTD reconstructions give similar monotonically
decreasing DTDs, consistent with $\sim t^{-1}$. For example,
\cite{Maoz12c} find $\beta=-1.07\pm 0.07$. These DTD reconstructions
are included
in Fig.~\ref{fig:Maoz12}. 

\cite{Brandt10} used the ``stretch parameter'' of the SN
light curves, to divide their SN~Ia sample into a ``high-stretch''
subsample and a ``low-stretch'' one, and derived the DTD for
each subsample. They found that luminous, high-stretch,
 SNe~Ia have most
of their DTD signal at short delays, while low-stretch
SNe~Ia have a DTD that peaks in the longest-delay bin. However, this
trend largely disappears in the analysis of the more-complete SDSS-II
SN~Ia sample by \cite{Maoz12c}. \cite{Perrett12} similarly
fail to see any strong stretch dependence in their 
DTD analysis of volumetric SN~Ia rates. While such a dependence is
expected based on the observed trends between light curve stretch and
mean host galaxy stellar age (see Section~\ref{sec:correlations}), 
apparently the DTD analyses still lack
the precision needed to reveal it.

\subsubsection{DTD, downsizing, and the rate-mass relation}
\label{sec:ratesize}

Indirect evidence for a 
broad DTD comes also from an observed dependence of SN~Ia rate per
unit mass on galaxy mass.  \cite{Li11a} first noted that the SN~Ia
rate per unit mass in the LOSS sample decreases for more massive
galaxies of a given type.  The rate-size relation has been confirmed
by \cite{Smith12} for SDSS-II SNe~Ia, and by \cite{Graur13a} for the
SNe~Ia discovered in archival SDSS spectra.  \cite{Kistler13} showed
that this ``rate-size relation'' can be explained by a combined effect
of ``downsizing'' in galaxy formation -- more massive galaxies are
formed, on average, earlier and on shorter timescales
(e.g. \citealt{Gallazzi05}) -- combined with a $\sim t^{-1}$
DTD.
Downsizing produces different mean ages and therefore SNe in
more massive galaxies come from the low-rate, long-delay tail of the
DTD.  
\cite{Graur13a} have independently confirmed
that downsizing plus a $\sim t^{-1}$ DTD explain the effect for their
sample as well. \cite{Kistler13} investigated also the possible effect
of metallicity on the rate-mass relation.  Less massive galaxies have
lower metallicities \citep{Tremonti04,Mannucci10}. Stellar evolution
at low metallicity
can produce more massive WDs, and the SN~Ia rates could be affected
if such WDs are easier to bring to $M_{\rm Ch}$ via accretion.
However, for their  model's assumed influence of metallicity
on WD mass and SN~Ia production, \cite{Kistler13} find only a weak
effect of metallicity on the rate-mass relation.
 
We note in passing that the rate-mass relation permits a reliable
estimate of the Milky Way's SN Ia rate.  For Sbc spirals of Milky Way
mass ($(6.4\pm0.6)\times 10^{10} M_\odot$; \citealt{McMillan11}), the
specific SN Ia rate is $(1.12\pm 0.35)\times
10^{-13}$~yr$^{-1}~M_\odot^{-1}$ \citep{Li11a}, so if the Milky Way is
not atypical, the Galactic SN~Ia rate is $(7.2\pm 2.3)\times
10^{-3}$~yr$^{-1}$.  The mean time between Galactic SNe~Ia is thus
100-200~yr.

\subsubsection{Element abundances in stars}
\label{sec:dtdmetals}

The different timescales of production of CC SNe and SNe~Ia, coupled
with the different chemical yields of the two populations of SNe,
produce diverse abundance ratios in galaxies having different SFHs
(see \citealt{Nomoto13} for a review).
If a galaxy forms most of its stars in a single, brief episode of star
formation, the Fe-peak-element yields from most SNe~Ia will not be
incorporated into subsequent generations of stars, while $\alpha$
elements, such as O, Mg, Si, and Ca from core-collapse SNe, with their
much smaller delay time, will do so. As a consequence, the
distribution of [$\alpha$/Fe] (typically [O/Fe] vs. [Fe/H]) is
sensitive to the DTD (\citealt{Tinsley79b,Matteucci86,Matteucci01b,
De-Donder04}). 
\cite{Matteucci06} find that 
DTDs with 
$\sim 10-30\%$ of the explosions occurring at  
$< 100$~Myr, are consistent with the
abundance ratios of stars in the solar neighborhood, but is it not
clear if such a fraction of short delays is required by the data.  

\cite{Mennekens12a} have used BPS calculations
 combined with a Galactic chemical
evolution model to investigate which combinations of progenitor
scenarios, binary evolution parameters, star-formation histories, and
initial mass functions, can simultaneously match the observed DTD and 
the distribution of [Fe/H] in solar-neighborhood G-dwarfs. Their main
 conclusion is that reproducing 
the observed [Fe/H] distribution requires a high DTD
 normalization (i.e. models with efficient SN~Ia production), 
as indicated independently by SN-Ia-rate-based DTD
 estimates (see Section~\ref{sec:ratenorm}, below), {\it and} a significant
early contribution to the DTD. This can be achieved with a combination
 of SD and DD progenitors, with binary parameters tuned to boost their 
SN Ia production efficiency.
Recent applications of chemical evolution modeling 
by \cite{kobayashi09}, \cite{matteucci09}, and
\cite{Tsujimoto12} take the inverse approach, and 
use the DTD from SN~Ia rate measurements to constrain chemical evolution. 
With improved stellar abundance data, chemical mapping
 could provide stronger constraints on
the early-time DTD.

\cite{Seitenzahl13b} 
have recently used the [Mn/Fe] abundance ratios in solar-neighborhood stars to
constrain the explosion scenario directly (i.e. not via the
DTD). They find that manganese is produced efficiently only in the
high-density conditions of near-$M_{\rm Ch}$ SN Ia explosions -- 
sub-$M_{\rm Ch}$ SNe Ia and CC-SNe have yields that make [Mn/Fe]
ratios about 3 times below Solar. Consequently, 
the observed distribution of stellar [Mn/Fe] requires that roughly
50\% of SNe Ia come from near-$M_{\rm Ch}$ explosions, be they from SD, DD,
or other, progenitors.

\subsubsection{DTD from SN remnants} 
\label{sec:dtdSNR}

An unconventional observational DTD reconstruction has been carried
out by \cite{Maoz10b}, by applying the \cite{Maoz11} algorithm to a
sample of SN remnants in the Magellanic Clouds \citep{Badenes10}, and
treating this as an effective SN sample, one from a survey in which the SNe
are detectable for $\sim 10$~kyr (the visibility time of a SN
remnant).  The
advantage of this approach is that the nearness of these galaxies permits
derivation of detailed and accurate stellar age distributions of
the individual galactic regions that host the SNe. 
The stellar age distributions are found by comparing the
densities of the resolved stars in color-magnitude diagrams to model
stellar isochrones \citep{Harris04,Harris09}.  The small number of
remnants, with the attendant large statistical errors, allows only the
coarsest DTD resolution, with two SN~Ia time bins. Nevertheless,
\cite{Maoz10b} find a significant detection of a prompt (this time
$35<\tau<330$~Myr) SN~Ia component, and an upper limit on the DTD
level at longer delays that is consistent with the long-delay DTD
levels measured with other methods.  Larger samples can be produced in
the future via ongoing and proposed deep radio surveys for the SN
remnant populations in additional nearby galaxies, such as M33 and
M31, and by using their spatially specific stellar age distributions,
again based on the resolved stellar populations.

\subsubsection{Normalization of the DTD}
\label{sec:ratenorm}

Figure~\ref{fig:Maoz12} compiles a number  
of the observational DTDs
described above. Despite the consistent $\sim t^{-1}$ DTD shape,
obtained by numerous different methods and samples, there is a clear
tension among the normalizations of the DTD, i.e., among the
Hubble-time-integrated (``total'') numbers of SN~Ia formed by a stellar
generation. The DTDs based on volumetric SN Ia rates (which are
not biased toward bright galaxies and contain a significant number of SNe
in dwarf galaxies, \citealt{Leaman11,Childress13a}) typically
indicate a time-integrated SN~Ia production efficiency of about 1
SN~Ia per $1000~ M_\odot$ formed [for an assumed ``diet-Salpeter''
initial mass function (IMF), Bell et al. (2003)]. 
Galaxy-targeted SN surveys seem to
find values about 2 times higher than this. Finally, DTDs based on
galaxy clusters are higher yet, perhaps by another factor of $2-3$. 
In fact, there have been previous indications for higher
rates in cluster ellipticals than in field ellipticals
\citep{Mannucci08a,Dilday10b}.
The highest observed
DTDs are a factor $3-10$ higher than the lowest ones.
While the range in normalizations appears real, more work is
required to confirm that some or all of it is not due to
systematic errors, such as incorrect global or individual galaxy SFHs
for particular galaxy types, surveys, or redshifts. 

The lower-normalization DTDs are within reach of some of the
DD-scenario DTDs from some recent BPS models
\citep{Ruiter13,Toonen12,Mennekens12b}. 
For example, \cite{Toonen12}
find a time-integrated number of SNe~Ia in their models of
$2-3.3\times 10^{-4}~ M_\odot^{-1}$, i.e., an order of magnitude lower
than observed. However, this comes from considering as SNe~Ia
progenitors only CO-CO WD mergers with total masses above $M_{\rm
  Ch}$. Considering {\it all} CO-CO WD mergers, they find a
time-integrated rate of $\sim 0.8 \times 10^{-3}~ M_\odot^{-1}$, and
this assumes 50\% binarity of the initial stellar population. A higher
binary fraction would give a correspondingly higher result, one quite
similar to some of the observed DTD normalizations above, as apparent in 
Fig.~\ref{fig:Maoz12}. 

As also seen in 
Fig.~\ref{fig:Maoz12}, the BPS predictions for SD models tend to be
much lower than DD 
predictions, both at short delays (although there is a large diversity
among models), and certainly beyond the cutoffs that SD models generically
have at  delays of a few Gyr. Their normalizations thus
miss the observed ones by at least an
order of magnitude, and often by many.

 A higher SN~Ia number per formed stellar
mass observed in the most massive galaxies could perhaps be related to the
recently noted dependence of IMF on galaxy mass
(\citealt{Cappellari12,Conroy12,Geha13}), or to the well know excess
of metals in clusters, compared to expectations from SN yields and
the present-day
stellar population (e.g. \citealt{Maoz10c,Loewenstein13}).
On the other hand, it is unclear
how the IMF variations, which are seen in the sub-solar-mass region of
the IMF, could affect the production of SNe~Ia whose progenitors are
WDs descended from super-solar-mass stars, on the side of the IMF
where variations have not been claimed.

As already noted in Section~\ref{sec:models}, above, head-on
collisions of WDs in dense environments such as globular clusters and
galactic centers have been proposed as a way of boosting SN~Ia
production, which would raise the DTD normalization.
Apart from collisions between unbound cluster WDs, dynamical
encounters between binaries and other cluster stars will harden the
binaries \citep{Shara02}; such an effect has been used to explain the
large enhancement in the number of low-mass X-ray binaries
observed in globular clusters
\citep{Sarazin03}.  \cite{Rosswog09} showed that, even if such an
enhancement of WD collisions occurs, the small fraction of all WDs
that are in globular clusters means that this could explain
only of order 1\% of all SN~Ia events. 
Observationally, the lack of globular clusters at the
locations of SNe~Ia observed with HST 
\citep{Voss12,Washabaugh13}, 
rules out globulars as a significant overall
rate-enhancement mechanism for SNe~Ia.

\subsubsection{Summary of the observed DTD}

In summary, SN~Ia rate measurements quite consistently indicate a $\sim
t^{-1}$ DTD. The coarse time resolution of many studies, the
systematic uncertainties discussed above, and the sparse information
below $\sim 1$ Gyr, presently allow more structured shapes, but the
overall behavior of the DTD appears to be similar to that generic to the DD
model.  Long delay times are possible in SD models with red-giant
donors, but BPS models generally produce very few of them.  On the
other hand, there is observational room for a SD contribution at
delays $\lesssim 1$Gyr, where the DTDs are not well-constrained.  As a
result, the DTD data presently cannot rule out comparable SD and DD
contributions. However, in terms of total numbers, 
BPS predictions for SD models are generally
lower than DD predictions, which themselves barely reach some of the
observed DTDs.

\section{Summary}
\label{sec:Summary}

\subsection{The emerging picture}

The SN~Ia progenitor question is far from resolved, despite the major
efforts to address it (e.g., there are over 400 papers in NASA-ADS
over the last six years with the words ``Type-Ia supernova
progenitor'' in their abstracts).  Nevertheless, this large body of
work has provided a wealth of relevant information. Progress has been
made in the observational domain, but also in theoretical work where,
e.g., recent 3D hydrodynamical simulations of mergers have revived the
possibility that double-detonation merger scenarios could produce
normal SNe Ia. To summarize some of the main observational results
that have emerged:\\
\begin{itemize}
\item Censii in the Milky Way and nearby galaxies of the potential
  progenitor populations of SNe Ia in the standard SD model
  --  accreting WDs growing in mass toward $M_{\rm
    Ch}$ -- show that they cannot be recurrent novae or supersoft
  X-ray sources for more than a small fraction of their growth time. A
  hidden population of rapidly accreting WDs is also limited to be no
  more than a
  minority of the progenitors, based on the apparent absence of their expected
  ionizing radiation. A population of differentially spun-up super-$M_{\rm Ch}$ WDs,
  waiting to spin down and explode, may be out there, but has
    theoretical problems and seems at odds with the very slow spins
observed generically in WDs.  
As for close double WD binaries that could be DD-model progenitors,
BPS calculations and surveys for such systems both indicate that their
total merger rate is of the same order of magnitude as the SN Ia
rate. The traditional $M_{\rm tot}>M_{\rm Ch}$, CO+CO, mergers are
probably too rare, by a factor of at least a few, but if lower-mass WD
mergers, or even CO+He WD mergers, lead to a normal SN~Ia, then their
numbers are likely sufficient. Which ones, if any, actually lead to SNe~Ia, 
 and how, is still unclear.

\item SN~2011fe, the best-ever studied SN~Ia, and a very typical event
  to boot, has provided a wealth of 
progenitor constraints, essentially all of them based on null
results. This event's immediate environment was remarkably ``clean''.
There are strong limits on a pre-explosion companion star, on any
signatures of ejecta interaction with a donor star or with a CSM, on any 
X-ray upscattering or optical absorption signatures by a CSM, 
and on any hydrogen that was stripped from a companion and 
entrained by the ejecta. All of these are consistent with a DD
  origin, and challenge the SD picture for this event, unless spin-up is
   invoked.

\item Variable Na~I absorption, which could come from a CSM, has been
  seen, but only in three normal SNe Ia and one
  CSM-interacting SN~Ia. The 2/31 or 2/16 discovery statistics in 
  well-defined samples suggest a $\sim 5-15\%$
  occurrence fraction. Statistical studies using single-epoch
  high-resolution spectra deduce, based on an excess of systems with
  blueshifted versus redshifted Na~I absorption, a $\sim 20\%$
  fraction of progenitors with a CSM, but these estimates still suffer
  from small numbers and systematics in defining samples and zero
 reference velocities. While a CSM producing the variable Na absorption is
  expected in the SD scenario, some DD merger models also predict such
  CSM absorptions.

\item No normal SN~Ia has shown signs of any entrained hydrogen  from a
  companion, sometimes to stringent limits. However, a rare subclass
  constituting $\sim 0.1-1\%$ (maybe somewhat more) of SNe Ia, do show
  prominent variable hydrogen emission, and other signs of CSM
  interaction, such as kinetic energy input to the light
  curves. On the other hand, the
  implied CSM masses may be of order a solar mass, and it is
  presently unclear how such CSM masses fit into any of the traditional
  progenitor pictures. 
 
\item Among SN remnants known or suspected to have been SNe Ia, none
  have thus far revealed any obvious surviving SD-scenario donors, in
  several cases with strict upper limits. Hydrodynamic models of
  remnants having detailed X-ray observations almost always indicate
  ejecta expanding through a uniform-density medium, rather than
  through a low-density cavity from a pre-explosion SD-scenario wind.      
     
\item Derivations of the SN Ia delay-time distribution from a variety
  of SN~Ia rate measurements and techniques, 
  in different environments and redshifts, are all consistent with a
  $t^{-1}$ dependence at delays $\gtrsim 1$~Gyr  (although DTDs from
  different samples may be showing somewhat different normalizations
  of this distribution). Such a dependence is typically found for
  DD models, while most BPS SD models predict few or no SNe~Ia at long
  delays. 
  A SD contribution to the DTD at short delays is not yet constrained.
   
\item From the diversity and correlations among observed
  spectral properties of SNe Ia and of their host galaxies, a rough
  general trend has emerged. At one extreme are events with high
  luminosities and slow light-curve evolution, occurring in
  star-forming galaxies, with high-velocity spectral features, no
  signs of unburnt carbon, higher symmetry indicators, and a relatively
  high chance of showing intervening absorption. At the other extreme
are events with the opposite properties. As some of the former
  properties can be plausibly associated with SD-model explosions, it
  is tempting to suggest the existence of two ``families'' of
  progenitors, SD progenitors that produce the first type of events
and DD progenitors that make the second type of SNe Ia. There are some
  problems with this picture. First, as reviewed in
  Section~\ref{sec:correlations}, the observed trends are weak, with
  large scatter, and sometimes with conflicting claims on the sense
  of the relations, and on selection effects. 
Some combination of progenitor properties, e.g., age, 
metallicity, and others, may explain the trend, rather than
 two separate families. Second, if two very diverse
  progenitor channels, e.g. a $M_{\rm Ch}$ SD channel and a
  double-detonation sub-$M_{\rm Ch}$ DD channel, were operating in
  parallel with comparable numbers, it would be remarkable that observed SN Ia properties are
  so uniform and continuous, the very features of SNe~Ia that have made them so
  useful for cosmology. 
\end{itemize}

 From a recent-historical point of view, it is noteworthy that there
 has been something of a paradigm shift. The DD
 scenario has traditionally been the underdog, with review articles
 (e.g. \citealt{Livio00})
 often concluding that the SD model will eventually be confirmed.
 The main argument against the DD scenario has been theoretical --
 that a DD merger would never lead to anything akin to a SN~Ia
 explosion.  This theoretical objection is no longer very strong, as some
   merger simulations now seem to successfully produce SNe~Ia. 
  And, for several SN~Ia events and remnants, the observations suggest that, among
  the various models, only DD progenitors are not ruled
   out. Thus, not only can the DD model no longer be rejected offhand,
   but rather it could turn out to be the main, or even the only,
 scenario behind SNe~Ia.

\subsection{Future outlook}
\label{sec:future}

Future observational data, and progress on the theoretical issues, will
undoubtedly provide new clues to the progenitor problem. Apart from
obvious breakthroughs, such as the direct detection of a progenitor, or
the explosion of a previously known candidate progenitor, there are several
developments that, in the near or far future, could significantly
advance the field.

On the theoretical side, one open issue 
is still how exactly does the accretion onto a WD proceed 
in an SD model, and
what are the parameter ranges (mass transfer rate, composition,
rotation, accretion history) that lead a  C/O WD to grow in mass.
Ignition in DD mergers is another
unsolved problem. Detailed
3D hydrodynamical models, including nucleosynthesis, exploring the run
of
merger-model parameter space, are now being published.  
However, the finding that rather subtle effects may be
important, such as residual hydrogen or
helium on the WD, or residual carbon in the centers of oxygen-neon
WDs, suggest that robust results will require higher-resolution
calculations that take the previous stellar evolution into account in
detail. 

Progress is within reach also in the area
of progenitor populations and the theoretical DTD. A recent 
detailed comparison of four different BPS
codes (Toonen et al. 2013, submitted) shows that their outputs 
differ not because of
numerical accuracy or treatment, but because of different physical assumptions.
One clear conclusion, in particular for the DD and He-accreting
  SD models, is that the phase where one of the two stars is a helium
star needs more detailed investigation. In parallel,
comparison of the BPS predictions with observed local populations of binaries
could calibrate out many of the uncertainties in BPS codes.

Turning to observations,
the understanding of potential progenitor populations has advanced
thanks to  wide-field 
spectroscopic surveys, such as SDSS, that led to large and
homogeneous samples of observed binaries. However, proper
understanding of selection effects is still a challenge.  The launch of
{\it Gaia} will mark the next big improvement, with the discovery of
 large samples of WD binaries, for which distances and proper motions are also known.

The DTD has proved to be a powerful tool for testing models. A
 better knowledge of the parent stellar populations, and
site-specific measurements of SN Ia rate vs. the stellar age
distributions at the SN locations, 
can improve the accuracy of observational DTDs. The 
multivariate distribution of delay times, explosion energies, host
metallicities, and more, (as opposed to the single-variable DTD), 
can provide physical links between these
parameters, which will provide sharper discrimination among models. 

SN~2011fe was a watershed event, not only because of its nearness but
also because of the wealth of pre-explosion data and the very early and
multi-wavelength coverage. With the rising prominence of time-domain,
multi-messenger surveys, it is likely that similar or even better data
will soon be available for additional SNe Ia.  

Gamma-ray observations of SNe Ia by future MeV-range missions
should provide excellent diagnostics not available at other energies.
\cite{Horiuchi10} have simulated
the gamma-ray spectra and light curves of SNe Ia, out to Virgo
distances, that could be obtained by the
proposed ACT  mission, showing that various SD models would have
distinct signatures in such gamma-ray data. \cite{Summa13} have
done this exercise for the proposed GRIPS and ASTRO-H missions. 
They show that the same two SD
and DD models, compared by \cite{Ropke12} to the optical data for
SN~2011fe (see Section~\ref{sec:spectra}, above), could be
distinguished by gamma-ray data from these missions.
 
In the more distant future, a space-based gravitational wave
interferometer will be able to detect signals individually 
from several thousand Galactic
double WD systems with orbital periods below 10 min, i.e.
that are within $\sim 1$~Myr of merging \citep{Amaro-Seoane13}. 
This will directly probe the WD merger rate. Together with the
millions of longer-period binaries that are detectable as a
foreground, this will provide quantitative information about the
binary WD population and on its relevance as SN~Ia progenitors. The
gravitational signal of an actual merger will be detectable
only for a Galactic (or perhaps Magellanic Cloud) event.

Indeed, we are arguably overdue for the privilege that Tycho, Kepler,
and Galileo enjoyed, of a Galactic SN~Ia. Of course, we may well have to
wait a few more centuries for this occasion, but when it comes, it
will surely be another transformational event, this time for the understanding
of the workings of SNe~Ia.\\

Acknowledgements: We thank R. Aloisio, C. Badenes, R. Foley,  O. Graur, 
E. Nakar, P. Tozzi, and J. Vink for
valuable discussions and input. DM acknowledges support by a grant
from the ISF, and by I-Core Program grant
1829/12 of the PDC and the ISF.

\bibliographystyle{Astronomy}
\bibliography{/Users/filippo/Dropbox/bibdesk/Bibliography}

\begin{thebibliography}{}
\expandafter\ifx\csname natexlab\endcsname\relax\def\natexlab#1{#1}\fi

\bibitem[{{Abazajian} et~al.(2009){Abazajian}, {Adelman-McCarthy},
  {Ag{\"u}eros}, {Allam}, {Allende Prieto} et~al.}]{Abazajian09}
{Abazajian} KN, {Adelman-McCarthy} JK, {Ag{\"u}eros} MA, {Allam} SS, {Allende
  Prieto} C, et~al. 2009.
\newblock \textit{\apjs} 182:543--558

\bibitem[{{Aldering} et~al.(2006){Aldering}, {Antilogus}, {Bailey}, {Baltay},
  {Bauer} et~al.}]{Aldering06}
{Aldering} G, {Antilogus} P, {Bailey} S, {Baltay} C, {Bauer} A, et~al. 2006.
\newblock \textit{\apj} 650:510--527

\bibitem[{{Alexander} et~al.(2011){Alexander}, {Wynn}, {King} \&
  {Pringle}}]{Alexander11}
{Alexander} RD, {Wynn} GA, {King} AR, {Pringle} JE. 2011.
\newblock \textit{\mnras} 418:2576--2583

\bibitem[{{Amaro-Seoane} et~al.(2013){Amaro-Seoane}, {Aoudia}, {Babak},
  {Bin{\'e}truy}, {Berti} et~al.}]{Amaro-Seoane13}
{Amaro-Seoane} P, {Aoudia} S, {Babak} S, {Bin{\'e}truy} P, {Berti} E, et~al.
  2013.
\newblock \textit{GW Notes, Vol.~6, p.~4-110} 6:4--110

\bibitem[{{Anupama}(2013)}]{Anupama13}
{Anupama} GC. 2013.
\newblock In \textit{IAU Symposium}, eds. R~{Di Stefano}, M~{Orio}, M~{Moe},
  vol. 281 of \textit{IAU Symposium}

\bibitem[{{Arnett}(1969)}]{Arnett69}
{Arnett} WD. 1969.
\newblock \textit{\apss} 5:180--212

\bibitem[{{Arnett}(1982)}]{Arnett82}
{Arnett} WD. 1982.
\newblock \textit{\apj} 253:785--797

\bibitem[{{Ashok} \& {Banerjee}(2003)}]{Ashok03}
{Ashok} NM, {Banerjee} DPK. 2003.
\newblock \textit{\aap} 409:1007--1015

\bibitem[{{Aubourg} et~al.(2008){Aubourg}, {Tojeiro}, {Jimenez}, {Heavens},
  {Strauss} \& {Spergel}}]{Aubourg08}
{Aubourg} {\'E}, {Tojeiro} R, {Jimenez} R, {Heavens} A, {Strauss} MA, {Spergel}
  DN. 2008.
\newblock \textit{\aap} 492:631--636

\bibitem[{{Badenes} et~al.(2006){Badenes}, {Borkowski}, {Hughes}, {Hwang} \&
  {Bravo}}]{Badenes06}
{Badenes} C, {Borkowski} KJ, {Hughes} JP, {Hwang} U, {Bravo} E. 2006.
\newblock \textit{\apj} 645:1373--1391

\bibitem[{{Badenes} et~al.(2007){Badenes}, {Hughes}, {Bravo} \&
  {Langer}}]{Badenes07}
{Badenes} C, {Hughes} JP, {Bravo} E, {Langer} N. 2007.
\newblock \textit{\apj} 662:472--486

\bibitem[{{Badenes} et~al.(2008){Badenes}, {Hughes}, {Cassam-Chena{\"\i}} \&
  {Bravo}}]{Badenes08}
{Badenes} C, {Hughes} JP, {Cassam-Chena{\"\i}} G, {Bravo} E. 2008.
\newblock \textit{\apj} 680:1149--1157

\bibitem[{{Badenes} \& {Maoz}(2012)}]{Badenes12}
{Badenes} C, {Maoz} D. 2012.
\newblock \textit{\apjl} 749:L11

\bibitem[{{Badenes}, {Maoz} \& {Draine}(2010)}]{Badenes10}
{Badenes} C, {Maoz} D, {Draine} BT. 2010.
\newblock \textit{\mnras} 407:1301--1313

\bibitem[{{Badenes} et~al.(2009){Badenes}, {Mullally}, {Thompson} \&
  {Lupton}}]{Badenes09d}
{Badenes} C, {Mullally} F, {Thompson} SE, {Lupton} RH. 2009.
\newblock \textit{\apj} 707:971--978

\bibitem[{{Barbary} et~al.(2012){Barbary}, {Aldering}, {Amanullah}, {Brodwin},
  {Connolly} et~al.}]{Barbary12a}
{Barbary} K, {Aldering} G, {Amanullah} R, {Brodwin} M, {Connolly} N, et~al.
  2012.
\newblock \textit{\apj} 745:32

\bibitem[{{Bearda} et~al.(2002){Bearda}, {Hartmann}, {Ebisawa}, {Heise},
  {Kaastra} et~al.}]{Bearda02}
{Bearda} H, {Hartmann} W, {Ebisawa} K, {Heise} J, {Kaastra} J, et~al. 2002.
\newblock \textit{\aap} 385:511--516

\bibitem[{{Bedin} et~al.(2013){Bedin}, {Ruiz-Lapuente}, {Gonzalez Hernandez},
  {Canal}, {Filippenko} et~al.}]{Bedin13}
{Bedin} LR, {Ruiz-Lapuente} P, {Gonzalez Hernandez} JI, {Canal} R, {Filippenko}
  AV, et~al. 2013.
\newblock \textit{ArXiv e-prints}

\bibitem[{{Benetti} et~al.(2005){Benetti}, {Cappellaro}, {Mazzali}, {Turatto},
  {Altavilla} et~al.}]{Benetti05}
{Benetti} S, {Cappellaro} E, {Mazzali} PA, {Turatto} M, {Altavilla} G, et~al.
  2005.
\newblock \textit{\apj} 623:1011--1016

\bibitem[{{Benetti} et~al.(2006){Benetti}, {Cappellaro}, {Turatto},
  {Taubenberger}, {Harutyunyan} \& {Valenti}}]{Benetti06}
{Benetti} S, {Cappellaro} E, {Turatto} M, {Taubenberger} S, {Harutyunyan} A,
  {Valenti} S. 2006.
\newblock \textit{\apjl} 653:L129--L132

\bibitem[{{Benz}, {Thielemann} \& {Hills}(1989)}]{Benz89}
{Benz} W, {Thielemann} FK, {Hills} JG. 1989.
\newblock \textit{\apj} 342:986--998

\bibitem[{{Berger} et~al.(2005){Berger}, {Koester}, {Napiwotzki}, {Reid} \&
  {Zuckerman}}]{Berger05}
{Berger} L, {Koester} D, {Napiwotzki} R, {Reid} IN, {Zuckerman} B. 2005.
\newblock \textit{\aap} 444:565--571

\bibitem[{{Beuermann} \& {Reinsch}(2002)}]{Beuermann02}
{Beuermann} K, {Reinsch} K. 2002.
\newblock \textit{\aap} 381:487--490

\bibitem[{{Bianco} et~al.(2011){Bianco}, {Howell}, {Sullivan}, {Conley},
  {Kasen} et~al.}]{Bianco11}
{Bianco} FB, {Howell} DA, {Sullivan} M, {Conley} A, {Kasen} D, et~al. 2011.
\newblock \textit{\apj} 741:20

\bibitem[{{Bildsten} et~al.(2007){Bildsten}, {Shen}, {Weinberg} \&
  {Nelemans}}]{Bildsten07}
{Bildsten} L, {Shen} KJ, {Weinberg} NN, {Nelemans} G. 2007.
\newblock \textit{\apjl} 662:L95--L98

\bibitem[{{Blair} et~al.(2007){Blair}, {Ghavamian}, {Long}, {Williams},
  {Borkowski} et~al.}]{Blair07}
{Blair} WP, {Ghavamian} P, {Long} KS, {Williams} BJ, {Borkowski} KJ, et~al.
  2007.
\newblock \textit{\apj} 662:998--1013

\bibitem[{{Blanc} \& {Greggio}(2008)}]{Blanc08b}
{Blanc} G, {Greggio} L. 2008.
\newblock \textit{New Astronomy} 13:606--618

\bibitem[{{Blondin} et~al.(2013){Blondin}, {Dessart}, {Hillier} \&
  {Khokhlov}}]{Blondin13}
{Blondin} S, {Dessart} L, {Hillier} DJ, {Khokhlov} AM. 2013.
\newblock \textit{\mnras} 429:2127--2142

\bibitem[{{Blondin} et~al.(2011){Blondin}, {Kasen}, {R{\"o}pke}, {Kirshner} \&
  {Mandel}}]{Blondin11b}
{Blondin} S, {Kasen} D, {R{\"o}pke} FK, {Kirshner} RP, {Mandel} KS. 2011.
\newblock \textit{\mnras} 417:1280--1302

\bibitem[{{Blondin} et~al.(2012){Blondin}, {Matheson}, {Kirshner}, {Mandel},
  {Berlind} et~al.}]{Blondin12}
{Blondin} S, {Matheson} T, {Kirshner} RP, {Mandel} KS, {Berlind} P, et~al.
  2012.
\newblock \textit{\aj} 143:126

\bibitem[{{Blondin} et~al.(2009){Blondin}, {Prieto}, {Patat}, {Challis},
  {Hicken} et~al.}]{Blondin09}
{Blondin} S, {Prieto} JL, {Patat} F, {Challis} P, {Hicken} M, et~al. 2009.
\newblock \textit{\apj} 693:207--215

\bibitem[{{Bloom} et~al.(2012){Bloom}, {Kasen}, {Shen}, {Nugent}, {Butler}
  et~al.}]{Bloom12}
{Bloom} JS, {Kasen} D, {Shen} KJ, {Nugent} PE, {Butler} NR, et~al. 2012.
\newblock \textit{\apjl} 744:L17

\bibitem[{{Bogomazov} \& {Tutukov}(2011)}]{Bogomazov11}
{Bogomazov} AI, {Tutukov} AV. 2011.
\newblock \textit{Astronomy Reports} 55:497--504

\bibitem[{{Borkowski}, {Blondin} \& {Sarazin}(1992)}]{Borkowski92}
{Borkowski} KJ, {Blondin} JM, {Sarazin} CL. 1992.
\newblock \textit{\apj} 400:222--237

\bibitem[{{Borkowski}, {Hendrick} \& {Reynolds}(2006)}]{Borkowski06}
{Borkowski} KJ, {Hendrick} SP, {Reynolds} SP. 2006.
\newblock \textit{\apj} 652:1259--1267

\bibitem[{{Borkowski}, {Sarazin} \& {Blondin}(1994)}]{Borkowski94}
{Borkowski} KJ, {Sarazin} CL, {Blondin} JM. 1994.
\newblock \textit{\apj} 429:710--725

\bibitem[{{Boshkayev} et~al.(2013){Boshkayev}, {Izzo}, {Rueda Hernandez} \&
  {Ruffini}}]{Boshkayev13}
{Boshkayev} K, {Izzo} L, {Rueda Hernandez} JA, {Ruffini} R. 2013.
\newblock \textit{\aap} 555:A151

\bibitem[{{Bours}, {Toonen} \& {Nelemans}(2013)}]{Bours13}
{Bours} MCP, {Toonen} S, {Nelemans} G. 2013.
\newblock \textit{\aap} 552:A24

\bibitem[{{Branch} \& {van den Bergh}(1993)}]{Branch93}
{Branch} D, {van den Bergh} S. 1993.
\newblock \textit{\aj} 105:2231--2235

\bibitem[{{Brandt} et~al.(2010){Brandt}, {Tojeiro}, {Aubourg}, {Heavens},
  {Jimenez} \& {Strauss}}]{Brandt10}
{Brandt} TD, {Tojeiro} R, {Aubourg} {\'E}, {Heavens} A, {Jimenez} R, {Strauss}
  MA. 2010.
\newblock \textit{\aj} 140:804--816

\bibitem[{{Brown} et~al.(2012{\natexlab{a}}){Brown}, {Dawson}, {de Pasquale},
  {Gronwall}, {Holland} et~al.}]{Brown12b}
{Brown} PJ, {Dawson} KS, {de Pasquale} M, {Gronwall} C, {Holland} S, et~al.
  2012{\natexlab{a}}.
\newblock \textit{\apj} 753:22

\bibitem[{{Brown} et~al.(2012{\natexlab{b}}){Brown}, {Dawson}, {Harris},
  {Olmstead}, {Milne} \& {Roming}}]{Brown12a}
{Brown} PJ, {Dawson} KS, {Harris} DW, {Olmstead} M, {Milne} P, {Roming} PWA.
  2012{\natexlab{b}}.
\newblock \textit{\apj} 749:18

\bibitem[{{Brown} et~al.(2011){Brown}, {Kilic}, {Hermes}, {Allende Prieto},
  {Kenyon} \& {Winget}}]{Brown11}
{Brown} WR, {Kilic} M, {Hermes} JJ, {Allende Prieto} C, {Kenyon} SJ, {Winget}
  DE. 2011.
\newblock \textit{\apjl} 737:L23

\bibitem[{{Burkey} et~al.(2013){Burkey}, {Reynolds}, {Borkowski} \&
  {Blondin}}]{Burkey13}
{Burkey} MT, {Reynolds} SP, {Borkowski} KJ, {Blondin} JM. 2013.
\newblock \textit{\apj} 764:63

\bibitem[{{Calura}, {Matteucci} \& {Tozzi}(2007)}]{Calura07}
{Calura} F, {Matteucci} F, {Tozzi} P. 2007.
\newblock \textit{\mnras} 378:L11--L15

\bibitem[{{Canal}, {M{\'e}ndez} \& {Ruiz-Lapuente}(2001)}]{Canal01}
{Canal} R, {M{\'e}ndez} J, {Ruiz-Lapuente} P. 2001.
\newblock \textit{\apjl} 550:L53--L56

\bibitem[{{Capetti}(2002)}]{Capetti02}
{Capetti} A. 2002.
\newblock \textit{\apjl} 574:L25--L27

\bibitem[{{Cappellari} et~al.(2012){Cappellari}, {McDermid}, {Alatalo},
  {Blitz}, {Bois} et~al.}]{Cappellari12}
{Cappellari} M, {McDermid} RM, {Alatalo} K, {Blitz} L, {Bois} M, et~al. 2012.
\newblock \textit{\nat} 484:485--488

\bibitem[{{Cappellaro}, {Evans} \& {Turatto}(1999)}]{Cappellaro99}
{Cappellaro} E, {Evans} R, {Turatto} M. 1999.
\newblock \textit{\aap} 351:459--466

\bibitem[{{Cappellaro} et~al.(2001){Cappellaro}, {Patat}, {Mazzali}, {Benetti},
  {Danziger} et~al.}]{Cappellaro01}
{Cappellaro} E, {Patat} F, {Mazzali} PA, {Benetti} S, {Danziger} JI, et~al.
  2001.
\newblock \textit{\apjl} 549:L215--L218

\bibitem[{{Cartier} et~al.(2011){Cartier}, {F{\"o}rster}, {Coppi}, {Hamuy},
  {Maeda} et~al.}]{Cartier11}
{Cartier} R, {F{\"o}rster} F, {Coppi} P, {Hamuy} M, {Maeda} K, et~al. 2011.
\newblock \textit{\aap} 534:L15

\bibitem[{{Chamel}, {Fantina} \& {Davis}(2013)}]{Chamel13}
{Chamel} N, {Fantina} AF, {Davis} PJ. 2013.
\newblock \textit{\prd} 88:081301

\bibitem[{{Charpinet}, {Fontaine} \& {Brassard}(2009)}]{Charpinet09}
{Charpinet} S, {Fontaine} G, {Brassard} P. 2009.
\newblock \textit{\nat} 461:501--503

\bibitem[{{Chevalier}(1982)}]{Chevalier82}
{Chevalier} RA. 1982.
\newblock \textit{\apj} 258:790--797

\bibitem[{{Chevalier}(1998)}]{Chevalier98}
{Chevalier} RA. 1998.
\newblock \textit{\apj} 499:810

\bibitem[{{Chevalier} \& {Fransson}(2006)}]{Chevalier06}
{Chevalier} RA, {Fransson} C. 2006.
\newblock \textit{\apj} 651:381--391

\bibitem[{{Chevalier}, {Kirshner} \& {Raymond}(1980)}]{Chevalier80}
{Chevalier} RA, {Kirshner} RP, {Raymond} JC. 1980.
\newblock \textit{\apj} 235:186--195

\bibitem[{{Childress} et~al.(2013){Childress}, {Aldering}, {Antilogus},
  {Aragon}, {Bailey} et~al.}]{Childress13a}
{Childress} M, {Aldering} G, {Antilogus} P, {Aragon} C, {Bailey} S, et~al.
  2013.
\newblock \textit{\apj} 770:107

\bibitem[{{Childress} et~al.(2014){Childress}, {Filippenko}, {Ganeshalingam} \&
  {Schmidt}}]{Childress14}
{Childress} MJ, {Filippenko} AV, {Ganeshalingam} M, {Schmidt} BP. 2014.
\newblock \textit{\mnras} 437:338--350

\bibitem[{{Chiotellis}, {Schure} \& {Vink}(2012)}]{Chiotellis12}
{Chiotellis} A, {Schure} KM, {Vink} J. 2012.
\newblock \textit{\aap} 537:A139

\bibitem[{{Chomiuk}(2013)}]{Chomiuk13}
{Chomiuk} L. 2013.
\newblock \textit{\pasa} 30:46

\bibitem[{{Chomiuk} et~al.(2012{\natexlab{a}}){Chomiuk}, {Krauss}, {Rupen},
  {Nelson}, {Roy} et~al.}]{Chomiuk12b}
{Chomiuk} L, {Krauss} MI, {Rupen} MP, {Nelson} T, {Roy} N, et~al.
  2012{\natexlab{a}}.
\newblock \textit{\apj} 761:173

\bibitem[{{Chomiuk} et~al.(2012{\natexlab{b}}){Chomiuk}, {Soderberg}, {Moe},
  {Chevalier}, {Rupen} et~al.}]{Chomiuk12a}
{Chomiuk} L, {Soderberg} AM, {Moe} M, {Chevalier} RA, {Rupen} MP, et~al.
  2012{\natexlab{b}}.
\newblock \textit{\apj} 750:164

\bibitem[{{Chugai}(2008)}]{Chugai08}
{Chugai} NN. 2008.
\newblock \textit{Astronomy Letters} 34:389--396

\bibitem[{{Chugai} \& {Yungelson}(2004)}]{Chugai04}
{Chugai} NN, {Yungelson} LR. 2004.
\newblock \textit{Astronomy Letters} 30:65--72

\bibitem[{{Colgate} \& {McKee}(1969)}]{Colgate69}
{Colgate} SA, {McKee} C. 1969.
\newblock \textit{\apj} 157:623

\bibitem[{{Conroy} \& {van Dokkum}(2012)}]{Conroy12}
{Conroy} C, {van Dokkum} P. 2012.
\newblock \textit{\apj} 747:69

\bibitem[{{Cooper}, {Newman} \& {Yan}(2009)}]{Cooper09}
{Cooper} MC, {Newman} JA, {Yan} R. 2009.
\newblock \textit{\apj} 704:687--704

\bibitem[{{C{\'o}rsico} et~al.(2011){C{\'o}rsico}, {Althaus}, {Kawaler},
  {Miller Bertolami}, {Garc{\'{\i}}a-Berro} \& {Kepler}}]{Corsico11}
{C{\'o}rsico} AH, {Althaus} LG, {Kawaler} SD, {Miller Bertolami} MM,
  {Garc{\'{\i}}a-Berro} E, {Kepler} SO. 2011.
\newblock \textit{\mnras} 418:2519--2526

\bibitem[{{Crotts} \& {Yourdon}(2008)}]{Crotts08}
{Crotts} APS, {Yourdon} D. 2008.
\newblock \textit{\apj} 689:1186--1190

\bibitem[{{Dahlen}, {Strolger} \& {Riess}(2008)}]{Dahlen08}
{Dahlen} T, {Strolger} LG, {Riess} AG. 2008.
\newblock \textit{\apj} 681:462--469

\bibitem[{{Dahlen} et~al.(2004){Dahlen}, {Strolger}, {Riess}, {Mobasher},
  {Chary} et~al.}]{Dahlen04}
{Dahlen} T, {Strolger} LG, {Riess} AG, {Mobasher} B, {Chary} RR, et~al. 2004.
\newblock \textit{\apj} 613:189--199

\bibitem[{{Dan} et~al.(2013){Dan}, {Rosswog}, {Brueggen} \&
  {Podsiadlowski}}]{Dan13}
{Dan} M, {Rosswog} S, {Brueggen} M, {Podsiadlowski} P. 2013.
\newblock \textit{ArXiv e-prints}

\bibitem[{{Dan} et~al.(2012){Dan}, {Rosswog}, {Guillochon} \&
  {Ramirez-Ruiz}}]{Dan12}
{Dan} M, {Rosswog} S, {Guillochon} J, {Ramirez-Ruiz} E. 2012.
\newblock \textit{\mnras} 422:2417--2428

\bibitem[{{D'Andrea} et~al.(2011){D'Andrea}, {Gupta}, {Sako}, {Morris},
  {Nichol} et~al.}]{DAndrea11}
{D'Andrea} CB, {Gupta} RR, {Sako} M, {Morris} M, {Nichol} RC, et~al. 2011.
\newblock \textit{\apj} 743:172

\bibitem[{{Darbha} et~al.(2010){Darbha}, {Metzger}, {Quataert}, {Kasen},
  {Nugent} \& {Thomas}}]{Darbha10}
{Darbha} S, {Metzger} BD, {Quataert} E, {Kasen} D, {Nugent} P, {Thomas} R.
  2010.
\newblock \textit{\mnras} 409:846--854

\bibitem[{{Darnley} et~al.(2006){Darnley}, {Bode}, {Kerins}, {Newsam}, {An}
  et~al.}]{Darnley06}
{Darnley} MJ, {Bode} MF, {Kerins} E, {Newsam} AM, {An} J, et~al. 2006.
\newblock \textit{\mnras} 369:257--271

\bibitem[{{Dav{\'e}}, {Oppenheimer} \& {Finlator}(2011)}]{Dave11b}
{Dav{\'e}} R, {Oppenheimer} BD, {Finlator} K. 2011.
\newblock \textit{\mnras} 415:11--31

\bibitem[{{Dawson} et~al.(2009){Dawson}, {Aldering}, {Amanullah}, {Barbary},
  {Barrientos} et~al.}]{Dawson09}
{Dawson} KS, {Aldering} G, {Amanullah} R, {Barbary} K, {Barrientos} LF, et~al.
  2009.
\newblock \textit{\aj} 138:1271--1283

\bibitem[{{De Donder} \& {Vanbeveren}(2004)}]{De-Donder04}
{De Donder} E, {Vanbeveren} D. 2004.
\newblock \textit{\nar} 48:861--975

\bibitem[{{della Valle} \& {Livio}(1996)}]{Della-Valle96}
{della Valle} M, {Livio} M. 1996.
\newblock \textit{\apj} 473:240

\bibitem[{{Della Valle} et~al.(2005){Della Valle}, {Panagia}, {Padovani},
  {Cappellaro}, {Mannucci} \& {Turatto}}]{Della-Valle05}
{Della Valle} M, {Panagia} N, {Padovani} P, {Cappellaro} E, {Mannucci} F,
  {Turatto} M. 2005.
\newblock \textit{\apj} 629:750--756

\bibitem[{{della Valle} et~al.(1994){della Valle}, {Rosino}, {Bianchini} \&
  {Livio}}]{Della-Valle94}
{della Valle} M, {Rosino} L, {Bianchini} A, {Livio} M. 1994.
\newblock \textit{\aap} 287:403--409

\bibitem[{{Dessart} et~al.(2013{\natexlab{a}}){Dessart}, {Blondin}, {Hillier}
  \& {Khokhlov}}]{Dessart13b}
{Dessart} L, {Blondin} S, {Hillier} DJ, {Khokhlov} A. 2013{\natexlab{a}}.
\newblock \textit{ArXiv e-prints}

\bibitem[{{Dessart} et~al.(2013{\natexlab{b}}){Dessart}, {Hillier}, {Blondin}
  \& {Khokhlov}}]{Dessart13a}
{Dessart} L, {Hillier} DJ, {Blondin} S, {Khokhlov} A. 2013{\natexlab{b}}.
\newblock \textit{ArXiv e-prints}

\bibitem[{{Di Stefano}(2010)}]{Di-Stefano10}
{Di Stefano} R. 2010.
\newblock \textit{\apj} 712:728--733

\bibitem[{{Di Stefano}, {Voss} \& {Claeys}(2011)}]{Di-Stefano11}
{Di Stefano} R, {Voss} R, {Claeys} JSW. 2011.
\newblock \textit{\apjl} 738:L1+

\bibitem[{{Diaz} et~al.(2010){Diaz}, {Williams}, {Luna}, {Moraes} \&
  {Takeda}}]{Diaz10}
{Diaz} MP, {Williams} RE, {Luna} GJ, {Moraes} M, {Takeda} L. 2010.
\newblock \textit{\aj} 140:1860--1867

\bibitem[{{Dilday} et~al.(2010{\natexlab{a}}){Dilday}, {Bassett}, {Becker},
  {Bender}, {Castander} et~al.}]{Dilday10b}
{Dilday} B, {Bassett} B, {Becker} A, {Bender} R, {Castander} F, et~al.
  2010{\natexlab{a}}.
\newblock \textit{\apj} 715:1021--1035

\bibitem[{{Dilday} et~al.(2012){Dilday}, {Howell}, {Cenko}, {Silverman},
  {Nugent} et~al.}]{Dilday12}
{Dilday} B, {Howell} DA, {Cenko} SB, {Silverman} JM, {Nugent} PE, et~al. 2012.
\newblock \textit{Science} 337:942--

\bibitem[{{Dilday} et~al.(2010{\natexlab{b}}){Dilday}, {Smith}, {Bassett},
  {Becker}, {Bender} et~al.}]{Dilday10a}
{Dilday} B, {Smith} M, {Bassett} B, {Becker} A, {Bender} R, et~al.
  2010{\natexlab{b}}.
\newblock \textit{\apj} 713:1026--1036

\bibitem[{{Drury}(2012)}]{Drury12}
{Drury} LO. 2012.
\newblock \textit{Astroparticle Physics} 39:52--60

\bibitem[{{Duch{\^e}ne} \& {Kraus}(2013)}]{Duchene13}
{Duch{\^e}ne} G, {Kraus} A. 2013.
\newblock \textit{\araa} 51:269--310

\bibitem[{{Edwards}, {Pagnotta} \& {Schaefer}(2012)}]{Edwards12}
{Edwards} ZI, {Pagnotta} A, {Schaefer} BE. 2012.
\newblock \textit{\apjl} 747:L19

\bibitem[{{Filippenko}(1997)}]{Filippenko97}
{Filippenko} AV. 1997.
\newblock \textit{\araa} 35:309--355

\bibitem[{{Fink}, {Hillebrandt} \& {R{\"o}pke}(2007)}]{Fink07}
{Fink} M, {Hillebrandt} W, {R{\"o}pke} FK. 2007.
\newblock \textit{\aap} 476:1133--1143

\bibitem[{{Fink} et~al.(2013){Fink}, {Kromer}, {Seitenzahl},
  {Ciaraldi-Schoolmann}, {Roepke} et~al.}]{Fink13}
{Fink} M, {Kromer} M, {Seitenzahl} IR, {Ciaraldi-Schoolmann} F, {Roepke} FK,
  et~al. 2013.
\newblock \textit{ArXiv e-prints}

\bibitem[{{Fink} et~al.(2010){Fink}, {R{\"o}pke}, {Hillebrandt}, {Seitenzahl},
  {Sim} \& {Kromer}}]{Fink10}
{Fink} M, {R{\"o}pke} FK, {Hillebrandt} W, {Seitenzahl} IR, {Sim} SA, {Kromer}
  M. 2010.
\newblock \textit{\aap} 514:A53

\bibitem[{{Folatelli} et~al.(2012){Folatelli}, {Phillips}, {Morrell}, {Tanaka},
  {Maeda} et~al.}]{Folatelli12}
{Folatelli} G, {Phillips} MM, {Morrell} N, {Tanaka} M, {Maeda} K, et~al. 2012.
\newblock \textit{\apj} 745:74

\bibitem[{{Foley}(2012)}]{Foley12a}
{Foley} RJ. 2012.
\newblock \textit{\apj} 748:127

\bibitem[{{Foley}(2013)}]{Foley13d}
{Foley} RJ. 2013.
\newblock \textit{\mnras} 435:273--288

\bibitem[{{Foley} et~al.(2010{\natexlab{a}}){Foley}, {Brown}, {Rest},
  {Challis}, {Kirshner} \& {Wood-Vasey}}]{Foley10c}
{Foley} RJ, {Brown} PJ, {Rest} A, {Challis} PJ, {Kirshner} RP, {Wood-Vasey} WM.
  2010{\natexlab{a}}.
\newblock \textit{\apjl} 708:L61--L65

\bibitem[{{Foley} et~al.(2013){Foley}, {Challis}, {Chornock}, {Ganeshalingam},
  {Li} et~al.}]{Foley13b}
{Foley} RJ, {Challis} PJ, {Chornock} R, {Ganeshalingam} M, {Li} W, et~al. 2013.
\newblock \textit{\apj} 767:57

\bibitem[{{Foley} et~al.(2012{\natexlab{a}}){Foley}, {Challis}, {Filippenko},
  {Ganeshalingam}, {Landsman} et~al.}]{Foley12e}
{Foley} RJ, {Challis} PJ, {Filippenko} AV, {Ganeshalingam} M, {Landsman} W,
  et~al. 2012{\natexlab{a}}.
\newblock \textit{\apj} 744:38

\bibitem[{{Foley} \& {Kasen}(2011)}]{Foley11a}
{Foley} RJ, {Kasen} D. 2011.
\newblock \textit{\apj} 729:55

\bibitem[{{Foley} \& {Kirshner}(2013)}]{Foley13a}
{Foley} RJ, {Kirshner} RP. 2013.
\newblock \textit{\apjl} 769:L1

\bibitem[{{Foley} et~al.(2010{\natexlab{b}}){Foley}, {Rest}, {Stritzinger},
  {Pignata}, {Anderson} et~al.}]{Foley10b}
{Foley} RJ, {Rest} A, {Stritzinger} M, {Pignata} G, {Anderson} JP, et~al.
  2010{\natexlab{b}}.
\newblock \textit{\aj} 140:1321--1328

\bibitem[{{Foley}, {Sanders} \& {Kirshner}(2011)}]{Foley11b}
{Foley} RJ, {Sanders} NE, {Kirshner} RP. 2011.
\newblock \textit{\apj} 742:89

\bibitem[{{Foley} et~al.(2012{\natexlab{b}}){Foley}, {Simon}, {Burns},
  {Gal-Yam}, {Hamuy} et~al.}]{Foley12c}
{Foley} RJ, {Simon} JD, {Burns} CR, {Gal-Yam} A, {Hamuy} M, et~al.
  2012{\natexlab{b}}.
\newblock \textit{\apj} 752:101

\bibitem[{{Fontaine}, {Brassard} \& {Charpinet}(2013)}]{Fontaine13}
{Fontaine} G, {Brassard} P, {Charpinet} S. 2013.
\newblock In \textit{European Physical Journal Web of Conferences}, vol.~43 of
  \textit{European Physical Journal Web of Conferences}

\bibitem[{{F{\"o}rster} et~al.(2013){F{\"o}rster}, {Gonz{\'a}lez-Gait{\'a}n},
  {Folatelli} \& {Morrell}}]{Forster13}
{F{\"o}rster} F, {Gonz{\'a}lez-Gait{\'a}n} S, {Folatelli} G, {Morrell} N. 2013.
\newblock \textit{\apj} 772:19

\bibitem[{{Fox} \& {Filippenko}(2013)}]{Fox13a}
{Fox} OD, {Filippenko} AV. 2013.
\newblock \textit{\apjl} 772:L6

\bibitem[{{Fox} et~al.(2013){Fox}, {Filippenko}, {Skrutskie}, {Silverman},
  {Ganeshalingam} et~al.}]{Fox13b}
{Fox} OD, {Filippenko} AV, {Skrutskie} MF, {Silverman} JM, {Ganeshalingam} M,
  et~al. 2013.
\newblock \textit{\aj} 146:2

\bibitem[{{Friedrich} \& {Jordan}(2001)}]{Friedrich01}
{Friedrich} S, {Jordan} S. 2001.
\newblock \textit{\aap} 367:577--581

\bibitem[{{Frieman} et~al.(2008){Frieman}, {Bassett}, {Becker}, {Choi},
  {Cinabro} et~al.}]{Frieman08}
{Frieman} JA, {Bassett} B, {Becker} A, {Choi} C, {Cinabro} D, et~al. 2008.
\newblock \textit{\aj} 135:338--347

\bibitem[{{Fryer} et~al.(2010){Fryer}, {Ruiter}, {Belczynski}, {Brown},
  {Bufano} et~al.}]{Fryer10}
{Fryer} CL, {Ruiter} AJ, {Belczynski} K, {Brown} PJ, {Bufano} F, et~al. 2010.
\newblock \textit{\apj} 725:296--308

\bibitem[{{Fuhrmann}(2005)}]{Fuhrmann05}
{Fuhrmann} K. 2005.
\newblock \textit{\mnras} 359:L35--L36

\bibitem[{{Fujimoto}(1982)}]{Fujimoto82}
{Fujimoto} MY. 1982.
\newblock \textit{\apj} 257:767

\bibitem[{{Gal-Yam} et~al.(2008){Gal-Yam}, {Maoz}, {Guhathakurta} \&
  {Filippenko}}]{Gal-Yam08}
{Gal-Yam} A, {Maoz} D, {Guhathakurta} P, {Filippenko} AV. 2008.
\newblock \textit{\apj} 680:550--567

\bibitem[{{Gal-Yam}, {Maoz} \& {Sharon}(2002)}]{Gal-Yam02}
{Gal-Yam} A, {Maoz} D, {Sharon} K. 2002.
\newblock \textit{\mnras} 332:37--48

\bibitem[{{Gallazzi} et~al.(2005){Gallazzi}, {Charlot}, {Brinchmann}, {White}
  \& {Tremonti}}]{Gallazzi05}
{Gallazzi} A, {Charlot} S, {Brinchmann} J, {White} SDM, {Tremonti} CA. 2005.
\newblock \textit{\mnras} 362:41--58

\bibitem[{{Ganeshalingam}, {Li} \& {Filippenko}(2011)}]{Ganeshalingam11}
{Ganeshalingam} M, {Li} W, {Filippenko} AV. 2011.
\newblock \textit{\mnras} 416:2607--2622

\bibitem[{{Garc{\'{\i}}a-Berro} et~al.(2012){Garc{\'{\i}}a-Berro},
  {Lor{\'e}n-Aguilar}, {Aznar-Sigu{\'a}n}, {Torres}, {Camacho}
  et~al.}]{Garcia-Berro12}
{Garc{\'{\i}}a-Berro} E, {Lor{\'e}n-Aguilar} P, {Aznar-Sigu{\'a}n} G, {Torres}
  S, {Camacho} J, et~al. 2012.
\newblock \textit{\apj} 749:25

\bibitem[{{Garc{\'{\i}}a-Senz} et~al.(2013){Garc{\'{\i}}a-Senz}, {Cabez{\'o}n},
  {Arcones}, {Rela{\~n}o} \& {Thielemann}}]{Garcia-Senz13}
{Garc{\'{\i}}a-Senz} D, {Cabez{\'o}n} RM, {Arcones} A, {Rela{\~n}o} A,
  {Thielemann} FK. 2013.
\newblock \textit{\mnras}

\bibitem[{{Garnavich} et~al.(2001){Garnavich}, {Kirshner}, {Challis}, {Jha},
  {Branch} et~al.}]{Garnavich01}
{Garnavich} PM, {Kirshner} RP, {Challis} P, {Jha} S, {Branch} D, et~al. 2001.
\newblock In \textit{American Astronomical Society Meeting Abstracts}, vol.~33
  of \textit{Bulletin of the American Astronomical Society}

\bibitem[{{Geha} et~al.(2013){Geha}, {Brown}, {Tumlinson}, {Kalirai}, {Simon}
  et~al.}]{Geha13}
{Geha} M, {Brown} TM, {Tumlinson} J, {Kalirai} JS, {Simon} JD, et~al. 2013.
\newblock \textit{\apj} 771:29

\bibitem[{{Geier} et~al.(2010){Geier}, {Heber}, {Kupfer} \&
  {Napiwotzki}}]{Geier10}
{Geier} S, {Heber} U, {Kupfer} T, {Napiwotzki} R. 2010.
\newblock \textit{\aap} 515:A37

\bibitem[{{Geier} et~al.(2013){Geier}, {Marsh}, {Wang}, {Dunlap}, {Barlow}
  et~al.}]{Geier13}
{Geier} S, {Marsh} TR, {Wang} B, {Dunlap} B, {Barlow} BN, et~al. 2013.
\newblock \textit{\aap} 554:A54

\bibitem[{{Geier} et~al.(2007){Geier}, {Nesslinger}, {Heber}, {Przybilla},
  {Napiwotzki} \& {Kudritzki}}]{Geier07}
{Geier} S, {Nesslinger} S, {Heber} U, {Przybilla} N, {Napiwotzki} R,
  {Kudritzki} RP. 2007.
\newblock \textit{\aap} 464:299--307

\bibitem[{{Gerardy} et~al.(2004){Gerardy}, {H{\"o}flich}, {Fesen}, {Marion},
  {Nomoto} et~al.}]{Gerardy04}
{Gerardy} CL, {H{\"o}flich} P, {Fesen} RA, {Marion} GH, {Nomoto} K, et~al.
  2004.
\newblock \textit{\apj} 607:391--405

\bibitem[{{Germany} et~al.(2004){Germany}, {Reiss}, {Schmidt}, {Stubbs} \&
  {Suntzeff}}]{Germany04}
{Germany} LM, {Reiss} DJ, {Schmidt} BP, {Stubbs} CW, {Suntzeff} NB. 2004.
\newblock \textit{\aap} 415:863--878

\bibitem[{{Ghavamian} et~al.(2003){Ghavamian}, {Rakowski}, {Hughes} \&
  {Williams}}]{Ghavamian03}
{Ghavamian} P, {Rakowski} CE, {Hughes} JP, {Williams} TB. 2003.
\newblock \textit{\apj} 590:833--845

\bibitem[{{Ghavamian} et~al.(2001){Ghavamian}, {Raymond}, {Smith} \&
  {Hartigan}}]{Ghavamian01}
{Ghavamian} P, {Raymond} J, {Smith} RC, {Hartigan} P. 2001.
\newblock \textit{\apj} 547:995--1009

\bibitem[{{Gilfanov} \& {Bogd{\'a}n}(2010)}]{Gilfanov10}
{Gilfanov} M, {Bogd{\'a}n} {\'A}. 2010.
\newblock \textit{\nat} 463:924--925

\bibitem[{{Gonz{\'a}lez Hern{\'a}ndez} et~al.(2009){Gonz{\'a}lez
  Hern{\'a}ndez}, {Ruiz-Lapuente}, {Filippenko}, {Foley}, {Gal-Yam} \&
  {Simon}}]{Gonzalez-Hernandez09}
{Gonz{\'a}lez Hern{\'a}ndez} JI, {Ruiz-Lapuente} P, {Filippenko} AV, {Foley}
  RJ, {Gal-Yam} A, {Simon} JD. 2009.
\newblock \textit{\apj} 691:1--15

\bibitem[{{Gonz{\'a}lez Hern{\'a}ndez} et~al.(2012){Gonz{\'a}lez
  Hern{\'a}ndez}, {Ruiz-Lapuente}, {Tabernero}, {Montes}, {Canal}
  et~al.}]{Gonzalez-Hernandez12}
{Gonz{\'a}lez Hern{\'a}ndez} JI, {Ruiz-Lapuente} P, {Tabernero} HM, {Montes} D,
  {Canal} R, et~al. 2012.
\newblock \textit{\nat} 489:533--536

\bibitem[{{Graham} et~al.(2008){Graham}, {Pritchet}, {Sullivan}, {Gwyn},
  {Neill} et~al.}]{Graham08}
{Graham} ML, {Pritchet} CJ, {Sullivan} M, {Gwyn} SDJ, {Neill} JD, et~al. 2008.
\newblock \textit{\aj} 135:1343--1349

\bibitem[{{Graham} et~al.(2010){Graham}, {Pritchet}, {Sullivan}, {Howell},
  {Gwyn} et~al.}]{Graham10}
{Graham} ML, {Pritchet} CJ, {Sullivan} M, {Howell} DA, {Gwyn} SDJ, et~al. 2010.
\newblock \textit{\aj} 139:594--605

\bibitem[{{Graur} \& {Maoz}(2012{\natexlab{a}})}]{Graur12a}
{Graur} O, {Maoz} D. 2012{\natexlab{a}}.
\newblock \textit{The Astronomer's Telegram} 4226:1

\bibitem[{{Graur} \& {Maoz}(2012{\natexlab{b}})}]{Graur12b}
{Graur} O, {Maoz} D. 2012{\natexlab{b}}.
\newblock \textit{The Astronomer's Telegram} 4535:1

\bibitem[{{Graur} \& {Maoz}(2013)}]{Graur13a}
{Graur} O, {Maoz} D. 2013.
\newblock \textit{\mnras} 430:1746--1763

\bibitem[{{Graur} et~al.(2011){Graur}, {Poznanski}, {Maoz}, {Yasuda}, {Totani}
  et~al.}]{Graur11}
{Graur} O, {Poznanski} D, {Maoz} D, {Yasuda} N, {Totani} T, et~al. 2011.
\newblock \textit{\mnras} 417:916--940

\bibitem[{{Graur} et~al.(2013){Graur}, {Rodney}, {Maoz}, {Riess}, {Jha}
  et~al.}]{Graur13b}
{Graur} O, {Rodney} SA, {Maoz} D, {Riess} AG, {Jha} SW, et~al. 2013.
\newblock \textit{ArXiv e-prints}

\bibitem[{{Greggio}(2005)}]{Greggio05}
{Greggio} L. 2005.
\newblock \textit{\aap} 441:1055--1078

\bibitem[{{Greggio}, {Renzini} \& {Daddi}(2008)}]{Greggio08}
{Greggio} L, {Renzini} A, {Daddi} E. 2008.
\newblock \textit{\mnras} 388:829--837

\bibitem[{{Gruyters} et~al.(2012){Gruyters}, {Exter}, {Roberts} \&
  {Rappaport}}]{Gruyters12}
{Gruyters} P, {Exter} K, {Roberts} TP, {Rappaport} S. 2012.
\newblock \textit{\aap} 544:A86

\bibitem[{{Guillochon} et~al.(2010){Guillochon}, {Dan}, {Ramirez-Ruiz} \&
  {Rosswog}}]{Guillochon10}
{Guillochon} J, {Dan} M, {Ramirez-Ruiz} E, {Rosswog} S. 2010.
\newblock \textit{\apjl} 709:L64--L69

\bibitem[{{Hachinger} et~al.(2013){Hachinger}, {Mazzali}, {Sullivan}, {Ellis},
  {Maguire} et~al.}]{Hachinger13}
{Hachinger} S, {Mazzali} PA, {Sullivan} M, {Ellis} RS, {Maguire} K, et~al.
  2013.
\newblock \textit{\mnras} 429:2228--2248

\bibitem[{{Hachisu} \& {Kato}(2003{\natexlab{a}})}]{Hachisu03b}
{Hachisu} I, {Kato} M. 2003{\natexlab{a}}.
\newblock \textit{\apj} 598:527--544

\bibitem[{{Hachisu} \& {Kato}(2003{\natexlab{b}})}]{Hachisu03a}
{Hachisu} I, {Kato} M. 2003{\natexlab{b}}.
\newblock \textit{\apj} 590:445--459

\bibitem[{{Hachisu} \& {Kato}(2012)}]{Hachisu12c}
{Hachisu} I, {Kato} M. 2012.
\newblock \textit{Baltic Astronomy} 21:68--75

\bibitem[{{Hachisu} et~al.(2006){Hachisu}, {Kato}, {Kiyota}, {Kubotera},
  {Maehara} et~al.}]{Hachisu06}
{Hachisu} I, {Kato} M, {Kiyota} S, {Kubotera} K, {Maehara} H, et~al. 2006.
\newblock \textit{\apjl} 651:L141--L144

\bibitem[{{Hachisu}, {Kato} \& {Nomoto}(1996)}]{Hachisu96}
{Hachisu} I, {Kato} M, {Nomoto} K. 1996.
\newblock \textit{\apjl} 470:L97

\bibitem[{{Hachisu}, {Kato} \& {Nomoto}(1999)}]{Hachisu99}
{Hachisu} I, {Kato} M, {Nomoto} K. 1999.
\newblock \textit{\apj} 522:487--503

\bibitem[{{Hachisu}, {Kato} \& {Nomoto}(2008{\natexlab{a}})}]{Hachisu08b}
{Hachisu} I, {Kato} M, {Nomoto} K. 2008{\natexlab{a}}.
\newblock \textit{\apjl} 683:L127--L130

\bibitem[{{Hachisu}, {Kato} \& {Nomoto}(2008{\natexlab{b}})}]{Hachisu08a}
{Hachisu} I, {Kato} M, {Nomoto} K. 2008{\natexlab{b}}.
\newblock \textit{\apj} 679:1390--1404

\bibitem[{{Hachisu}, {Kato} \& {Nomoto}(2010)}]{Hachisu10}
{Hachisu} I, {Kato} M, {Nomoto} K. 2010.
\newblock \textit{\apjl} 724:L212--L216

\bibitem[{{Hachisu}, {Kato} \& {Nomoto}(2012)}]{Hachisu12b}
{Hachisu} I, {Kato} M, {Nomoto} K. 2012.
\newblock \textit{\apjl} 756:L4

\bibitem[{{Hachisu} et~al.(2012){Hachisu}, {Kato}, {Saio} \&
  {Nomoto}}]{Hachisu12a}
{Hachisu} I, {Kato} M, {Saio} H, {Nomoto} K. 2012.
\newblock \textit{\apj} 744:69

\bibitem[{{Hakobyan} et~al.(2008){Hakobyan}, {Petrosian}, {McLean}, {Kunth},
  {Allen} et~al.}]{Hakobyan08}
{Hakobyan} AA, {Petrosian} AR, {McLean} B, {Kunth} D, {Allen} RJ, et~al. 2008.
\newblock \textit{\aap} 488:523--531

\bibitem[{{Hamann} et~al.(2003){Hamann}, {Pe{\~n}a}, {Gr{\"a}fener} \&
  {Ruiz}}]{Hamann03}
{Hamann} WR, {Pe{\~n}a} M, {Gr{\"a}fener} G, {Ruiz} MT. 2003.
\newblock \textit{\aap} 409:969--982

\bibitem[{{Hamers} et~al.(2013){Hamers}, {Pols}, {Claeys} \&
  {Nelemans}}]{Hamers13}
{Hamers} AS, {Pols} OR, {Claeys} JSW, {Nelemans} G. 2013.
\newblock \textit{\mnras} 430:2262--2280

\bibitem[{{Hamilton} \& {Sarazin}(1984)}]{Hamilton84}
{Hamilton} AJS, {Sarazin} CL. 1984.
\newblock \textit{\apj} 287:282--294

\bibitem[{{Hamuy} et~al.(2003){Hamuy}, {Phillips}, {Suntzeff}, {Maza},
  {Gonz{\'a}lez} et~al.}]{Hamuy03}
{Hamuy} M, {Phillips} MM, {Suntzeff} NB, {Maza} J, {Gonz{\'a}lez} LE, et~al.
  2003.
\newblock \textit{\nat} 424:651--654

\bibitem[{{Hamuy} et~al.(2000){Hamuy}, {Trager}, {Pinto}, {Phillips},
  {Schommer} et~al.}]{Hamuy00}
{Hamuy} M, {Trager} SC, {Pinto} PA, {Phillips} MM, {Schommer} RA, et~al. 2000.
\newblock \textit{\aj} 120:1479--1486

\bibitem[{{Han}(1998)}]{Han98}
{Han} Z. 1998.
\newblock \textit{\mnras} 296:1019--1040

\bibitem[{{Han} \& {Podsiadlowski}(2004)}]{Han04}
{Han} Z, {Podsiadlowski} P. 2004.
\newblock \textit{\mnras} 350:1301--1309

\bibitem[{{Han} \& {Podsiadlowski}(2006)}]{Han06}
{Han} Z, {Podsiadlowski} P. 2006.
\newblock \textit{\mnras} 368:1095--1100

\bibitem[{{Hancock}, {Gaensler} \& {Murphy}(2011)}]{Hancock11}
{Hancock} PJ, {Gaensler} BM, {Murphy} T. 2011.
\newblock \textit{\apjl} 735:L35

\bibitem[{{Harris} \& {Zaritsky}(2004)}]{Harris04}
{Harris} J, {Zaritsky} D. 2004.
\newblock \textit{\aj} 127:1531--1544

\bibitem[{{Harris} \& {Zaritsky}(2009)}]{Harris09}
{Harris} J, {Zaritsky} D. 2009.
\newblock \textit{\aj} 138:1243--1260

\bibitem[{{Hawley}, {Athanassiadou} \& {Timmes}(2012)}]{Hawley12}
{Hawley} WP, {Athanassiadou} T, {Timmes} FX. 2012.
\newblock \textit{\apj} 759:39

\bibitem[{{Hayden} et~al.(2010{\natexlab{a}}){Hayden}, {Garnavich}, {Kasen},
  {Dilday}, {Frieman} et~al.}]{Hayden10b}
{Hayden} BT, {Garnavich} PM, {Kasen} D, {Dilday} B, {Frieman} JA, et~al.
  2010{\natexlab{a}}.
\newblock \textit{\apj} 722:1691--1698

\bibitem[{{Hayden} et~al.(2010{\natexlab{b}}){Hayden}, {Garnavich}, {Kessler},
  {Frieman}, {Jha} et~al.}]{Hayden10a}
{Hayden} BT, {Garnavich} PM, {Kessler} R, {Frieman} JA, {Jha} SW, et~al.
  2010{\natexlab{b}}.
\newblock \textit{\apj} 712:350--366

\bibitem[{{Hayden} et~al.(2013){Hayden}, {Gupta}, {Garnavich}, {Mannucci},
  {Nichol} \& {Sako}}]{Hayden12}
{Hayden} BT, {Gupta} RR, {Garnavich} PM, {Mannucci} F, {Nichol} RC, {Sako} M.
  2013.
\newblock \textit{\apj} 764:191

\bibitem[{{Heng}(2010)}]{Heng10}
{Heng} K. 2010.
\newblock \textit{\pasa} 27:23--44

\bibitem[{{Hermes} et~al.(2012){Hermes}, {Kilic}, {Brown}, {Winget}, {Allende
  Prieto} et~al.}]{Hermes12}
{Hermes} JJ, {Kilic} M, {Brown} WR, {Winget} DE, {Allende Prieto} C, et~al.
  2012.
\newblock \textit{\apjl} 757:L21

\bibitem[{{Hicken} et~al.(2009){Hicken}, {Challis}, {Jha}, {Kirshner},
  {Matheson} et~al.}]{Hicken09b}
{Hicken} M, {Challis} P, {Jha} S, {Kirshner} RP, {Matheson} T, et~al. 2009.
\newblock \textit{\apj} 700:331--357

\bibitem[{{Hillebrandt} et~al.(2013){Hillebrandt}, {Kromer}, {R{\"o}pke} \&
  {Ruiter}}]{Hillebrandt13}
{Hillebrandt} W, {Kromer} M, {R{\"o}pke} FK, {Ruiter} AJ. 2013.
\newblock \textit{Frontiers of Physics} 8:116--143

\bibitem[{{Hillebrandt} \& {Niemeyer}(2000)}]{Hillebrandt00}
{Hillebrandt} W, {Niemeyer} JC. 2000.
\newblock \textit{\araa} 38:191--230

\bibitem[{{Hillebrandt}, {Sim} \& {R{\"o}pke}(2007)}]{Hillebrandt07}
{Hillebrandt} W, {Sim} SA, {R{\"o}pke} FK. 2007.
\newblock \textit{\aap} 465:L17--L20

\bibitem[{{Hoeflich} \& {Khokhlov}(1996)}]{Hoeflich96}
{Hoeflich} P, {Khokhlov} A. 1996.
\newblock \textit{\apj} 457:500

\bibitem[{{H{\"o}flich} \& {Schaefer}(2009)}]{Hoeflich09b}
{H{\"o}flich} P, {Schaefer} BE. 2009.
\newblock \textit{\apj} 705:483--495

\bibitem[{{Hole}, {Kasen} \& {Nordsieck}(2010)}]{Hole10}
{Hole} KT, {Kasen} D, {Nordsieck} KH. 2010.
\newblock \textit{\apj} 720:1500--1512

\bibitem[{{Horesh} et~al.(2012){Horesh}, {Kulkarni}, {Fox}, {Carpenter},
  {Kasliwal} et~al.}]{Horesh12}
{Horesh} A, {Kulkarni} SR, {Fox} DB, {Carpenter} J, {Kasliwal} MM, et~al. 2012.
\newblock \textit{\apj} 746:21

\bibitem[{{Horiuchi} \& {Beacom}(2010)}]{Horiuchi10}
{Horiuchi} S, {Beacom} JF. 2010.
\newblock \textit{\apj} 723:329--341

\bibitem[{{Howell}(2011)}]{Howell11}
{Howell} DA. 2011.
\newblock \textit{Nature Communications} 2

\bibitem[{{Howell} et~al.(2009{\natexlab{a}}){Howell}, {Conley}, {Della Valle},
  {Nugent}, {Perlmutter} et~al.}]{Howell09b}
{Howell} DA, {Conley} A, {Della Valle} M, {Nugent} PE, {Perlmutter} S, et~al.
  2009{\natexlab{a}}.
\newblock \textit{ArXiv e-prints (0903.1086)}

\bibitem[{{Howell} et~al.(2001){Howell}, {H{\"o}flich}, {Wang} \&
  {Wheeler}}]{Howell01}
{Howell} DA, {H{\"o}flich} P, {Wang} L, {Wheeler} JC. 2001.
\newblock \textit{\apj} 556:302--321

\bibitem[{{Howell} et~al.(2009{\natexlab{b}}){Howell}, {Sullivan}, {Brown},
  {Conley}, {LeBorgne} et~al.}]{Howell09a}
{Howell} DA, {Sullivan} M, {Brown} EF, {Conley} A, {LeBorgne} D, et~al.
  2009{\natexlab{b}}.
\newblock \textit{\apj} 691:661--671

\bibitem[{{Howell} et~al.(2006){Howell}, {Sullivan}, {Nugent}, {Ellis},
  {Conley} et~al.}]{Howell06}
{Howell} DA, {Sullivan} M, {Nugent} PE, {Ellis} RS, {Conley} AJ, et~al. 2006.
\newblock \textit{\nat} 443:308--311

\bibitem[{{Hoyle} \& {Fowler}(1960)}]{Hoyle60}
{Hoyle} F, {Fowler} WA. 1960.
\newblock \textit{\apj} 132:565

\bibitem[{{Hughes} et~al.(2007){Hughes}, {Chugai}, {Chevalier}, {Lundqvist} \&
  {Schlegel}}]{Hughes07}
{Hughes} JP, {Chugai} N, {Chevalier} R, {Lundqvist} P, {Schlegel} E. 2007.
\newblock \textit{\apj} 670:1260--1274

\bibitem[{{Iben} \& {Renzini}(1983)}]{Iben83}
{Iben} Jr. I, {Renzini} A. 1983.
\newblock \textit{\araa} 21:271--342

\bibitem[{{Iben} \& {Tutukov}(1984)}]{Iben84a}
{Iben} Jr. I, {Tutukov} AV. 1984.
\newblock \textit{\apjs} 54:335--372

\bibitem[{{Iben}, {Tutukov} \& {Yungelson}(1997)}]{Iben97}
{Iben} Jr. I, {Tutukov} AV, {Yungelson} LR. 1997.
\newblock \textit{\apj} 475:291

\bibitem[{{Idan}, {Shaviv} \& {Shaviv}(2013)}]{Idan13}
{Idan} I, {Shaviv} N, {Shaviv} G. 2013.
\newblock \textit{\mnras} 433:2884--2892

\bibitem[{{Ihara} et~al.(2007){Ihara}, {Ozaki}, {Doi}, {Shigeyama}, {Kashikawa}
  et~al.}]{Ihara07}
{Ihara} Y, {Ozaki} J, {Doi} M, {Shigeyama} T, {Kashikawa} N, et~al. 2007.
\newblock \textit{\pasj} 59:811--826

\bibitem[{{Ilkov} \& {Soker}(2011)}]{Ilkov11}
{Ilkov} M, {Soker} N. 2011.
\newblock \textit{\mnras} :1798

\bibitem[{{Ivanova} et~al.(2013){Ivanova}, {Justham}, {Chen}, {De Marco},
  {Fryer} et~al.}]{Ivanova13}
{Ivanova} N, {Justham} S, {Chen} X, {De Marco} O, {Fryer} CL, et~al. 2013.
\newblock \textit{\aapr} 21:59

\bibitem[{{Iwamoto} et~al.(1999){Iwamoto}, {Brachwitz}, {Nomoto}, {Kishimoto},
  {Umeda} et~al.}]{Iwamoto99}
{Iwamoto} K, {Brachwitz} F, {Nomoto} K, {Kishimoto} N, {Umeda} H, et~al. 1999.
\newblock \textit{\apjs} 125:439--462

\bibitem[{{Jimenez} et~al.(2007){Jimenez}, {Bernardi}, {Haiman}, {Panter} \&
  {Heavens}}]{Jimenez07}
{Jimenez} R, {Bernardi} M, {Haiman} Z, {Panter} B, {Heavens} AF. 2007.
\newblock \textit{\apj} 669:947--951

\bibitem[{{Johansson}, {Amanullah} \& {Goobar}(2013)}]{Johansson13}
{Johansson} J, {Amanullah} R, {Goobar} A. 2013.
\newblock \textit{\mnras} 431:L43--L47

\bibitem[{{Jordan} et~al.(2012){Jordan}, {Perets}, {Fisher} \& {van
  Rossum}}]{Jordan12}
{Jordan} IV GC, {Perets} HB, {Fisher} RT, {van Rossum} DR. 2012.
\newblock \textit{\apjl} 761:L23

\bibitem[{{Jordan} et~al.(2007){Jordan}, {Aznar Cuadrado}, {Napiwotzki},
  {Schmid} \& {Solanki}}]{Jordan07}
{Jordan} S, {Aznar Cuadrado} R, {Napiwotzki} R, {Schmid} HM, {Solanki} SK.
  2007.
\newblock \textit{\aap} 462:1097--1101

\bibitem[{{Jorgensen} et~al.(1997){Jorgensen}, {Lipunov}, {Panchenko},
  {Postnov} \& {Prokhorov}}]{Jorgensen97}
{Jorgensen} HE, {Lipunov} VM, {Panchenko} IE, {Postnov} KA, {Prokhorov} ME.
  1997.
\newblock \textit{\apj} 486:110--+

\bibitem[{{Justham}(2011)}]{Justham11}
{Justham} S. 2011.
\newblock \textit{\apjl} 730:L34+

\bibitem[{{Kafka}, {Honeycutt} \& {Williams}(2012)}]{Kafka12}
{Kafka} S, {Honeycutt} RK, {Williams} R. 2012.
\newblock \textit{\mnras} 425:1585--1590

\bibitem[{{Kafka} \& {Williams}(2011)}]{Kafka11}
{Kafka} S, {Williams} R. 2011.
\newblock \textit{\aap} 526:A83

\bibitem[{{Kamiya} et~al.(2012){Kamiya}, {Tanaka}, {Nomoto}, {Blinnikov},
  {Sorokina} \& {Suzuki}}]{Kamiya12}
{Kamiya} Y, {Tanaka} M, {Nomoto} K, {Blinnikov} SI, {Sorokina} EI, {Suzuki} T.
  2012.
\newblock \textit{\apj} 756:191

\bibitem[{{Kaplan}, {Bildsten} \& {Steinfadt}(2012)}]{Kaplan12}
{Kaplan} DL, {Bildsten} L, {Steinfadt} JDR. 2012.
\newblock \textit{\apj} 758:64

\bibitem[{{Kasen}(2010)}]{Kasen10}
{Kasen} D. 2010.
\newblock \textit{\apj} 708:1025--1031

\bibitem[{{Kasen} \& {Nugent}(2013)}]{Kasen13}
{Kasen} D, {Nugent} P. 2013.
\newblock \textit{Annual Review of Nuclear and Particle Science} 63:153--174

\bibitem[{{Kasen} et~al.(2003){Kasen}, {Nugent}, {Wang}, {Howell}, {Wheeler}
  et~al.}]{Kasen03}
{Kasen} D, {Nugent} P, {Wang} L, {Howell} DA, {Wheeler} JC, et~al. 2003.
\newblock \textit{\apj} 593:788--808

\bibitem[{{Kasen}, {R{\"o}pke} \& {Woosley}(2009)}]{Kasen09}
{Kasen} D, {R{\"o}pke} FK, {Woosley} SE. 2009.
\newblock \textit{\nat} 460:869--872

\bibitem[{{Kasen} \& {Woosley}(2007)}]{Kasen07}
{Kasen} D, {Woosley} SE. 2007.
\newblock \textit{\apj} 656:661--665

\bibitem[{{Kashi} \& {Soker}(2011)}]{Kashi11}
{Kashi} A, {Soker} N. 2011.
\newblock \textit{\mnras} 417:1466--1479

\bibitem[{{Kato} \& {Hachisu}(2012)}]{Kato12}
{Kato} M, {Hachisu} I. 2012.
\newblock \textit{Bulletin of the Astronomical Society of India} 40:393

\bibitem[{{Kato} et~al.(2008){Kato}, {Hachisu}, {Kiyota} \& {Saio}}]{Kato08}
{Kato} M, {Hachisu} I, {Kiyota} S, {Saio} H. 2008.
\newblock \textit{\apj} 684:1366--1373

\bibitem[{{Kattner} et~al.(2012){Kattner}, {Leonard}, {Burns}, {Phillips},
  {Folatelli} et~al.}]{Kattner12}
{Kattner} S, {Leonard} DC, {Burns} CR, {Phillips} MM, {Folatelli} G, et~al.
  2012.
\newblock \textit{\pasp} 124:114--127

\bibitem[{{Katz} \& {Dong}(2012)}]{Katz12}
{Katz} B, {Dong} S. 2012.
\newblock \textit{ArXiv e-prints}

\bibitem[{{Kawaler}(2004)}]{Kawaler04}
{Kawaler} SD. 2004.
\newblock In \textit{Stellar Rotation}, eds. A~{Maeder}, P~{Eenens}, vol. 215
  of \textit{IAU Symposium}

\bibitem[{{Kawka} et~al.(2007){Kawka}, {Vennes}, {Schmidt}, {Wickramasinghe} \&
  {Koch}}]{Kawka07}
{Kawka} A, {Vennes} S, {Schmidt} GD, {Wickramasinghe} DT, {Koch} R. 2007.
\newblock \textit{\apj} 654:499--520

\bibitem[{{Kepler} et~al.(2013){Kepler}, {Pelisoli}, {Jordan}, {Kleinman},
  {Koester} et~al.}]{Kepler13}
{Kepler} SO, {Pelisoli} I, {Jordan} S, {Kleinman} SJ, {Koester} D, et~al. 2013.
\newblock \textit{\mnras} 429:2934--2944

\bibitem[{{Kerzendorf} et~al.(2013{\natexlab{a}}){Kerzendorf}, {Childress},
  {Scharwachter} \& {Schmidt}}]{Kerzendorf13}
{Kerzendorf} WE, {Childress} M, {Scharwachter} TJ, {Schmidt} BPK.
  2013{\natexlab{a}}.
\newblock \textit{submitted}

\bibitem[{{Kerzendorf} et~al.(2009){Kerzendorf}, {Schmidt}, {Asplund},
  {Nomoto}, {Podsiadlowski} et~al.}]{Kerzendorf09}
{Kerzendorf} WE, {Schmidt} BP, {Asplund} M, {Nomoto} K, {Podsiadlowski} P,
  et~al. 2009.
\newblock \textit{\apj} 701:1665--1672

\bibitem[{{Kerzendorf} et~al.(2012){Kerzendorf}, {Schmidt}, {Laird},
  {Podsiadlowski} \& {Bessell}}]{Kerzendorf12b}
{Kerzendorf} WE, {Schmidt} BP, {Laird} JB, {Podsiadlowski} P, {Bessell} MS.
  2012.
\newblock \textit{\apj} 759:7

\bibitem[{{Kerzendorf} et~al.(2013{\natexlab{b}}){Kerzendorf}, {Yong},
  {Schmidt}, {Simon}, {Jeffery} et~al.}]{Kerzendorf12a}
{Kerzendorf} WE, {Yong} D, {Schmidt} BP, {Simon} JD, {Jeffery} CS, et~al.
  2013{\natexlab{b}}.
\newblock \textit{\apj} 774:99

\bibitem[{{Khokhlov}(1991)}]{Khokhlov91}
{Khokhlov} AM. 1991.
\newblock \textit{\aap} 245:114--128

\bibitem[{{Kilic} et~al.(2012){Kilic}, {Brown}, {Allende Prieto}, {Kenyon},
  {Heinke} et~al.}]{Kilic12}
{Kilic} M, {Brown} WR, {Allende Prieto} C, {Kenyon} SJ, {Heinke} CO, et~al.
  2012.
\newblock \textit{\apj} 751:141

\bibitem[{{Kistler} et~al.(2013){Kistler}, {Stanek}, {Kochanek}, {Prieto} \&
  {Thompson}}]{Kistler13}
{Kistler} MD, {Stanek} KZ, {Kochanek} CS, {Prieto} JL, {Thompson} TA. 2013.
\newblock \textit{\apj} 770:88

\bibitem[{{Kobayashi} \& {Nomoto}(2009)}]{kobayashi09}
{Kobayashi} C, {Nomoto} K. 2009.
\newblock \textit{\apj} 707:1466--1484

\bibitem[{{Kosenko}, {Blinnikov} \& {Vink}(2011)}]{Kosenko11}
{Kosenko} D, {Blinnikov} SI, {Vink} J. 2011.
\newblock \textit{\aap} 532:A114

\bibitem[{{Kosenko} et~al.(2008){Kosenko}, {Vink}, {Blinnikov} \&
  {Rasmussen}}]{Kosenko08}
{Kosenko} D, {Vink} J, {Blinnikov} S, {Rasmussen} A. 2008.
\newblock \textit{\aap} 490:223--230

\bibitem[{{Kozai}(1962)}]{Kozai62}
{Kozai} Y. 1962.
\newblock \textit{\aj} 67:591

\bibitem[{{Krause} et~al.(2008){Krause}, {Tanaka}, {Usuda}, {Hattori}, {Goto}
  et~al.}]{Krause08}
{Krause} O, {Tanaka} M, {Usuda} T, {Hattori} T, {Goto} M, et~al. 2008.
\newblock \textit{\nat} 456:617--619

\bibitem[{{Kromer} et~al.(2013){Kromer}, {Pakmor}, {Taubenberger}, {Pignata},
  {Fink} et~al.}]{Kromer13}
{Kromer} M, {Pakmor} R, {Taubenberger} S, {Pignata} G, {Fink} M, et~al. 2013.
\newblock \textit{\apjl} 778:L18

\bibitem[{{Kushnir} et~al.(2013){Kushnir}, {Katz}, {Dong}, {Livne} \&
  {Fern{\'a}ndez}}]{Kushnir13a}
{Kushnir} D, {Katz} B, {Dong} S, {Livne} E, {Fern{\'a}ndez} R. 2013.
\newblock \textit{ArXiv e-prints}

\bibitem[{{Kuznetsova} \& {Connolly}(2007)}]{Kuznetsova07}
{Kuznetsova} NV, {Connolly} BM. 2007.
\newblock \textit{\apj} 659:530--540

\bibitem[{{Lampeitl} et~al.(2010){Lampeitl}, {Smith}, {Nichol}, {Bassett},
  {Cinabro} et~al.}]{Lampeitl10}
{Lampeitl} H, {Smith} M, {Nichol} RC, {Bassett} B, {Cinabro} D, et~al. 2010.
\newblock \textit{\apj} 722:566--576

\bibitem[{{Landstreet} et~al.(2012){Landstreet}, {Bagnulo}, {Valyavin},
  {Fossati}, {Jordan} et~al.}]{Landstreet12}
{Landstreet} JD, {Bagnulo} S, {Valyavin} GG, {Fossati} L, {Jordan} S, et~al.
  2012.
\newblock \textit{\aap} 545:A30

\bibitem[{{Langer} et~al.(2000){Langer}, {Deutschmann}, {Wellstein} \&
  {H{\"o}flich}}]{Langer00}
{Langer} N, {Deutschmann} A, {Wellstein} S, {H{\"o}flich} P. 2000.
\newblock \textit{\aap} 362:1046--1064

\bibitem[{{Lanz} et~al.(2005){Lanz}, {Telis}, {Audard}, {Paerels}, {Rasmussen}
  \& {Hubeny}}]{Lanz05}
{Lanz} T, {Telis} GA, {Audard} M, {Paerels} F, {Rasmussen} AP, {Hubeny} I.
  2005.
\newblock \textit{\apj} 619:517--526

\bibitem[{{Leaman} et~al.(2011){Leaman}, {Li}, {Chornock} \&
  {Filippenko}}]{Leaman11}
{Leaman} J, {Li} W, {Chornock} R, {Filippenko} AV. 2011.
\newblock \textit{\mnras} 412:1419--1440

\bibitem[{{Leigh} \& {Geller}(2013)}]{Leigh13}
{Leigh} NWC, {Geller} AM. 2013.
\newblock \textit{\mnras} 432:2474--2479

\bibitem[{{Leonard}(2007)}]{Leonard07}
{Leonard} DC. 2007.
\newblock \textit{\apj} 670:1275--1282

\bibitem[{{Leonard} et~al.(2005){Leonard}, {Li}, {Filippenko}, {Foley} \&
  {Chornock}}]{Leonard05}
{Leonard} DC, {Li} W, {Filippenko} AV, {Foley} RJ, {Chornock} R. 2005.
\newblock \textit{\apj} 632:450--475

\bibitem[{{Lepo} \& {van Kerkwijk}(2013)}]{Lepo13b}
{Lepo} K, {van Kerkwijk} M. 2013.
\newblock \textit{\apj} 771:13

\bibitem[{{Li} et~al.(2011{\natexlab{a}}){Li}, {Bloom}, {Podsiadlowski},
  {Miller}, {Cenko} et~al.}]{Li11c}
{Li} W, {Bloom} JS, {Podsiadlowski} P, {Miller} AA, {Cenko} SB, et~al.
  2011{\natexlab{a}}.
\newblock \textit{\nat} 480:348--350

\bibitem[{{Li} et~al.(2011{\natexlab{b}}){Li}, {Chornock}, {Leaman},
  {Filippenko}, {Poznanski} et~al.}]{Li11b}
{Li} W, {Chornock} R, {Leaman} J, {Filippenko} AV, {Poznanski} D, et~al.
  2011{\natexlab{b}}.
\newblock \textit{\mnras} 412:1473--1507

\bibitem[{{Li} et~al.(2011{\natexlab{c}}){Li}, {Leaman}, {Chornock},
  {Filippenko}, {Poznanski} et~al.}]{Li11a}
{Li} W, {Leaman} J, {Chornock} R, {Filippenko} AV, {Poznanski} D, et~al.
  2011{\natexlab{c}}.
\newblock \textit{\mnras} 412:1441--1472

\bibitem[{{Li} \& {van den Heuvel}(1997)}]{Li97}
{Li} XD, {van den Heuvel} EPJ. 1997.
\newblock \textit{\aap} 322:L9--L12

\bibitem[{{Lidov}(1962)}]{Lidov62}
{Lidov} ML. 1962.
\newblock \textit{\planss} 9:719--759

\bibitem[{{Lipunov}, {Panchenko} \& {Pruzhinskaya}(2011)}]{Lipunov11}
{Lipunov} VM, {Panchenko} IE, {Pruzhinskaya} MV. 2011.
\newblock \textit{\na} 16:250--252

\bibitem[{{Liu} et~al.(2012{\natexlab{a}}){Liu}, {Di Stefano}, {Wang} \&
  {Moe}}]{Liu12a}
{Liu} J, {Di Stefano} R, {Wang} T, {Moe} M. 2012{\natexlab{a}}.
\newblock \textit{\apj} 749:141

\bibitem[{{Liu} et~al.(2013{\natexlab{a}}){Liu}, {Kromer}, {Fink}, {Pakmor},
  {R{\"o}pke} et~al.}]{Liu13c}
{Liu} ZW, {Kromer} M, {Fink} M, {Pakmor} R, {R{\"o}pke} FK, et~al.
  2013{\natexlab{a}}.
\newblock \textit{\apj} 778:121

\bibitem[{{Liu} et~al.(2013{\natexlab{b}}){Liu}, {Pakmor}, {R{\"o}pke},
  {Edelmann}, {Hillebrandt} et~al.}]{Liu13a}
{Liu} ZW, {Pakmor} R, {R{\"o}pke} FK, {Edelmann} P, {Hillebrandt} W, et~al.
  2013{\natexlab{b}}.
\newblock \textit{\aap} 554:A109

\bibitem[{{Liu} et~al.(2012{\natexlab{b}}){Liu}, {Pakmor}, {R{\"o}pke},
  {Edelmann}, {Wang} et~al.}]{Liu12b}
{Liu} ZW, {Pakmor} R, {R{\"o}pke} FK, {Edelmann} P, {Wang} B, et~al.
  2012{\natexlab{b}}.
\newblock \textit{\aap} 548:A2

\bibitem[{{Livio}(2000)}]{Livio00}
{Livio} M. 2000.
\newblock In \textit{Type Ia Supernovae, Theory and Cosmology}, eds.
  JC~{Niemeyer}, JW~{Truran}

\bibitem[{{Livio}, {Riess} \& {Sparks}(2002)}]{Livio02}
{Livio} M, {Riess} A, {Sparks} W. 2002.
\newblock \textit{\apjl} 571:L99--L102

\bibitem[{{Livio} \& {Riess}(2003)}]{Livio03}
{Livio} M, {Riess} AG. 2003.
\newblock \textit{\apjl} 594:L93--L94

\bibitem[{{Livne}(1990)}]{Livne90}
{Livne} E. 1990.
\newblock \textit{\apjl} 354:L53--L55

\bibitem[{{Loewenstein}(2013)}]{Loewenstein13}
{Loewenstein} M. 2013.
\newblock \textit{\apj} 773:52

\bibitem[{{Lor{\'e}n-Aguilar}, {Isern} \&
  {Garc{\'{\i}}a-Berro}(2009)}]{Loren-Aguilar09}
{Lor{\'e}n-Aguilar} P, {Isern} J, {Garc{\'{\i}}a-Berro} E. 2009.
\newblock \textit{\aap} 500:1193--1205

\bibitem[{{Lor{\'e}n-Aguilar}, {Isern} \&
  {Garc{\'{\i}}a-Berro}(2010)}]{Loren-Aguilar10}
{Lor{\'e}n-Aguilar} P, {Isern} J, {Garc{\'{\i}}a-Berro} E. 2010.
\newblock \textit{\mnras} 406:2749--2763

\bibitem[{{Luna}, {Sokoloski} \& {Mukai}(2008)}]{Luna08}
{Luna} GJM, {Sokoloski} JL, {Mukai} K. 2008.
\newblock In \textit{RS Ophiuchi (2006) and the Recurrent Nova Phenomenon},
  eds. A~{Evans}, MF~{Bode}, TJ~{O'Brien}, MJ~{Darnley}, vol. 401 of
  \textit{Astronomical Society of the Pacific Conference Series}

\bibitem[{{Lundqvist} et~al.(2013){Lundqvist}, {Mattila}, {Sollerman}, {Kozma},
  {Baron} et~al.}]{Lundqvist13}
{Lundqvist} P, {Mattila} S, {Sollerman} J, {Kozma} C, {Baron} E, et~al. 2013.
\newblock \textit{\mnras} 435:329--345

\bibitem[{{Lyman} et~al.(2013){Lyman}, {James}, {Perets}, {Anderson}, {Gal-Yam}
  et~al.}]{Lyman13}
{Lyman} JD, {James} PA, {Perets} HB, {Anderson} JP, {Gal-Yam} A, et~al. 2013.
\newblock \textit{\mnras} 434:527--541

\bibitem[{{Ma} et~al.(2013){Ma}, {Woosley}, {Malone}, {Almgren} \&
  {Bell}}]{Ma13b}
{Ma} H, {Woosley} SE, {Malone} CM, {Almgren} A, {Bell} J. 2013.
\newblock \textit{\apj} 771:58

\bibitem[{{Maeda} et~al.(2010){Maeda}, {Benetti}, {Stritzinger}, {R{\"o}pke},
  {Folatelli} et~al.}]{Maeda10}
{Maeda} K, {Benetti} S, {Stritzinger} M, {R{\"o}pke} FK, {Folatelli} G, et~al.
  2010.
\newblock \textit{\nat} 466:82--85

\bibitem[{{Maeda} et~al.(2011){Maeda}, {Leloudas}, {Taubenberger},
  {Stritzinger}, {Sollerman} et~al.}]{Maeda11}
{Maeda} K, {Leloudas} G, {Taubenberger} S, {Stritzinger} M, {Sollerman} J,
  et~al. 2011.
\newblock \textit{\mnras} 413:3075--3094

\bibitem[{{Maguire} et~al.(2012){Maguire}, {Sullivan}, {Ellis}, {Nugent},
  {Howell} et~al.}]{Maguire12}
{Maguire} K, {Sullivan} M, {Ellis} RS, {Nugent} PE, {Howell} DA, et~al. 2012.
\newblock \textit{\mnras} 426:2359--2379

\bibitem[{{Maguire} et~al.(2013){Maguire}, {Sullivan}, {Patat}, {Gal-Yam},
  {Hook} et~al.}]{Maguire13}
{Maguire} K, {Sullivan} M, {Patat} F, {Gal-Yam} A, {Hook} IM, et~al. 2013.
\newblock \textit{\mnras} 436:222--240

\bibitem[{{Mannucci}(2008)}]{Mannucci08b}
{Mannucci} F. 2008.
\newblock \textit{Chinese Journal of Astronomy and Astrophysics Supplement}
  8:143--154

\bibitem[{{Mannucci}(2009)}]{Mannucci09a}
{Mannucci} F. 2009.
\newblock In \textit{American Institute of Physics Conference Series}, eds.
  G~{Giobbi}, A~{Tornambe}, G~{Raimondo}, M~{Limongi}, LA~{Antonelli},
  N~{Menci}, E~{Brocato}, vol. 1111 of \textit{American Institute of Physics
  Conference Series}

\bibitem[{{Mannucci} et~al.(2010){Mannucci}, {Cresci}, {Maiolino}, {Marconi} \&
  {Gnerucci}}]{Mannucci10}
{Mannucci} F, {Cresci} G, {Maiolino} R, {Marconi} A, {Gnerucci} A. 2010.
\newblock \textit{\mnras} 408:2115--2127

\bibitem[{{Mannucci}, {Della Valle} \& {Panagia}(2006)}]{Mannucci06}
{Mannucci} F, {Della Valle} M, {Panagia} N. 2006.
\newblock \textit{\mnras} 370:773--783

\bibitem[{{Mannucci}, {Della Valle} \& {Panagia}(2007)}]{Mannucci07b}
{Mannucci} F, {Della Valle} M, {Panagia} N. 2007.
\newblock \textit{\mnras} 377:1229--1235

\bibitem[{{Mannucci} et~al.(2005){Mannucci}, {Della Valle}, {Panagia},
  {Cappellaro}, {Cresci} et~al.}]{Mannucci05}
{Mannucci} F, {Della Valle} M, {Panagia} N, {Cappellaro} E, {Cresci} G, et~al.
  2005.
\newblock \textit{\aap} 433:807--814

\bibitem[{{Mannucci} et~al.(2008){Mannucci}, {Maoz}, {Sharon}, {Botticella},
  {Della Valle} et~al.}]{Mannucci08a}
{Mannucci} F, {Maoz} D, {Sharon} K, {Botticella} MT, {Della Valle} M, et~al.
  2008.
\newblock \textit{\mnras} 383:1121--1130

\bibitem[{{Maoz} \& {Badenes}(2010)}]{Maoz10b}
{Maoz} D, {Badenes} C. 2010.
\newblock \textit{\mnras} 407:1314--1327

\bibitem[{{Maoz}, {Badenes} \& {Bickerton}(2012)}]{Maoz12b}
{Maoz} D, {Badenes} C, {Bickerton} SJ. 2012.
\newblock \textit{\apj} 751:143

\bibitem[{{Maoz} \& {Gal-Yam}(2004)}]{Maoz04}
{Maoz} D, {Gal-Yam} A. 2004.
\newblock \textit{\mnras} 347:951--956

\bibitem[{{Maoz} \& {Mannucci}(2008)}]{Maoz08b}
{Maoz} D, {Mannucci} F. 2008.
\newblock \textit{\mnras} 388:421--428

\bibitem[{{Maoz} \& {Mannucci}(2012)}]{Maoz12a}
{Maoz} D, {Mannucci} F. 2012.
\newblock \textit{\pasa} 29:447--465

\bibitem[{{Maoz}, {Mannucci} \& {Brandt}(2012)}]{Maoz12c}
{Maoz} D, {Mannucci} F, {Brandt} TD. 2012.
\newblock \textit{\mnras} 426:3282--3294

\bibitem[{{Maoz} et~al.(2011){Maoz}, {Mannucci}, {Li}, {Filippenko}, {Della
  Valle} \& {Panagia}}]{Maoz11}
{Maoz} D, {Mannucci} F, {Li} W, {Filippenko} AV, {Della Valle} M, {Panagia} N.
  2011.
\newblock \textit{\mnras} 412:1508--1521

\bibitem[{{Maoz}, {Sharon} \& {Gal-Yam}(2010)}]{Maoz10c}
{Maoz} D, {Sharon} K, {Gal-Yam} A. 2010.
\newblock \textit{\apj} 722:1879--1894

\bibitem[{{Margutti} et~al.(2012){Margutti}, {Soderberg}, {Chomiuk},
  {Chevalier}, {Hurley} et~al.}]{Margutti12}
{Margutti} R, {Soderberg} AM, {Chomiuk} L, {Chevalier} R, {Hurley} K, et~al.
  2012.
\newblock \textit{\apj} 751:134

\bibitem[{{Marietta}, {Burrows} \& {Fryxell}(2000)}]{Marietta00}
{Marietta} E, {Burrows} A, {Fryxell} B. 2000.
\newblock \textit{\apjs} 128:615--650

\bibitem[{{Marsh}(2011)}]{Marsh11}
{Marsh} TR. 2011.
\newblock \textit{Classical and Quantum Gravity} 28:094019

\bibitem[{{Mason}(2013)}]{Mason13}
{Mason} E. 2013.
\newblock \textit{\aap} 556:C2

\bibitem[{{Matteucci}(2001)}]{Matteucci01b}
{Matteucci} F. 2001.
\newblock \textit{{The chemical evolution of the Galaxy}}, vol. 253 of
  \textit{Astrophysics and Space Science Library}.
\newblock Kluwer

\bibitem[{{Matteucci} \& {Greggio}(1986)}]{Matteucci86}
{Matteucci} F, {Greggio} L. 1986.
\newblock \textit{\aap} 154:279--287

\bibitem[{{Matteucci} et~al.(2006){Matteucci}, {Panagia}, {Pipino}, {Mannucci},
  {Recchi} \& {Della Valle}}]{Matteucci06}
{Matteucci} F, {Panagia} N, {Pipino} A, {Mannucci} F, {Recchi} S, {Della Valle}
  M. 2006.
\newblock \textit{\mnras} 372:265--275

\bibitem[{{Matteucci} et~al.(2009){Matteucci}, {Spitoni}, {Recchi} \&
  {Valiante}}]{matteucci09}
{Matteucci} F, {Spitoni} E, {Recchi} S, {Valiante} R. 2009.
\newblock \textit{\aap} 501:531--538

\bibitem[{{Mattila} et~al.(2005){Mattila}, {Lundqvist}, {Sollerman}, {Kozma},
  {Baron} et~al.}]{Mattila05}
{Mattila} S, {Lundqvist} P, {Sollerman} J, {Kozma} C, {Baron} E, et~al. 2005.
\newblock \textit{\aap} 443:649--662

\bibitem[{{Maund} et~al.(2010){Maund}, {H{\"o}flich}, {Patat}, {Wheeler},
  {Zelaya} et~al.}]{Maund10}
{Maund} JR, {H{\"o}flich} P, {Patat} F, {Wheeler} JC, {Zelaya} P, et~al. 2010.
\newblock \textit{\apjl} 725:L167--L171

\bibitem[{{Maund} et~al.(2013){Maund}, {Spyromilio}, {H{\"o}flich}, {Wheeler},
  {Baade} et~al.}]{Maund13}
{Maund} JR, {Spyromilio} J, {H{\"o}flich} PA, {Wheeler} JC, {Baade} D, et~al.
  2013.
\newblock \textit{\mnras} 433:L20--L24

\bibitem[{{Maxted}, {Marsh} \& {North}(2000)}]{Maxted00}
{Maxted} PFL, {Marsh} TR, {North} RC. 2000.
\newblock \textit{\mnras} 317:L41--L44

\bibitem[{{Mazeh} \& {Shaham}(1979)}]{Mazeh79}
{Mazeh} T, {Shaham} J. 1979.
\newblock \textit{\aap} 77:145--151

\bibitem[{{Mazzali} et~al.(2013){Mazzali}, {Sullivan}, {Hachinger}, {Ellis},
  {Nugent} et~al.}]{Mazzali13}
{Mazzali} P, {Sullivan} M, {Hachinger} S, {Ellis} R, {Nugent} PE, et~al. 2013.
\newblock \textit{ArXiv e-prints}

\bibitem[{{Mazzali} et~al.(2005){Mazzali}, {Benetti}, {Altavilla}, {Blanc},
  {Cappellaro} et~al.}]{Mazzali05}
{Mazzali} PA, {Benetti} S, {Altavilla} G, {Blanc} G, {Cappellaro} E, et~al.
  2005.
\newblock \textit{\apjl} 623:L37--L40

\bibitem[{{Mazzali} et~al.(1998){Mazzali}, {Cappellaro}, {Danziger}, {Turatto}
  \& {Benetti}}]{Mazzali98}
{Mazzali} PA, {Cappellaro} E, {Danziger} IJ, {Turatto} M, {Benetti} S. 1998.
\newblock \textit{\apjl} 499:L49

\bibitem[{{Mazzali} \& {Podsiadlowski}(2006)}]{Mazzali06}
{Mazzali} PA, {Podsiadlowski} P. 2006.
\newblock \textit{\mnras} 369:L19--L22

\bibitem[{{Mazzali} et~al.(2007){Mazzali}, {R{\"o}pke}, {Benetti} \&
  {Hillebrandt}}]{Mazzali07}
{Mazzali} PA, {R{\"o}pke} FK, {Benetti} S, {Hillebrandt} W. 2007.
\newblock \textit{Science} 315:825--

\bibitem[{{McCully} et~al.(2013){McCully}, {Jha}, {Foley}, {Chornock},
  {Holtzman} et~al.}]{McCully13}
{McCully} C, {Jha} SW, {Foley} RJ, {Chornock} R, {Holtzman} JA, et~al. 2013.
\newblock \textit{ArXiv e-prints}

\bibitem[{{McMillan}(2011)}]{McMillan11}
{McMillan} PJ. 2011.
\newblock \textit{\mnras} 414:2446--2457

\bibitem[{{Meng} \& {Yang}(2010)}]{Meng10c}
{Meng} X, {Yang} W. 2010.
\newblock \textit{\apj} 710:1310--1323

\bibitem[{{Meng} \& {Yang}(2011)}]{Meng11b}
{Meng} XC, {Yang} WM. 2011.
\newblock \textit{Research in Astronomy and Astrophysics} 11:965--973

\bibitem[{{Mennekens}, {Vanbeveren} \& {De Greve}(2012)}]{Mennekens12b}
{Mennekens} N, {Vanbeveren} D, {De Greve} JP. 2012.
\newblock \textit{ArXiv e-prints}

\bibitem[{{Mennekens} et~al.(2010){Mennekens}, {Vanbeveren}, {De Greve} \& {De
  Donder}}]{Mennekens10}
{Mennekens} N, {Vanbeveren} D, {De Greve} JP, {De Donder} E. 2010.
\newblock \textit{\aap} 515:A89

\bibitem[{{Mennekens} et~al.(2013){Mennekens}, {Vanbeveren}, {De Greve} \& {De
  Donder}}]{Mennekens12a}
{Mennekens} N, {Vanbeveren} D, {De Greve} JP, {De Donder} E. 2013.
\newblock In \textit{IAU Symposium}, eds. R~{Di Stefano}, M~{Orio}, M~{Moe},
  vol. 281 of \textit{IAU Symposium}

\bibitem[{{Moll} et~al.(2013){Moll}, {Raskin}, {Kasen} \& {Woosley}}]{Moll13}
{Moll} R, {Raskin} C, {Kasen} D, {Woosley} S. 2013.
\newblock \textit{ArXiv e-prints}

\bibitem[{{Nakar} \& {Sari}(2010)}]{Nakar10}
{Nakar} E, {Sari} R. 2010.
\newblock \textit{\apj} 725:904--921

\bibitem[{{Nakar} \& {Sari}(2012)}]{Nakar12}
{Nakar} E, {Sari} R. 2012.
\newblock \textit{\apj} 747:88

\bibitem[{{Napiwotzki} et~al.(2004){Napiwotzki}, {Karl}, {Lisker}, {Heber},
  {Christlieb} et~al.}]{Napiwotzki04}
{Napiwotzki} R, {Karl} CA, {Lisker} T, {Heber} U, {Christlieb} N, et~al. 2004.
\newblock \textit{\apss} 291:321--328

\bibitem[{{Napiwotzki} et~al.(2007){Napiwotzki}, {Karl}, {Nelemans},
  {Yungelson}, {Christlieb} et~al.}]{Napiwotzki07}
{Napiwotzki} R, {Karl} CA, {Nelemans} G, {Yungelson} L, {Christlieb} N, et~al.
  2007.
\newblock In \textit{15th European Workshop on White Dwarfs}, eds.
  R~{Napiwotzki}, MR~{Burleigh}, vol. 372 of \textit{Astronomical Society of
  the Pacific Conference Series}

\bibitem[{{Nelemans} et~al.(2005){Nelemans}, {Napiwotzki}, {Karl}, {Marsh},
  {Voss} et~al.}]{Nelemans05b}
{Nelemans} G, {Napiwotzki} R, {Karl} C, {Marsh} TR, {Voss} B, et~al. 2005.
\newblock \textit{\aap} 440:1087--1095

\bibitem[{{Nelemans}, {Toonen} \& {Bours}(2013)}]{Nelemans13}
{Nelemans} G, {Toonen} S, {Bours} M. 2013.
\newblock In \textit{IAU Symposium}, eds. R~{Di Stefano}, M~{Orio}, M~{Moe},
  vol. 281 of \textit{IAU Symposium}

\bibitem[{{Nelemans} et~al.(2008){Nelemans}, {Voss}, {Roelofs} \&
  {Bassa}}]{Nelemans08}
{Nelemans} G, {Voss} R, {Roelofs} G, {Bassa} C. 2008.
\newblock \textit{\mnras} 388:487--494

\bibitem[{{Nelemans}, {Yungelson} \& {Portegies Zwart}(2001)}]{Nelemans01}
{Nelemans} G, {Yungelson} LR, {Portegies Zwart} SF. 2001.
\newblock \textit{\aap} 375:890--898

\bibitem[{{Nelson} et~al.(2012){Nelson}, {Donato}, {Mukai}, {Sokoloski} \&
  {Chomiuk}}]{Nelson12b}
{Nelson} T, {Donato} D, {Mukai} K, {Sokoloski} J, {Chomiuk} L. 2012.
\newblock \textit{\apj} 748:43

\bibitem[{{Newsham}, {Starrfield} \& {Timmes}(2013)}]{Newsham13}
{Newsham} G, {Starrfield} S, {Timmes} F. 2013.
\newblock \textit{ArXiv e-prints}

\bibitem[{{Nielsen} et~al.(2013{\natexlab{a}}){Nielsen}, {Dominik}, {Nelemans}
  \& {Voss}}]{Nielsen13a}
{Nielsen} MTB, {Dominik} C, {Nelemans} G, {Voss} R. 2013{\natexlab{a}}.
\newblock \textit{\aap} 549:A32

\bibitem[{{Nielsen} et~al.(2013{\natexlab{b}}){Nielsen}, {Nelemans}, {Voss} \&
  {Toonen}}]{Nielsen13b}
{Nielsen} MTB, {Nelemans} G, {Voss} R, {Toonen} S. 2013{\natexlab{b}}.
\newblock \textit{ArXiv e-prints}

\bibitem[{{Nielsen}, {Voss} \& {Nelemans}(2012)}]{Nielsen12}
{Nielsen} MTB, {Voss} R, {Nelemans} G. 2012.
\newblock \textit{\mnras} 426:2668--2676

\bibitem[{{Nielsen}, {Voss} \& {Nelemans}(2013)}]{Nielsen13c}
{Nielsen} MTB, {Voss} R, {Nelemans} G. 2013.
\newblock \textit{\mnras} 435:187--193

\bibitem[{{Nomoto}(1982)}]{Nomoto82}
{Nomoto} K. 1982.
\newblock \textit{\apj} 253:798--810

\bibitem[{{Nomoto} \& {Iben}(1985)}]{Nomoto85}
{Nomoto} K, {Iben} Jr. I. 1985.
\newblock \textit{\apj} 297:531--537

\bibitem[{{Nomoto}, {Kobayashi} \& {Tominaga}(2013)}]{Nomoto13}
{Nomoto} K, {Kobayashi} C, {Tominaga} N. 2013.
\newblock \textit{\araa} 51:457--509

\bibitem[{{Nomoto} et~al.(2007){Nomoto}, {Saio}, {Kato} \&
  {Hachisu}}]{Nomoto07}
{Nomoto} K, {Saio} H, {Kato} M, {Hachisu} I. 2007.
\newblock \textit{\apj} 663:1269--1276

\bibitem[{{Nomoto}, {Thielemann} \& {Yokoi}(1984)}]{Nomoto84}
{Nomoto} K, {Thielemann} FK, {Yokoi} K. 1984.
\newblock \textit{\apj} 286:644--658

\bibitem[{{Nordhaus} et~al.(2011){Nordhaus}, {Wellons}, {Spiegel}, {Metzger} \&
  {Blackman}}]{Nordhaus11}
{Nordhaus} J, {Wellons} S, {Spiegel} DS, {Metzger} BD, {Blackman} EG. 2011.
\newblock \textit{Proceedings of the National Academy of Science}
  108:3135--3140

\bibitem[{{Nugent} et~al.(2011){Nugent}, {Sullivan}, {Cenko}, {Thomas}, {Kasen}
  et~al.}]{Nugent11}
{Nugent} PE, {Sullivan} M, {Cenko} SB, {Thomas} RC, {Kasen} D, et~al. 2011.
\newblock \textit{\nat} 480:344--347

\bibitem[{{Orio} et~al.(2010){Orio}, {Nelson}, {Bianchini}, {Di Mille} \&
  {Harbeck}}]{Orio10}
{Orio} M, {Nelson} T, {Bianchini} A, {Di Mille} F, {Harbeck} D. 2010.
\newblock \textit{\apj} 717:739--765

\bibitem[{{Pakmor} et~al.(2012){Pakmor}, {Kromer}, {Taubenberger}, {Sim},
  {R{\"o}pke} \& {Hillebrandt}}]{Pakmor12}
{Pakmor} R, {Kromer} M, {Taubenberger} S, {Sim} SA, {R{\"o}pke} FK,
  {Hillebrandt} W. 2012.
\newblock \textit{\apjl} 747:L10

\bibitem[{{Pakmor} et~al.(2013){Pakmor}, {Kromer}, {Taubenberger} \&
  {Springel}}]{Pakmor13}
{Pakmor} R, {Kromer} M, {Taubenberger} S, {Springel} V. 2013.
\newblock \textit{\apjl} 770:L8

\bibitem[{{Pan}, {Ricker} \& {Taam}(2013)}]{Pan13a}
{Pan} KC, {Ricker} PM, {Taam} RE. 2013.
\newblock \textit{\apj} 773:49

\bibitem[{{Pan} et~al.(2013){Pan}, {Sullivan}, {Maguire}, {Hook}, {Nugent}
  et~al.}]{Pan13}
{Pan} YC, {Sullivan} M, {Maguire} K, {Hook} IM, {Nugent} PE, et~al. 2013.
\newblock \textit{ArXiv e-prints}

\bibitem[{{Panagia} et~al.(2006){Panagia}, {Van Dyk}, {Weiler}, {Sramek},
  {Stockdale} \& {Murata}}]{Panagia06}
{Panagia} N, {Van Dyk} SD, {Weiler} KW, {Sramek} RA, {Stockdale} CJ, {Murata}
  KP. 2006.
\newblock \textit{\apj} 646:369--377

\bibitem[{{Parikh} et~al.(2013){Parikh}, {Jos{\'e}}, {Seitenzahl} \&
  {R{\"o}pke}}]{Parikh13}
{Parikh} A, {Jos{\'e}} J, {Seitenzahl} IR, {R{\"o}pke} FK. 2013.
\newblock \textit{\aap} 557:A3

\bibitem[{{Parrent} et~al.(2012){Parrent}, {Howell}, {Friesen}, {Thomas},
  {Fesen} et~al.}]{Parrent12}
{Parrent} JT, {Howell} DA, {Friesen} B, {Thomas} RC, {Fesen} RA, et~al. 2012.
\newblock \textit{\apjl} 752:L26

\bibitem[{{Parrent} et~al.(2011){Parrent}, {Thomas}, {Fesen}, {Marion},
  {Challis} et~al.}]{Parrent11}
{Parrent} JT, {Thomas} RC, {Fesen} RA, {Marion} GH, {Challis} P, et~al. 2011.
\newblock \textit{\apj} 732:30

\bibitem[{{Patat}(2005)}]{Patat05}
{Patat} F. 2005.
\newblock \textit{\mnras} 357:1161--1177

\bibitem[{{Patat} et~al.(2009){Patat}, {Baade}, {H{\"o}flich}, {Maund}, {Wang}
  \& {Wheeler}}]{Patat09a}
{Patat} F, {Baade} D, {H{\"o}flich} P, {Maund} JR, {Wang} L, {Wheeler} JC.
  2009.
\newblock \textit{\aap} 508:229--246

\bibitem[{{Patat} et~al.(2006){Patat}, {Benetti}, {Cappellaro} \&
  {Turatto}}]{Patat06}
{Patat} F, {Benetti} S, {Cappellaro} E, {Turatto} M. 2006.
\newblock \textit{\mnras} 369:1949--1960

\bibitem[{{Patat} et~al.(2007){Patat}, {Chandra}, {Chevalier}, {Justham},
  {Podsiadlowski} et~al.}]{Patat07}
{Patat} F, {Chandra} P, {Chevalier} R, {Justham} S, {Podsiadlowski} P, et~al.
  2007.
\newblock \textit{Science} 317:924--

\bibitem[{{Patat} et~al.(2011){Patat}, {Chugai}, {Podsiadlowski}, {Mason},
  {Melo} \& {Pasquini}}]{Patat11}
{Patat} F, {Chugai} NN, {Podsiadlowski} P, {Mason} E, {Melo} C, {Pasquini} L.
  2011.
\newblock \textit{\aap} 530:A63

\bibitem[{{Patat} et~al.(2013){Patat}, {Cordiner}, {Cox}, {Anderson},
  {Harutyunyan} et~al.}]{Patat13}
{Patat} F, {Cordiner} MA, {Cox} NLJ, {Anderson} RI, {Harutyunyan} A, et~al.
  2013.
\newblock \textit{\aap} 549:A62

\bibitem[{{Patat} et~al.(2012){Patat}, {H{\"o}flich}, {Baade}, {Maund}, {Wang}
  \& {Wheeler}}]{Patat12}
{Patat} F, {H{\"o}flich} P, {Baade} D, {Maund} JR, {Wang} L, {Wheeler} JC.
  2012.
\newblock \textit{\aap} 545:A7

\bibitem[{{Patnaude} et~al.(2012){Patnaude}, {Badenes}, {Park} \&
  {Laming}}]{Patnaude12}
{Patnaude} DJ, {Badenes} C, {Park} S, {Laming} JM. 2012.
\newblock \textit{\apj} 756:6

\bibitem[{{Patterson} et~al.(2013){Patterson}, {Oksanen}, {Monard}, {Rea},
  {Hambsch} et~al.}]{Patterson13}
{Patterson} J, {Oksanen} A, {Monard} B, {Rea} R, {Hambsch} FJ, et~al. 2013.
\newblock \textit{ArXiv e-prints}

\bibitem[{{Perlmutter} et~al.(1999){Perlmutter}, {Aldering}, {Goldhaber},
  {Knop}, {Nugent} et~al.}]{Perlmutter99}
{Perlmutter} S, {Aldering} G, {Goldhaber} G, {Knop} RA, {Nugent} P, et~al.
  1999.
\newblock \textit{\apj} 517:565--586

\bibitem[{{Perrett} et~al.(2012){Perrett}, {Sullivan}, {Conley},
  {Gonz{\'a}lez-Gait{\'a}n}, {Carlberg} et~al.}]{Perrett12}
{Perrett} K, {Sullivan} M, {Conley} A, {Gonz{\'a}lez-Gait{\'a}n} S, {Carlberg}
  R, et~al. 2012.
\newblock \textit{\aj} 144:59

\bibitem[{{Phillips}(1993)}]{Phillips93}
{Phillips} MM. 1993.
\newblock \textit{\apjl} 413:L105--L108

\bibitem[{{Phillips} et~al.(2013){Phillips}, {Simon}, {Morrell}, {Burns}, {Cox}
  et~al.}]{Phillips13}
{Phillips} MM, {Simon} JD, {Morrell} N, {Burns} CR, {Cox} NLJ, et~al. 2013.
\newblock \textit{\apj} 779:38

\bibitem[{{Piersanti} et~al.(2000){Piersanti}, {Cassisi}, {Iben} \&
  {Tornamb{\'e}}}]{Piersanti00}
{Piersanti} L, {Cassisi} S, {Iben} Jr. I, {Tornamb{\'e}} A. 2000.
\newblock \textit{\apj} 535:932--942

\bibitem[{{Piersanti} et~al.(2003){Piersanti}, {Gagliardi}, {Iben} \&
  {Tornamb{\'e}}}]{Piersanti03}
{Piersanti} L, {Gagliardi} S, {Iben} Jr. I, {Tornamb{\'e}} A. 2003.
\newblock \textit{\apj} 583:885--901

\bibitem[{{Pinto} \& {Eastman}(2000)}]{Pinto00}
{Pinto} PA, {Eastman} RG. 2000.
\newblock \textit{\apj} 530:757--776

\bibitem[{{Piro}(2008)}]{Piro08b}
{Piro} AL. 2008.
\newblock \textit{\apj} 679:616--625

\bibitem[{{Piro} \& {Bildsten}(2008)}]{Piro08a}
{Piro} AL, {Bildsten} L. 2008.
\newblock \textit{\apj} 673:1009--1013

\bibitem[{{Piro}, {Chang} \& {Weinberg}(2010)}]{Piro10}
{Piro} AL, {Chang} P, {Weinberg} NN. 2010.
\newblock \textit{\apj} 708:598--604

\bibitem[{{Piro} \& {Kulkarni}(2013)}]{Piro13e}
{Piro} AL, {Kulkarni} SR. 2013.
\newblock \textit{\apjl} 762:L17

\bibitem[{{Piro} \& {Nakar}(2012)}]{Piro13b}
{Piro} AL, {Nakar} E. 2012.
\newblock \textit{ArXiv e-prints}

\bibitem[{{Piro} \& {Nakar}(2013)}]{Piro13a}
{Piro} AL, {Nakar} E. 2013.
\newblock \textit{\apj} 769:67

\bibitem[{{Piro}, {Thompson} \& {Kochanek}(2013)}]{Piro13d}
{Piro} AL, {Thompson} TA, {Kochanek} CS. 2013.
\newblock \textit{ArXiv e-prints}

\bibitem[{{Pols} \& {Marinus}(1994)}]{Pols94}
{Pols} OR, {Marinus} M. 1994.
\newblock \textit{\aap} 288:475--501

\bibitem[{{Poznanski} et~al.(2007){Poznanski}, {Maoz}, {Yasuda}, {Foley}, {Doi}
  et~al.}]{Poznanski07b}
{Poznanski} D, {Maoz} D, {Yasuda} N, {Foley} RJ, {Doi} M, et~al. 2007.
\newblock \textit{\mnras} 382:1169--1186

\bibitem[{{Prieto} et~al.(2007){Prieto}, {Garnavich}, {Phillips}, {DePoy},
  {Parrent} et~al.}]{Prieto07}
{Prieto} JL, {Garnavich} PM, {Phillips} MM, {DePoy} DL, {Parrent} J, et~al.
  2007.
\newblock \textit{ArXiv e-prints (0706.4088)}

\bibitem[{{Pritchet}, {Howell} \& {Sullivan}(2008)}]{Pritchet08}
{Pritchet} CJ, {Howell} DA, {Sullivan} M. 2008.
\newblock \textit{\apjl} 683:L25--L28

\bibitem[{{Prodan}, {Murray} \& {Thompson}(2013)}]{Prodan13}
{Prodan} S, {Murray} N, {Thompson} TA. 2013.
\newblock \textit{ArXiv e-prints}

\bibitem[{{Quinn} et~al.(2006){Quinn}, {Garnavich}, {Li}, {Panagia}, {Riess}
  et~al.}]{Quinn06}
{Quinn} JL, {Garnavich} PM, {Li} W, {Panagia} N, {Riess} A, et~al. 2006.
\newblock \textit{\apj} 652:512--517

\bibitem[{{Rabinak}, {Livne} \& {Waxman}(2012)}]{Rabinak12}
{Rabinak} I, {Livne} E, {Waxman} E. 2012.
\newblock \textit{\apj} 757:35

\bibitem[{{Rabinak} \& {Waxman}(2011)}]{Rabinak11}
{Rabinak} I, {Waxman} E. 2011.
\newblock \textit{\apj} 728:63

\bibitem[{{Rajoelimanana} et~al.(2013){Rajoelimanana}, {Charles}, {Meintjes},
  {Odendaal} \& {Udalski}}]{Rajoelimanana13}
{Rajoelimanana} AF, {Charles} PA, {Meintjes} PJ, {Odendaal} A, {Udalski} A.
  2013.
\newblock \textit{\mnras} 432:2886--2894

\bibitem[{{Rappaport} et~al.(1994){Rappaport}, {Chiang}, {Kallman} \&
  {Malina}}]{Rappaport94}
{Rappaport} S, {Chiang} E, {Kallman} T, {Malina} R. 1994.
\newblock \textit{\apj} 431:237--246

\bibitem[{{Raskin} \& {Kasen}(2013)}]{Raskin13a}
{Raskin} C, {Kasen} D. 2013.
\newblock \textit{\apj} 772:1

\bibitem[{{Raskin} et~al.(2013){Raskin}, {Kasen}, {Moll}, {Schwab} \&
  {Woosley}}]{Raskin13b}
{Raskin} C, {Kasen} D, {Moll} R, {Schwab} J, {Woosley} S. 2013.
\newblock \textit{ArXiv e-prints}

\bibitem[{{Raskin} et~al.(2012){Raskin}, {Scannapieco}, {Fryer}, {Rockefeller}
  \& {Timmes}}]{Raskin12}
{Raskin} C, {Scannapieco} E, {Fryer} C, {Rockefeller} G, {Timmes} FX. 2012.
\newblock \textit{\apj} 746:62

\bibitem[{{Raskin} et~al.(2009){Raskin}, {Scannapieco}, {Rhoads} \& {Della
  Valle}}]{Raskin09}
{Raskin} C, {Scannapieco} E, {Rhoads} J, {Della Valle} M. 2009.
\newblock \textit{\apj} 707:74--78

\bibitem[{{Raskin} et~al.(2010){Raskin}, {Scannapieco}, {Rockefeller}, {Fryer},
  {Diehl} \& {Timmes}}]{Raskin10}
{Raskin} C, {Scannapieco} E, {Rockefeller} G, {Fryer} C, {Diehl} S, {Timmes}
  FX. 2010.
\newblock \textit{\apj} 724:111--125

\bibitem[{{Raymond} et~al.(2007){Raymond}, {Korreck}, {Sedlacek}, {Blair},
  {Ghavamian} \& {Sankrit}}]{Raymond07}
{Raymond} JC, {Korreck} KE, {Sedlacek} QC, {Blair} WP, {Ghavamian} P, {Sankrit}
  R. 2007.
\newblock \textit{\apj} 659:1257--1264

\bibitem[{{Remillard}, {Rappaport} \& {Macri}(1995)}]{Remillard95}
{Remillard} RA, {Rappaport} S, {Macri} LM. 1995.
\newblock \textit{\apj} 439:646--651

\bibitem[{{Rest} et~al.(2008{\natexlab{a}}){Rest}, {Matheson}, {Blondin},
  {Bergmann}, {Welch} et~al.}]{Rest08a}
{Rest} A, {Matheson} T, {Blondin} S, {Bergmann} M, {Welch} DL, et~al.
  2008{\natexlab{a}}.
\newblock \textit{\apj} 680:1137--1148

\bibitem[{{Rest}, {Sinnott} \& {Welch}(2012)}]{Rest12}
{Rest} A, {Sinnott} B, {Welch} DL. 2012.
\newblock \textit{\pasa} 29:466--481

\bibitem[{{Rest} et~al.(2005){Rest}, {Suntzeff}, {Olsen}, {Prieto}, {Smith}
  et~al.}]{Rest05}
{Rest} A, {Suntzeff} NB, {Olsen} K, {Prieto} JL, {Smith} RC, et~al. 2005.
\newblock \textit{\nat} 438:1132--1134

\bibitem[{{Rest} et~al.(2008{\natexlab{b}}){Rest}, {Welch}, {Suntzeff},
  {Oaster}, {Lanning} et~al.}]{Rest08b}
{Rest} A, {Welch} DL, {Suntzeff} NB, {Oaster} L, {Lanning} H, et~al.
  2008{\natexlab{b}}.
\newblock \textit{\apjl} 681:L81--L84

\bibitem[{{Rettura} et~al.(2011){Rettura}, {Mei}, {Stanford}, {Raichoor},
  {Moran} et~al.}]{Rettura11}
{Rettura} A, {Mei} S, {Stanford} SA, {Raichoor} A, {Moran} S, et~al. 2011.
\newblock \textit{\apj} 732:94

\bibitem[{{Reynolds} et~al.(2007){Reynolds}, {Borkowski}, {Hwang}, {Hughes},
  {Badenes} et~al.}]{Reynolds07}
{Reynolds} SP, {Borkowski} KJ, {Hwang} U, {Hughes} JP, {Badenes} C, et~al.
  2007.
\newblock \textit{\apjl} 668:L135--L138

\bibitem[{{Riess} et~al.(1998){Riess}, {Filippenko}, {Challis}, {Clocchiatti},
  {Diercks} et~al.}]{Riess98}
{Riess} AG, {Filippenko} AV, {Challis} P, {Clocchiatti} A, {Diercks} A, et~al.
  1998.
\newblock \textit{\aj} 116:1009--1038

\bibitem[{{Rigault} et~al.(2013){Rigault}, {Copin}, {Aldering}, {Antilogus},
  {Aragon} et~al.}]{Rigault13}
{Rigault} M, {Copin} Y, {Aldering} G, {Antilogus} P, {Aragon} C, et~al. 2013.
\newblock \textit{\aap} 560:A66

\bibitem[{{Rodney} et al.(2013)}]{Rodney13}
{Rodney} S, et al. 2013, 
\newblock \textit{\apj}, submitted

\bibitem[{{Rodr{\'{\i}}guez-Gil} et~al.(2010){Rodr{\'{\i}}guez-Gil},
  {Santander-Garc{\'{\i}}a}, {Knigge}, {Corradi}, {G{\"a}nsicke}
  et~al.}]{Rodriguez-Gil10}
{Rodr{\'{\i}}guez-Gil} P, {Santander-Garc{\'{\i}}a} M, {Knigge} C, {Corradi}
  RLM, {G{\"a}nsicke} BT, et~al. 2010.
\newblock \textit{\mnras} 407:L21--L25

\bibitem[{{Roelofs} et~al.(2008){Roelofs}, {Bassa}, {Voss} \&
  {Nelemans}}]{Roelofs08}
{Roelofs} G, {Bassa} C, {Voss} R, {Nelemans} G. 2008.
\newblock \textit{\mnras} 391:290--296

\bibitem[{{R{\"o}pke} et~al.(2012){R{\"o}pke}, {Kromer}, {Seitenzahl},
  {Pakmor}, {Sim} et~al.}]{Ropke12}
{R{\"o}pke} FK, {Kromer} M, {Seitenzahl} IR, {Pakmor} R, {Sim} SA, et~al. 2012.
\newblock \textit{\apjl} 750:L19

\bibitem[{{Rosswog} et~al.(2009){Rosswog}, {Kasen}, {Guillochon} \&
  {Ramirez-Ruiz}}]{Rosswog09}
{Rosswog} S, {Kasen} D, {Guillochon} J, {Ramirez-Ruiz} E. 2009.
\newblock \textit{\apjl} 705:L128--L132

\bibitem[{{Ruiter}, {Belczynski} \& {Fryer}(2009)}]{Ruiter09}
{Ruiter} AJ, {Belczynski} K, {Fryer} C. 2009.
\newblock \textit{\apj} 699:2026--2036

\bibitem[{{Ruiter} et~al.(2013){Ruiter}, {Sim}, {Pakmor}, {Kromer},
  {Seitenzahl} et~al.}]{Ruiter13}
{Ruiter} AJ, {Sim} SA, {Pakmor} R, {Kromer} M, {Seitenzahl} IR, et~al. 2013.
\newblock \textit{\mnras} 429:1425--1436

\bibitem[{{Ruiz-Lapuente}(1997)}]{Ruiz-Lapuente97}
{Ruiz-Lapuente} P. 1997.
\newblock \textit{Science} 276:1813--1814

\bibitem[{{Ruiz-Lapuente} \& {Canal}(1998)}]{Ruiz-Lapuente98}
{Ruiz-Lapuente} P, {Canal} R. 1998.
\newblock \textit{\apjl} 497:L57

\bibitem[{{Ruiz-Lapuente} et~al.(2004){Ruiz-Lapuente}, {Comeron}, {M{\'e}ndez},
  {Canal}, {Smartt} et~al.}]{Ruiz-Lapuente04}
{Ruiz-Lapuente} P, {Comeron} F, {M{\'e}ndez} J, {Canal} R, {Smartt} SJ, et~al.
  2004.
\newblock \textit{\nat} 431:1069--1072

\bibitem[{{Rupen}, {Mioduszewski} \& {Sokoloski}(2008)}]{Rupen08}
{Rupen} MP, {Mioduszewski} AJ, {Sokoloski} JL. 2008.
\newblock \textit{\apj} 688:559--567

\bibitem[{{Russell} \& {Immler}(2012)}]{Russell12}
{Russell} BR, {Immler} S. 2012.
\newblock \textit{\apjl} 748:L29

\bibitem[{{Sadat} et~al.(1998){Sadat}, {Blanchard}, {Guiderdoni} \&
  {Silk}}]{Sadat98}
{Sadat} R, {Blanchard} A, {Guiderdoni} B, {Silk} J. 1998.
\newblock \textit{\aap} 331:L69--L72

\bibitem[{{Sahman} et~al.(2013){Sahman}, {Dhillon}, {Marsh}, {Moll},
  {Thoroughgood} et~al.}]{Sahman13}
{Sahman} DI, {Dhillon} VS, {Marsh} TR, {Moll} S, {Thoroughgood} TD, et~al.
  2013.
\newblock \textit{\mnras} 433:1588--1598

\bibitem[{{Saio} \& {Nomoto}(1998)}]{Saio98}
{Saio} H, {Nomoto} K. 1998.
\newblock \textit{\apj} 500:388

\bibitem[{{Saio} \& {Nomoto}(2004)}]{Saio04}
{Saio} H, {Nomoto} K. 2004.
\newblock \textit{\apj} 615:444--449

\bibitem[{{Sako} et~al.(2008){Sako}, {Bassett}, {Becker}, {Cinabro}, {DeJongh}
  et~al.}]{Sako08}
{Sako} M, {Bassett} B, {Becker} A, {Cinabro} D, {DeJongh} F, et~al. 2008.
\newblock \textit{\aj} 135:348--373

\bibitem[{{Sand} et~al.(2012){Sand}, {Graham}, {Bildfell}, {Zaritsky},
  {Pritchet} et~al.}]{Sand12}
{Sand} DJ, {Graham} ML, {Bildfell} C, {Zaritsky} D, {Pritchet} C, et~al. 2012.
\newblock \textit{\apj} 746:163

\bibitem[{{Sand} et~al.(2008){Sand}, {Zaritsky}, {Herbert-Fort}, {Sivanandam}
  \& {Clowe}}]{Sand08}
{Sand} DJ, {Zaritsky} D, {Herbert-Fort} S, {Sivanandam} S, {Clowe} D. 2008.
\newblock \textit{\aj} 135:1917--1933

\bibitem[{{Sarazin} et~al.(2003){Sarazin}, {Kundu}, {Irwin}, {Sivakoff},
  {Blanton} \& {Randall}}]{Sarazin03}
{Sarazin} CL, {Kundu} A, {Irwin} JA, {Sivakoff} GR, {Blanton} EL, {Randall} SW.
  2003.
\newblock \textit{\apj} 595:743--759

\bibitem[{{Scalzo} et~al.(2012){Scalzo}, {Aldering}, {Antilogus}, {Aragon},
  {Bailey} et~al.}]{Scalzo12}
{Scalzo} R, {Aldering} G, {Antilogus} P, {Aragon} C, {Bailey} S, et~al. 2012.
\newblock \textit{\apj} 757:12

\bibitem[{{Scalzo} et~al.(2010){Scalzo}, {Aldering}, {Antilogus}, {Aragon},
  {Bailey} et~al.}]{Scalzo10}
{Scalzo} RA, {Aldering} G, {Antilogus} P, {Aragon} C, {Bailey} S, et~al. 2010.
\newblock \textit{\apj} 713:1073--1094

\bibitem[{{Scannapieco} \& {Bildsten}(2005)}]{Scannapieco05a}
{Scannapieco} E, {Bildsten} L. 2005.
\newblock \textit{\apjl} 629:L85--L88

\bibitem[{{Schaefer}(2010)}]{Schaefer10a}
{Schaefer} BE. 2010.
\newblock \textit{\apjs} 187:275--373

\bibitem[{{Schaefer}(2013)}]{Schaefer13a}
{Schaefer} BE. 2013.
\newblock In \textit{American Astronomical Society Meeting Abstracts}, vol. 221
  of \textit{American Astronomical Society Meeting Abstracts}

\bibitem[{{Schaefer} \& {Pagnotta}(2012)}]{Schaefer12}
{Schaefer} BE, {Pagnotta} A. 2012.
\newblock \textit{\nat} 481:164--166

\bibitem[{{Schaefer}, {Pagnotta} \& {Shara}(2010)}]{Schaefer10c}
{Schaefer} BE, {Pagnotta} A, {Shara} MM. 2010.
\newblock \textit{\apj} 708:381--402

\bibitem[{{Schaefer} et~al.(2010){Schaefer}, {Pagnotta}, {Xiao}, {Darnley},
  {Bode} et~al.}]{Schaefer10b}
{Schaefer} BE, {Pagnotta} A, {Xiao} L, {Darnley} MJ, {Bode} MF, et~al. 2010.
\newblock \textit{\aj} 140:925--932

\bibitem[{{Schawinski}(2009)}]{Schawinski09c}
{Schawinski} K. 2009.
\newblock \textit{\mnras} 397:717--725

\bibitem[{{Schmidt} et~al.(1994){Schmidt}, {Kirshner}, {Leibundgut}, {Wells},
  {Porter} et~al.}]{Schmidt94}
{Schmidt} BP, {Kirshner} RP, {Leibundgut} B, {Wells} LA, {Porter} AC, et~al.
  1994.
\newblock \textit{\apjl} 434:L19--L23

\bibitem[{{Schwab} et~al.(2012){Schwab}, {Shen}, {Quataert}, {Dan} \&
  {Rosswog}}]{Schwab12}
{Schwab} J, {Shen} KJ, {Quataert} E, {Dan} M, {Rosswog} S. 2012.
\newblock \textit{\mnras} 427:190--203

\bibitem[{{Scolnic} et~al.(2014){Scolnic}, {Riess}, {Foley}, {Rest}, {Rodney}
  et~al.}]{Scolnic14}
{Scolnic} DM, {Riess} AG, {Foley} RJ, {Rest} A, {Rodney} SA, et~al. 2014.
\newblock \textit{\apj} 780:37

\bibitem[{{Seitenzahl} et~al.(2013{\natexlab{a}}){Seitenzahl}, {Cescutti},
  {R{\"o}pke}, {Ruiter} \& {Pakmor}}]{Seitenzahl13b}
{Seitenzahl} IR, {Cescutti} G, {R{\"o}pke} FK, {Ruiter} AJ, {Pakmor} R.
  2013{\natexlab{a}}.
\newblock \textit{\aap} 559:L5

\bibitem[{{Seitenzahl} et~al.(2013{\natexlab{b}}){Seitenzahl},
  {Ciaraldi-Schoolmann}, {R{\"o}pke}, {Fink}, {Hillebrandt}
  et~al.}]{Seitenzahl13a}
{Seitenzahl} IR, {Ciaraldi-Schoolmann} F, {R{\"o}pke} FK, {Fink} M,
  {Hillebrandt} W, et~al. 2013{\natexlab{b}}.
\newblock \textit{\mnras} 429:1156--1172

\bibitem[{{Selvelli} et~al.(2008){Selvelli}, {Cassatella}, {Gilmozzi} \&
  {Gonz{\'a}lez-Riestra}}]{Selvelli08}
{Selvelli} P, {Cassatella} A, {Gilmozzi} R, {Gonz{\'a}lez-Riestra} R. 2008.
\newblock \textit{\aap} 492:787--803

\bibitem[{{Shappee}, {Kochanek} \& {Stanek}(2013)}]{Shappee13b}
{Shappee} BJ, {Kochanek} CS, {Stanek} KZ. 2013.
\newblock \textit{\apj} 765:150

\bibitem[{{Shappee} et~al.(2013){Shappee}, {Stanek}, {Pogge} \&
  {Garnavich}}]{Shappee13a}
{Shappee} BJ, {Stanek} KZ, {Pogge} RW, {Garnavich} PM. 2013.
\newblock \textit{\apjl} 762:L5

\bibitem[{{Shara} et~al.(2013){Shara}, {Bibby}, {Zurek}, {Crowther}, {Moffat}
  \& {Drissen}}]{Shara13}
{Shara} MM, {Bibby} JL, {Zurek} D, {Crowther} PA, {Moffat} AFJ, {Drissen} L.
  2013.
\newblock \textit{\aj} 146:162

\bibitem[{{Shara} \& {Hurley}(2002)}]{Shara02}
{Shara} MM, {Hurley} JR. 2002.
\newblock \textit{\apj} 571:830--842

\bibitem[{{Sharon} et~al.(2007){Sharon}, {Gal-Yam}, {Maoz}, {Filippenko} \&
  {Guhathakurta}}]{Sharon07}
{Sharon} K, {Gal-Yam} A, {Maoz} D, {Filippenko} AV, {Guhathakurta} P. 2007.
\newblock \textit{\apj} 660:1165--1175

\bibitem[{{Shen} \& {Bildsten}(2007)}]{Shen07}
{Shen} KJ, {Bildsten} L. 2007.
\newblock \textit{\apj} 660:1444--1450

\bibitem[{{Shen} \& {Bildsten}(2009)}]{Shen09}
{Shen} KJ, {Bildsten} L. 2009.
\newblock \textit{\apj} 699:1365--1373

\bibitem[{{Shen} \& {Bildsten}(2013)}]{Shen13a}
{Shen} KJ, {Bildsten} L. 2013.
\newblock \textit{ArXiv e-prints}

\bibitem[{{Shen} et~al.(2012){Shen}, {Bildsten}, {Kasen} \&
  {Quataert}}]{Shen12}
{Shen} KJ, {Bildsten} L, {Kasen} D, {Quataert} E. 2012.
\newblock \textit{\apj} 748:35

\bibitem[{{Shen}, {Guillochon} \& {Foley}(2013)}]{Shen13b}
{Shen} KJ, {Guillochon} J, {Foley} RJ. 2013.
\newblock \textit{\apjl} 770:L35

\bibitem[{{Shore} et~al.(2011){Shore}, {Augusteijn}, {Ederoclite} \&
  {Uthas}}]{Shore11b}
{Shore} SN, {Augusteijn} T, {Ederoclite} A, {Uthas} H. 2011.
\newblock \textit{\aap} 533:L8

\bibitem[{{Silverman} \& {Filippenko}(2012)}]{Silverman12}
{Silverman} JM, {Filippenko} AV. 2012.
\newblock \textit{\mnras} 425:1917--1933

\bibitem[{{Silverman}, {Ganeshalingam} \& {Filippenko}(2013)}]{Silverman13c}
{Silverman} JM, {Ganeshalingam} M, {Filippenko} AV. 2013.
\newblock \textit{\mnras} 430:1030--1041

\bibitem[{{Silverman} et~al.(2011){Silverman}, {Ganeshalingam}, {Li},
  {Filippenko}, {Miller} \& {Poznanski}}]{Silverman11}
{Silverman} JM, {Ganeshalingam} M, {Li} W, {Filippenko} AV, {Miller} AA,
  {Poznanski} D. 2011.
\newblock \textit{\mnras} 410:585--611

\bibitem[{{Silverman} et~al.(2013{\natexlab{a}}){Silverman}, {Nugent},
  {Gal-Yam}, {Sullivan}, {Howell} et~al.}]{Silverman13a}
{Silverman} JM, {Nugent} PE, {Gal-Yam} A, {Sullivan} M, {Howell} DA, et~al.
  2013{\natexlab{a}}.
\newblock \textit{\apj} 772:125

\bibitem[{{Silverman} et~al.(2013{\natexlab{b}}){Silverman}, {Nugent},
  {Gal-Yam}, {Sullivan}, {Howell} et~al.}]{Silverman13b}
{Silverman} JM, {Nugent} PE, {Gal-Yam} A, {Sullivan} M, {Howell} DA, et~al.
  2013{\natexlab{b}}.
\newblock \textit{\apjs} 207:3

\bibitem[{{Simon} et~al.(2009){Simon}, {Gal-Yam}, {Gnat}, {Quimby},
  {Ganeshalingam} et~al.}]{Simon09}
{Simon} JD, {Gal-Yam} A, {Gnat} O, {Quimby} RM, {Ganeshalingam} M, et~al. 2009.
\newblock \textit{\apj} 702:1157--1170

\bibitem[{{Smartt}(2009)}]{Smartt09}
{Smartt} SJ. 2009.
\newblock \textit{\araa} 47:63--106

\bibitem[{{Smith} et~al.(2012){Smith}, {Nichol}, {Dilday}, {Marriner},
  {Kessler} et~al.}]{Smith12}
{Smith} M, {Nichol} RC, {Dilday} B, {Marriner} J, {Kessler} R, et~al. 2012.
\newblock \textit{\apj} 755:61

\bibitem[{{Smith} et~al.(2011){Smith}, {Williams}, {Smith}, {Milne}, {Jannuzi}
  \& {Green}}]{Smith11c}
{Smith} PS, {Williams} GG, {Smith} N, {Milne} PA, {Jannuzi} BT, {Green} EM.
  2011.
\newblock \textit{ArXiv e-prints (1111.6626)}

\bibitem[{{Soderberg} et~al.(2012){Soderberg}, {Margutti}, {Zauderer},
  {Krauss}, {Katz} et~al.}]{Soderberg12}
{Soderberg} AM, {Margutti} R, {Zauderer} BA, {Krauss} M, {Katz} B, et~al. 2012.
\newblock \textit{\apj} 752:78

\bibitem[{{Soker}(2013)}]{Soker13a}
{Soker} N. 2013.
\newblock In \textit{IAU Symposium}, eds. R~{Di Stefano}, M~{Orio}, M~{Moe},
  vol. 281 of \textit{IAU Symposium}

\bibitem[{{Soker}, {Garcia-Berro} \& {Althaus}(2013)}]{Soker13c}
{Soker} N, {Garcia-Berro} E, {Althaus} LG. 2013.
\newblock \textit{ArXiv e-prints}

\bibitem[{{Soker} et~al.(2013){Soker}, {Kashi}, {Garc{\'{\i}}a-Berro}, {Torres}
  \& {Camacho}}]{Soker13b}
{Soker} N, {Kashi} A, {Garc{\'{\i}}a-Berro} E, {Torres} S, {Camacho} J. 2013.
\newblock \textit{\mnras} 431:1541--1546

\bibitem[{{Sokoloski} et~al.(2006){Sokoloski}, {Luna}, {Mukai} \&
  {Kenyon}}]{Sokoloski06}
{Sokoloski} JL, {Luna} GJM, {Mukai} K, {Kenyon} SJ. 2006.
\newblock \textit{\nat} 442:276--278

\bibitem[{{Solheim}(2010)}]{Solheim10}
{Solheim} JE. 2010.
\newblock \textit{\pasp} 122:1133--1163

\bibitem[{{Sorokina} et~al.(2004){Sorokina}, {Blinnikov}, {Kosenko} \&
  {Lundqvist}}]{Sorokina04}
{Sorokina} EI, {Blinnikov} SI, {Kosenko} DI, {Lundqvist} P. 2004.
\newblock \textit{Astronomy Letters} 30:737--750

\bibitem[{{Sparks} et~al.(1999){Sparks}, {Macchetto}, {Panagia}, {Boffi},
  {Branch} et~al.}]{Sparks99}
{Sparks} WB, {Macchetto} F, {Panagia} N, {Boffi} FR, {Branch} D, et~al. 1999.
\newblock \textit{\apj} 523:585--592

\bibitem[{{Spruit}(1998)}]{Spruit98}
{Spruit} HC. 1998.
\newblock \textit{\aap} 333:603--612

\bibitem[{{Starrfield}, {Sparks} \& {Truran}(1985)}]{Starrfield85}
{Starrfield} S, {Sparks} WM, {Truran} JW. 1985.
\newblock \textit{\apj} 291:136--146

\bibitem[{{Starrfield} et~al.(1972){Starrfield}, {Truran}, {Sparks} \&
  {Kutter}}]{Starrfield72}
{Starrfield} S, {Truran} JW, {Sparks} WM, {Kutter} GS. 1972.
\newblock \textit{\apj} 176:169

\bibitem[{{Stehle} et~al.(2005){Stehle}, {Mazzali}, {Benetti} \&
  {Hillebrandt}}]{Stehle05}
{Stehle} M, {Mazzali} PA, {Benetti} S, {Hillebrandt} W. 2005.
\newblock \textit{\mnras} 360:1231--1243

\bibitem[{{Steiner} \& {Diaz}(1998)}]{Steiner98}
{Steiner} JE, {Diaz} MP. 1998.
\newblock \textit{\pasp} 110:276--282

\bibitem[{{Stephenson} \& {Green}(2002)}]{Stephenson02}
{Stephenson} FR, {Green} DA. 2002.
\newblock \textit{Historical supernovae and their remnants, by F.~Richard
  Stephenson and David A.~Green.~International series in astronomy and
  astrophysics, vol.~5.~Oxford: Clarendon Press, 2002, ISBN 0198507666} 5

\bibitem[{{Sternberg} et~al.(2011){Sternberg}, {Gal-Yam}, {Simon}, {Leonard},
  {Quimby} et~al.}]{Sternberg11}
{Sternberg} A, {Gal-Yam} A, {Simon} D, {Leonard} DC, {Quimby} RM, et~al. 2011.
\newblock \textit{Science} 333:856--859

\bibitem[{{Sternberg} et~al.(2013){Sternberg}, {Gal Yam}, {Simon}, {Patat},
  {Hillebrandt} et~al.}]{Sternberg13}
{Sternberg} A, {Gal Yam} A, {Simon} JD, {Patat} F, {Hillebrandt} W, et~al.
  2013.
\newblock \textit{ArXiv e-prints}

\bibitem[{{Strolger}, {Dahlen} \& {Riess}(2010)}]{Strolger10}
{Strolger} LG, {Dahlen} T, {Riess} AG. 2010.
\newblock \textit{\apj} 713:32--40

\bibitem[{{Strolger} et~al.(2004){Strolger}, {Riess}, {Dahlen}, {Livio},
  {Panagia} et~al.}]{Strolger04}
{Strolger} LG, {Riess} AG, {Dahlen} T, {Livio} M, {Panagia} N, et~al. 2004.
\newblock \textit{\apj} 613:200--223

\bibitem[{{Suh} et~al.(2011){Suh}, {Yoon}, {Jeong} \& {Yi}}]{Suh11}
{Suh} H, {Yoon} Sc, {Jeong} H, {Yi} SK. 2011.
\newblock \textit{\apj} 730:110

\bibitem[{{Suijs} et~al.(2008){Suijs}, {Langer}, {Poelarends}, {Yoon}, {Heger}
  \& {Herwig}}]{Suijs08}
{Suijs} MPL, {Langer} N, {Poelarends} AJ, {Yoon} SC, {Heger} A, {Herwig} F.
  2008.
\newblock \textit{\aap} 481:L87--L90

\bibitem[{{Sullivan} et~al.(2006){Sullivan}, {Le Borgne}, {Pritchet},
  {Hodsman}, {Neill} et~al.}]{Sullivan06}
{Sullivan} M, {Le Borgne} D, {Pritchet} CJ, {Hodsman} A, {Neill} JD, et~al.
  2006.
\newblock \textit{\apj} 648:868--883

\bibitem[{{Summa} et~al.(2013){Summa}, {Ulyanov}, {Kromer}, {Boyer},
  {R{\"o}pke} et~al.}]{Summa13}
{Summa} A, {Ulyanov} A, {Kromer} M, {Boyer} S, {R{\"o}pke} FK, et~al. 2013.
\newblock \textit{\aap} 554:A67

\bibitem[{{Taam}(1980)}]{Taam80}
{Taam} RE. 1980.
\newblock \textit{\apj} 237:142--147

\bibitem[{{Taddia} et~al.(2012){Taddia}, {Stritzinger}, {Phillips}, {Burns},
  {Heinrich-Josties} et~al.}]{Taddia12}
{Taddia} F, {Stritzinger} MD, {Phillips} MM, {Burns} CR, {Heinrich-Josties} E,
  et~al. 2012.
\newblock \textit{\aap} 545:L7

\bibitem[{{Tanaka} et~al.(2010){Tanaka}, {Kawabata}, {Yamanaka}, {Maeda},
  {Hattori} et~al.}]{Tanaka10}
{Tanaka} M, {Kawabata} KS, {Yamanaka} M, {Maeda} K, {Hattori} T, et~al. 2010.
\newblock \textit{\apj} 714:1209--1216

\bibitem[{{Tanaka} et~al.(2011){Tanaka}, {Mazzali}, {Stanishev}, {Maurer},
  {Kerzendorf} \& {Nomoto}}]{Tanaka11}
{Tanaka} M, {Mazzali} PA, {Stanishev} V, {Maurer} I, {Kerzendorf} WE, {Nomoto}
  K. 2011.
\newblock \textit{\mnras} 410:1725--1738

\bibitem[{{Taubenberger} et~al.(2011){Taubenberger}, {Benetti}, {Childress},
  {Pakmor}, {Hachinger} et~al.}]{Taubenberger11}
{Taubenberger} S, {Benetti} S, {Childress} M, {Pakmor} R, {Hachinger} S, et~al.
  2011.
\newblock \textit{\mnras} 412:2735--2762

\bibitem[{{Taubenberger} et~al.(2013{\natexlab{a}}){Taubenberger}, {Kromer},
  {Hachinger}, {Mazzali}, {Benetti} et~al.}]{Taubenberger13a}
{Taubenberger} S, {Kromer} M, {Hachinger} S, {Mazzali} PA, {Benetti} S, et~al.
  2013{\natexlab{a}}.
\newblock \textit{\mnras} 432:3117--3130

\bibitem[{{Taubenberger} et~al.(2013{\natexlab{b}}){Taubenberger}, {Kromer},
  {Pakmor}, {Pignata}, {Maeda} et~al.}]{Taubenberger13b}
{Taubenberger} S, {Kromer} M, {Pakmor} R, {Pignata} G, {Maeda} K, et~al.
  2013{\natexlab{b}}.
\newblock \textit{\apjl} 775:L43

\bibitem[{{Thomas} et~al.(2011){Thomas}, {Aldering}, {Antilogus}, {Aragon},
  {Bailey} et~al.}]{Thomas11b}
{Thomas} RC, {Aldering} G, {Antilogus} P, {Aragon} C, {Bailey} S, et~al. 2011.
\newblock \textit{\apj} 743:27

\bibitem[{{Thompson}(2011)}]{Thompson11}
{Thompson} TA. 2011.
\newblock \textit{\apj} 741:82

\bibitem[{{Thomson} \& {Chary}(2011)}]{Thomson11}
{Thomson} MG, {Chary} RR. 2011.
\newblock \textit{\apj} 731:72

\bibitem[{{Thoroughgood} et~al.(2001){Thoroughgood}, {Dhillon}, {Littlefair},
  {Marsh} \& {Smith}}]{Thoroughgood01}
{Thoroughgood} TD, {Dhillon} VS, {Littlefair} SP, {Marsh} TR, {Smith} DA. 2001.
\newblock \textit{\mnras} 327:1323--1333

\bibitem[{{Timmes}, {Brown} \& {Truran}(2003)}]{Timmes03}
{Timmes} FX, {Brown} EF, {Truran} JW. 2003.
\newblock \textit{\apjl} 590:L83--L86

\bibitem[{{Tinsley}(1979)}]{Tinsley79b}
{Tinsley} BM. 1979.
\newblock \textit{\apj} 229:1046--1056

\bibitem[{{Tojeiro} et~al.(2009){Tojeiro}, {Wilkins}, {Heavens}, {Panter} \&
  {Jimenez}}]{Tojeiro09}
{Tojeiro} R, {Wilkins} S, {Heavens} AF, {Panter} B, {Jimenez} R. 2009.
\newblock \textit{\apjs} 185:1--19

\bibitem[{{Tominaga} et~al.(2011){Tominaga}, {Morokuma}, {Blinnikov},
  {Baklanov}, {Sorokina} \& {Nomoto}}]{Tominaga11}
{Tominaga} N, {Morokuma} T, {Blinnikov} SI, {Baklanov} P, {Sorokina} EI,
  {Nomoto} K. 2011.
\newblock \textit{\apjs} 193:20

\bibitem[{{Toonen}, {Nelemans} \& {Portegies Zwart}(2012)}]{Toonen12}
{Toonen} S, {Nelemans} G, {Portegies Zwart} S. 2012.
\newblock \textit{\aap} 546:A70

\bibitem[{{Tornamb{\'e}} \& {Piersanti}(2013)}]{Tornambe13}
{Tornamb{\'e}} A, {Piersanti} L. 2013.
\newblock \textit{\mnras} 431:1812--1822

\bibitem[{{Totani} et~al.(2008){Totani}, {Morokuma}, {Oda}, {Doi} \&
  {Yasuda}}]{Totani08}
{Totani} T, {Morokuma} T, {Oda} T, {Doi} M, {Yasuda} N. 2008.
\newblock \textit{\pasj} 60:1327--

\bibitem[{{Tout}(2005)}]{Tout05}
{Tout} CA. 2005.
\newblock In \textit{The Astrophysics of Cataclysmic Variables and Related
  Objects}, eds. JM~{Hameury}, JP~{Lasota}, vol. 330 of \textit{Astronomical
  Society of the Pacific Conference Series}

\bibitem[{{Tout} et~al.(2008){Tout}, {Wickramasinghe}, {Liebert}, {Ferrario} \&
  {Pringle}}]{Tout08}
{Tout} CA, {Wickramasinghe} DT, {Liebert} J, {Ferrario} L, {Pringle} JE. 2008.
\newblock \textit{\mnras} 387:897--901

\bibitem[{{Tovmassian} et~al.(2010){Tovmassian}, {Yungelson}, {Rauch},
  {Suleimanov}, {Napiwotzki} et~al.}]{Tovmassian10}
{Tovmassian} G, {Yungelson} L, {Rauch} T, {Suleimanov} V, {Napiwotzki} R,
  et~al. 2010.
\newblock \textit{\apj} 714:178--193

\bibitem[{{Tremonti} et~al.(2004){Tremonti}, {Heckman}, {Kauffmann},
  {Brinchmann}, {Charlot} et~al.}]{Tremonti04}
{Tremonti} CA, {Heckman} TM, {Kauffmann} G, {Brinchmann} J, {Charlot} S, et~al.
  2004.
\newblock \textit{\apj} 613:898--913

\bibitem[{{Trundle} et~al.(2008){Trundle}, {Kotak}, {Vink} \&
  {Meikle}}]{Trundle08}
{Trundle} C, {Kotak} R, {Vink} JS, {Meikle} WPS. 2008.
\newblock \textit{\aap} 483:L47--L50

\bibitem[{{Truran} \& {Livio}(1986)}]{Truran86}
{Truran} JW, {Livio} M. 1986.
\newblock \textit{\apj} 308:721--727

\bibitem[{{Tsujimoto} \& {Shigeyama}(2012)}]{Tsujimoto12}
{Tsujimoto} T, {Shigeyama} T. 2012.
\newblock \textit{\apjl} 760:L38

\bibitem[{{Tucker}(2011)}]{Tucker11}
{Tucker} BE. 2011.
\newblock \textit{\apss} 335:223--230

\bibitem[{{Tutukov} \& {Yungelson}(1996)}]{Tutukov96}
{Tutukov} A, {Yungelson} L. 1996.
\newblock \textit{\mnras} 280:1035--1045

\bibitem[{{Tutukov} \& {Yungelson}(1981)}]{Tutukov81}
{Tutukov} AV, {Yungelson} LR. 1981.
\newblock \textit{Nauchnye Informatsii} 49:3

\bibitem[{{Udalski}, {Kubiak} \& {Szymanski}(1997)}]{Udalski97}
{Udalski} A, {Kubiak} M, {Szymanski} M. 1997.
\newblock \textit{\actaa} 47:319--344

\bibitem[{{Umeda} et~al.(1999){Umeda}, {Nomoto}, {Kobayashi}, {Hachisu} \&
  {Kato}}]{Umeda99}
{Umeda} H, {Nomoto} K, {Kobayashi} C, {Hachisu} I, {Kato} M. 1999.
\newblock \textit{\apjl} 522:L43--L47

\bibitem[{{Uthas}, {Knigge} \& {Steeghs}(2010)}]{Uthas10}
{Uthas} H, {Knigge} C, {Steeghs} D. 2010.
\newblock \textit{\mnras} 409:237--246

\bibitem[{{{\v S}imon}(2003)}]{Simon03}
{{\v S}imon} V. 2003.
\newblock \textit{\aap} 406:613--621

\bibitem[{{Valenti} et~al.(2009){Valenti}, {Pastorello}, {Cappellaro},
  {Benetti}, {Mazzali} et~al.}]{Valenti09}
{Valenti} S, {Pastorello} A, {Cappellaro} E, {Benetti} S, {Mazzali} PA, et~al.
  2009.
\newblock \textit{\nat} 459:674--677

\bibitem[{{van den Heuvel} et~al.(1992){van den Heuvel}, {Bhattacharya},
  {Nomoto} \& {Rappaport}}]{van-den-Heuvel92}
{van den Heuvel} EPJ, {Bhattacharya} D, {Nomoto} K, {Rappaport} SA. 1992.
\newblock \textit{\aap} 262:97--105

\bibitem[{{van Kerkwijk}, {Chang} \& {Justham}(2010)}]{van-Kerkwijk10}
{van Kerkwijk} MH, {Chang} P, {Justham} S. 2010.
\newblock \textit{\apjl} 722:L157--L161

\bibitem[{{van Winckel} et~al.(2009){van Winckel}, {Lloyd Evans}, {Briquet},
  {De Cat}, {Degroote} et~al.}]{van-Winckel09}
{van Winckel} H, {Lloyd Evans} T, {Briquet} M, {De Cat} P, {Degroote} P, et~al.
  2009.
\newblock \textit{\aap} 505:1221--1232

\bibitem[{{Vink}(2008)}]{Vink08}
{Vink} J. 2008.
\newblock \textit{\apj} 689:231--241

\bibitem[{{Vink}(2012)}]{Vink12}
{Vink} J. 2012.
\newblock \textit{\aapr} 20:49

\bibitem[{{Vink} et~al.(2006){Vink}, {Bleeker}, {van der Heyden}, {Bykov},
  {Bamba} \& {Yamazaki}}]{Vink06}
{Vink} J, {Bleeker} J, {van der Heyden} K, {Bykov} A, {Bamba} A, {Yamazaki} R.
  2006.
\newblock \textit{\apjl} 648:L33--L37

\bibitem[{{Vink}, {Kaastra} \& {Bleeker}(1997)}]{Vink97}
{Vink} J, {Kaastra} JS, {Bleeker} JAM. 1997.
\newblock \textit{\aap} 328:628--633

\bibitem[{{Vink} et~al.(2003){Vink}, {Laming}, {Gu}, {Rasmussen} \&
  {Kaastra}}]{Vink03}
{Vink} J, {Laming} JM, {Gu} MF, {Rasmussen} A, {Kaastra} JS. 2003.
\newblock \textit{\apjl} 587:L31--L34

\bibitem[{{Voss} \& {Nelemans}(2008)}]{Voss08}
{Voss} R, {Nelemans} G. 2008.
\newblock \textit{\nat} 451:802--804

\bibitem[{{Voss} \& {Nelemans}(2012)}]{Voss12}
{Voss} R, {Nelemans} G. 2012.
\newblock \textit{\aap} 539:A77

\bibitem[{{Walker} et~al.(2012){Walker}, {Hachinger}, {Mazzali}, {Ellis},
  {Sullivan} et~al.}]{Walker12}
{Walker} ES, {Hachinger} S, {Mazzali} PA, {Ellis} RS, {Sullivan} M, et~al.
  2012.
\newblock \textit{\mnras} 427:103--113

\bibitem[{{Wang} \& {Han}(2012)}]{Wang12b}
{Wang} B, {Han} Z. 2012.
\newblock \textit{\nar} 56:122--141

\bibitem[{{Wang}, {Li} \& {Han}(2010)}]{Wang10a}
{Wang} B, {Li} XD, {Han} ZW. 2010.
\newblock \textit{\mnras} 401:2729--2738

\bibitem[{{Wang} et~al.(2009{\natexlab{a}}){Wang}, {Meng}, {Chen} \&
  {Han}}]{Wang09a}
{Wang} B, {Meng} X, {Chen} X, {Han} Z. 2009{\natexlab{a}}.
\newblock \textit{\mnras} 395:847--854

\bibitem[{{Wang} et~al.(2006){Wang}, {Baade}, {H{\"o}flich}, {Wheeler},
  {Kawabata} et~al.}]{Wang06b}
{Wang} L, {Baade} D, {H{\"o}flich} P, {Wheeler} JC, {Kawabata} K, et~al. 2006.
\newblock \textit{\apj} 653:490--502

\bibitem[{{Wang}, {Baade} \& {Patat}(2007)}]{Wang07}
{Wang} L, {Baade} D, {Patat} F. 2007.
\newblock \textit{Science} 315:212--

\bibitem[{{Wang} \& {Wheeler}(2008)}]{Wang08b}
{Wang} L, {Wheeler} JC. 2008.
\newblock \textit{\araa} 46:433--474

\bibitem[{{Wang} et~al.(2009{\natexlab{b}}){Wang}, {Filippenko},
  {Ganeshalingam}, {Li}, {Silverman} et~al.}]{Wang09i}
{Wang} X, {Filippenko} AV, {Ganeshalingam} M, {Li} W, {Silverman} JM, et~al.
  2009{\natexlab{b}}.
\newblock \textit{\apjl} 699:L139--L143

\bibitem[{{Wang} et~al.(2008){Wang}, {Li}, {Filippenko}, {Foley}, {Smith} \&
  {Wang}}]{Wang08c}
{Wang} X, {Li} W, {Filippenko} AV, {Foley} RJ, {Smith} N, {Wang} L. 2008.
\newblock \textit{\apj} 677:1060--1068

\bibitem[{{Wang} et~al.(2013){Wang}, {Wang}, {Filippenko}, {Zhang} \&
  {Zhao}}]{Wang13b}
{Wang} X, {Wang} L, {Filippenko} AV, {Zhang} T, {Zhao} X. 2013.
\newblock \textit{Science} 340:170--173

\bibitem[{{Warner}(2003)}]{Warner03}
{Warner} B. 2003.
\newblock \textit{{Cataclysmic Variable Stars}}.
\newblock Cambridge University Press

\bibitem[{{Washabaugh} \& {Bregman}(2013)}]{Washabaugh13}
{Washabaugh} PC, {Bregman} JN. 2013.
\newblock \textit{\apj} 762:1

\bibitem[{{Webbink}(1984)}]{Webbink84}
{Webbink} RF. 1984.
\newblock \textit{\apj} 277:355--360

\bibitem[{{Wheeler}(2012)}]{Wheeler12}
{Wheeler} JC. 2012.
\newblock \textit{\apj} 758:123

\bibitem[{{Wheeler} \& {Pooley}(2013)}]{Wheeler13}
{Wheeler} JC, {Pooley} D. 2013.
\newblock \textit{\apj} 762:75

\bibitem[{{Whelan} \& {Iben}(1973)}]{Whelan73}
{Whelan} J, {Iben} Jr. I. 1973.
\newblock \textit{\apj} 186:1007--1014

\bibitem[{{Wiersma}, {Schaye} \& {Theuns}(2011)}]{Wiersma11}
{Wiersma} RPC, {Schaye} J, {Theuns} T. 2011.
\newblock \textit{\mnras} 415:353--371

\bibitem[{{Williams} et~al.(2011){Williams}, {Blair}, {Blondin}, {Borkowski},
  {Ghavamian} et~al.}]{Williams11b}
{Williams} BJ, {Blair} WP, {Blondin} JM, {Borkowski} KJ, {Ghavamian} P, et~al.
  2011.
\newblock \textit{\apj} 741:96

\bibitem[{{Winkler} et~al.(2005){Winkler}, {Long}, {Hamilton} \&
  {Fesen}}]{Winkler05}
{Winkler} PF, {Long} KS, {Hamilton} AJS, {Fesen} RA. 2005.
\newblock \textit{\apj} 624:189--197

\bibitem[{{Wolf} et~al.(2013){Wolf}, {Bildsten}, {Brooks} \& {Paxton}}]{Wolf13}
{Wolf} WM, {Bildsten} L, {Brooks} J, {Paxton} B. 2013.
\newblock \textit{\apj} 777:136

\bibitem[{{Wood-Vasey} \& {Sokoloski}(2006)}]{Wood-Vasey06}
{Wood-Vasey} WM, {Sokoloski} JL. 2006.
\newblock \textit{\apjl} 645:L53--L56

\bibitem[{{Woods} \& {Gilfanov}(2013{\natexlab{a}})}]{Woods13b}
{Woods} TE, {Gilfanov} M. 2013{\natexlab{a}}.
\newblock \textit{ArXiv e-prints}

\bibitem[{{Woods} \& {Gilfanov}(2013{\natexlab{b}})}]{Woods13a}
{Woods} TE, {Gilfanov} M. 2013{\natexlab{b}}.
\newblock \textit{\mnras} 432:1640--1650

\bibitem[{{Woosley} \& {Kasen}(2011)}]{Woosley11}
{Woosley} SE, {Kasen} D. 2011.
\newblock \textit{\apj} 734:38

\bibitem[{{Woudt} et~al.(2009){Woudt}, {Steeghs}, {Karovska}, {Warner}, {Groot}
  et~al.}]{Woudt09}
{Woudt} PA, {Steeghs} D, {Karovska} M, {Warner} B, {Groot} PJ, et~al. 2009.
\newblock \textit{\apj} 706:738--746

\bibitem[{{Wu} et~al.(1993){Wu}, {Crenshaw}, {Fesen}, {Hamilton} \&
  {Sarazin}}]{Wu93}
{Wu} CC, {Crenshaw} DM, {Fesen} RA, {Hamilton} AJS, {Sarazin} CL. 1993.
\newblock \textit{\apj} 416:247

\bibitem[{{Xavier} et~al.(2013){Xavier}, {Gupta}, {Sako}, {D'Andrea}, {Frieman}
  et~al.}]{Xavier13}
{Xavier} HS, {Gupta} RR, {Sako} M, {D'Andrea} CB, {Frieman} JA, et~al. 2013.
\newblock \textit{\mnras} 434:1443--1459

\bibitem[{{Yamaguchi} et~al.(2008){Yamaguchi}, {Koyama}, {Katsuda}, {Nakajima},
  {Hughes} et~al.}]{Yamaguchi08}
{Yamaguchi} H, {Koyama} K, {Katsuda} S, {Nakajima} H, {Hughes} JP, et~al. 2008.
\newblock \textit{\pasj} 60:141

\bibitem[{{Yaron} et~al.(2005){Yaron}, {Prialnik}, {Shara} \&
  {Kovetz}}]{Yaron05}
{Yaron} O, {Prialnik} D, {Shara} MM, {Kovetz} A. 2005.
\newblock \textit{\apj} 623:398--410

\bibitem[{{Yoon} \& {Langer}(2004)}]{Yoon04}
{Yoon} SC, {Langer} N. 2004.
\newblock \textit{\aap} 419:623--644

\bibitem[{{Yoon} \& {Langer}(2005)}]{Yoon05}
{Yoon} SC, {Langer} N. 2005.
\newblock \textit{\aap} 435:967--985

\bibitem[{{Yungelson}(2005)}]{Yungelson05}
{Yungelson} LR. 2005.
\newblock In \textit{White dwarfs: cosmological and galactic probes}, eds.
  EM~{Sion}, S~{Vennes}, HL~{Shipman}, vol. 332 of \textit{Astrophysics and
  Space Science Library}

\bibitem[{{Yungelson}(2008)}]{Yungelson08}
{Yungelson} LR. 2008.
\newblock \textit{Astronomy Letters} 34:620--634

\bibitem[{{Yungelson}(2013)}]{Yungelson13}
{Yungelson} LR. 2013.
\newblock In \textit{IAU Symposium}, eds. R~{Di Stefano}, M~{Orio}, M~{Moe},
  vol. 281 of \textit{IAU Symposium}

\bibitem[{{Yungelson} \& {Livio}(2000)}]{Yungelson00}
{Yungelson} LR, {Livio} M. 2000.
\newblock \textit{\apj} 528:108--117

\bibitem[{{Zelaya} et~al.(2013){Zelaya}, {Quinn}, {Baade}, {Clocchiatti},
  {H{\"o}flich} et~al.}]{Zelaya13}
{Zelaya} P, {Quinn} JR, {Baade} D, {Clocchiatti} A, {H{\"o}flich} P, et~al.
  2013.
\newblock \textit{\aj} 145:27

\bibitem[{{Zheng} et~al.(2013){Zheng}, {Silverman}, {Filippenko}, {Kasen},
  {Nugent} et~al.}]{Zheng13}
{Zheng} W, {Silverman} JM, {Filippenko} AV, {Kasen} D, {Nugent} PE, et~al.
  2013.
\newblock \textit{\apjl} 778:L15

\bibitem[{{Zhu} et~al.(2013){Zhu}, {Chang}, {van Kerkwijk} \&
  {Wadsley}}]{Zhu13}
{Zhu} C, {Chang} P, {van Kerkwijk} MH, {Wadsley} J. 2013.
\newblock \textit{\apj} 767:164

\bibitem[{{Zhu}, {Blanton} \& {Moustakas}(2010)}]{Zhu10}
{Zhu} G, {Blanton} MR, {Moustakas} J. 2010.
\newblock \textit{\apj} 722:491--519

\bibitem[{{Zorotovic}, {Schreiber} \& {G{\"a}nsicke}(2011)}]{Zorotovic11}
{Zorotovic} M, {Schreiber} MR, {G{\"a}nsicke} BT. 2011.
\newblock \textit{\aap} 536:A42

\end{thebibliography}

\end{document}